\documentclass[11pt]{article}
\usepackage{mymacros}

\title{Krylov operator complexity in holographic CFTs:\\Smeared boundary reconstruction and the dual proper radial momentum}
\author[a]{Sergio E. Aguilar-Gutierrez,}
\author[b]{Hugo A. Camargo,}
\author[b]{Viktor Jahnke,}
\author[b,c]{Keun-Young Kim,}
\author[d]{and Mitsuhiro Nishida}
\affiliation[a]{Qubits and Spacetime Unit, Okinawa Institute of Science and Technology Graduate University,\\ 1919-1 Tancha, Onna, Okinawa 904 0495, Japan}
\affiliation[b]{Department of Physics and Photon Science, Gwangju Institute of Science and Technology,\\123 Cheomdan-gwagiro, Gwangju 61005, Republic of Korea}
\affiliation[c]{Research Center for Photon Science Technology,
Gwangju Institute of Science and Technology, Gwangju 61005, Korea}
\affiliation[d]{National Institute of Technology, Yuge College, Ehime 794-2593, Japan}
\emailAdd{sergio.ernesto.aguilar@gmail.com,  hugo.camargo@phys.ncts.ntu.edu.tw, v.jahnke@unesp.br, fortoe@gist.ac.kr, mnishida124@gmail.com }
\abstract{Motivated by bulk reconstruction of smeared boundary operators, we study the Krylov complexity of local and non-local primary CFT$_d$ operators from the local bulk-to-bulk propagator of a minimally-coupled massive scalar field in Rindler-AdS$_{d+1}$ space. We derive analytic and numerical evidence on how the degree of non-locality in the dual CFT$_d$ observable affects the evolution of Krylov complexity and the Lanczos coefficients. Curiously, the near-horizon limit matches with the same observable for conformally-coupled probe scalar fields inserted at the asymptotic boundary of AdS$_{d+1}$ space. Our results also show that the evolution of the growth rate of Krylov operator complexity in the CFT$_d$ takes the same form as to the proper radial momentum of a probe particle inside the bulk to a good approximation. The exact equality only occurs when the probe particle is inserted in the asymptotic boundary or in the horizon limit. Our results capture a prosperous interplay between Krylov complexity in the CFT, thermal ensembles at finite bulk locations and their role in the holographic dictionary. }

\begin{document}

\maketitle
\newpage

\section{Introduction}
A central challenge in AdS/CFT research is to define a precise and robust boundary theory quantity that effectively captures the growth dynamics of the black hole interior. This quantity has been proposed to embody a notion of quantum complexity in the boundary theory~\cite{Susskind:2014rva}\footnote{See~\cite{Susskind:2018pmk} for a pedagogical introduction to this topic and \cite{Chapman:2021jbh,Baiguera:2025dkc} for recent reviews.}. Although numerous measures on the gravity side have been proposed to measure the growth of the wormhole interior \cite{Belin:2021bga, Belin:2022xmt}, finding the corresponding notion in the conformal field theory (CFT) or boundary theory remains an open problem. Recent progress in the study of the double-scaled Sachdev--Ye--Kitaev (SYK) model \cite{Berkooz:2018qkz,Berkooz:2018jqr} suggests that the bulk geometric length is dual to the Krylov complexity of the boundary theory \cite{Rabinovici:2023yex,Ambrosini:2024sre,Heller:2024ldz,Xu:2024gfm,Aguilar-Gutierrez:2025pqp}. Moreover, increasing evidence indicates that the growth rate of Krylov complexity for thermal states perturbed by local operators in holographic CFTs corresponds to the proper radial momentum of a particle in AdS space ~\cite{Caputa:2024sux,Caputa:2025dep} (see related developments in \cite{Fan:2024iop,He:2024pox}).\footnote{For earlier work in this direction within the context of the SYK model and Jackiw--Teitelboim gravity, see~\cite{Jian:2020qpp}. {{Other arguments that this relation should hold for more generic notions of quantum complexity appears in \cite{Susskind:2014jwa,Lin:2019kpf,Susskind:2018tei,Susskind:2019ddc,Magan:2018nmu,Susskind:2020gnl}; which was more rigorously studied for the CV conjecture in \cite{Barbon:2020olv,Barbon:2019tuq,Barbon:2019wsy}.}}} Given these promising developments, it is now crucial to determine whether Krylov complexity can serve as a robust and reliable new entry in the AdS/CFT dictionary, extending beyond the scope of the double-scaled SYK model to more general holographic scenarios.

Krylov operator complexity provides a framework for quantifying operator growth in quantum systems. Under time evolution, Heisenberg operators do not explore the entire Hilbert space of operators but are constrained to a smaller subspace called the Krylov subspace. This measure captures how an initially local operator spreads within the Krylov basis, becoming progressively more non-local and complex. The dynamics of this process can be mapped onto a one-dimensional particle hopping on a chain, where the Krylov operator complexity corresponds to the particle's average position on the chain. Recently, this concept has received significant attention as a novel tool for characterizing quantum chaos. It bridges early-time diagnostics of chaos, such as out-of-time-order correlators (OTOCs) \cite{aleiner1996divergence,Rozenbaum:2016mmv}, with late-time measures like spectral statistics \cite{Dyson:1962es,Dyson:1962oir,wigner1993characteristic}. For a detailed review, including an introduction to Krylov state complexity (also known as spread complexity), we refer the reader to \cite{Nandy:2024htc}. Since quantum chaos has played an important role in recent developments in AdS/CFT, it is valuable to explore Krylov complexity in this context, as it offers a new perspective on it that can potentially give us valuable lessons about the inner-working mechanisms of AdS/CFT. However, care must be taken when discussing chaos in quantum field theories (QFTs), as there are subtleties and limitations to the applicability of Krylov complexity in this setting. See for instance~\cite{Dymarsky:2021bjq, Avdoshkin:2022xuw, Camargo:2022rnt, Kundu:2023hbk}.

In this work, we investigate the Krylov operator complexity of bulk fields in AdS, focusing on a minimally coupled scalar field in a Rindler-AdS$_{d+1}$ geometry. This particular foliation of AdS space admits a dual boundary description in terms of a $d$-dimensional CFT on hyperbolic space~\cite{Czech:2012be}. It provides a convenient setup to model the presence of a horizon while preserving the analytic tractability of pure AdS geometries, including, for instance, explicit knowledge of bulk-to-bulk and bulk-to-boundary propagators, which are essential for the calculations in this paper.
Specifically, we consider the bulk-to-bulk propagator of the scalar field as the autocorrelation function $\phi_0(t)$ in the Lanczos algorithm. Our study is aimed at exploring the proposed equivalence between the proper radial momentum of a bulk particle and the rate of change of Krylov complexity associated with boundary operators. 

The proposal that the rate of change of Krylov complexity of boundary operators corresponds to the proper momentum of a bulk particle was initially investigated in the context of AdS$_3$/CFT$_2$~\cite{Caputa:2024sux}. Although the proper momentum can be computed for higher-dimensional backgrounds, as demonstrated in~\cite{Fan:2024iop}, the corresponding CFT results have not yet been derived. In this work, we explicitly show that this proposal extends to higher-dimensional scenarios, specifically within the Rindler-AdS$_{d+1}$/CFT$_d$ correspondence, obtaining a precise match between calculations on both sides of the duality. 

Furthermore, we show that the relation between Krylov complexity and proper momentum of an infalling particle also holds exactly for bulk-fields in the near-horizon region, and holds approximately for bulk fields in any AdS radial location.  Additionally, investigating the Krylov complexity of local bulk fields is particularly interesting, as these fields correspond to non-local boundary operators, as established in the seminal work by Hamilton, Kabat, Lifschytz, and Lowe (HKLL)~\cite{Hamilton:2005ju,Hamilton:2006fh,Hamilton:2006az,Hamilton:2007wj}, which we review in App. \ref{app:appHKLL}. In this context, the bulk radial coordinate is interpreted as a scale that controls the degree of non-locality of the dual CFT observables\footnote{This consideration has also played an important role in other contexts, such as in holographic tensor networks, e.g. \cite{Pastawski:2015qua}, and, crucially, in the algebraic approach to quantum gravity observables, where smeared fields are constructed within the Gelfand--Naimark--Segal (GNS) formalism~\cite{Witten:2018zxz}.}.

\paragraph{Summary of Our Findings} 
We compute the Krylov complexity for bulk fields at an arbitrary AdS radial position $r$ in a Rindler-AdS$_{d+1}$ geometry.\footnote{As a side calculation, we also compare our results with those of T$\overline{\text{T}}$ deformed CFTs, dual to finite cutoff AdS space in App. \ref{app:figs}. The Krylov complexity of non-local primary CFT$_d$ operators shares a noteworthy similar structure to primaries in T$\overline{\text{T}}$-deformed CFTs.} We obtain analytical results for the near-boundary region ($r \rightarrow \infty$), which, as expected, match the corresponding boundary results for a conformal field theory in hyperbolic space. Similarly, for the near-horizon region ($r \rightarrow 1$), our results coincide with conformal field theory operators with scaling dimension $\Delta = (d-1)/2$. Interestingly, this scaling dimension corresponds to a conformally coupled scalar bulk field in AdS$_{d+1}$ with mass $m^2 = (1-d^2)/4$. For intermediate values of the AdS radial coordinate, $1 < r < \infty$, we compute the Krylov complexity numerically.

One of the main findings in this work is the relation between the growth rate of Krylov complexity of boundary operators and the proper momentum of a bulk particle remains valid in higher dimensions. Moreover, a similar relation exists for bulk fields, although it can only be approximately satisfied at for arbitrary finite radial locations due to breaking of conformal invariance in the corresponding two-point function. This suggests this duality can be extended for non-local operators {only approximately. However, we highlight that our results indicate the holographic dictionary between the momentum of the probe particle and the the rate of growth of Krylov complexity is not a simple proportionality relation when we consider an arbitrary radial location and locally probe the bulk matter fields.

Interestingly, although the duality between the growth rate of Krylov complexity and the proper momentum of a bulk particle holds only approximately at finite radial positions in AdS, it becomes exact again near the horizon. This behavior may point to an emergent conformal symmetry governing the bulk-to-bulk two-point function in this region.

\paragraph{Outline} In Sec. \ref{sec:Krylov bulk bulk} we investigate the evolution of Krylov complexity of holographic CFTs from the bulk-to-bulk propagator in Rindler-AdS space for local operators in the bulk, corresponding to non-local observables in the boundary theory. In Sec. \ref{sec:proper momentum}, we match the rate of growth of Krylov operator complexity for the bulk propagator with the proper radial momentum of an infalling bulk particle, near the boundary of AdS as well as in the near-horizon region. Lastly, in Sec. \ref{sec:dis} we comment further on the connection of our results and recent discussions on the relation between Krylov complexity in CFTs and holographic complexity. Additionally, we provide a summary of our results and point out directions for future developments, including the switchback effect in Krylov complexity and holographic complexity. We provide several appendices with technical details. In particular, we discuss the similarities between our results with finite cutoff AdS space dual to T$\overline{\text{T}}$-deformed CFTs in App. \ref{app:figs}.

\section{Krylov complexity from bulk-to-bulk propagators}\label{sec:Krylov bulk bulk}
In this section, we evaluate the Krylov complexity of local bulk fields (corresponding to smeared CFT operators) considering a free scalar field in Rindler-AdS$_{d+1}$ spacetime, which is the main setting for the rest of the work.

\subsection{Gravity set-up}
To begin with, we review the framework of Rindler-AdS/CFT~\cite{Czech:2012be}, which we employ in this work. The essential point of this framework is that a pure AdS geometry in arbitrary number of dimensions can be seen as a maximally extended `topological' black hole geometry, which is dual to a thermofield double (TFD) state constructed by entangling two copies of a CFT in hyperbolic space. This setup allows for great analytic control for many calculations, both in the gravity side and in the field theory side of the duality.

We start by considering Einstein gravity
\be
S=\frac{1}{16 \pi G_N} \int \textrm{d}^{d+1}x
\left[ R + \frac{d(d-1)}{\ell^2} \right]\,,
\ee
where $\ell$ denotes the AdS length scale and $d$ is the number of spatial dimensions. As a solution to the equations of motion following from the above action, we consider a pure AdS$_{d+1}$ geometry, which can be defined as the universal cover of the hyperboloid
\be
-T_1^2-T_2^2+X_1^2+\cdots +X_d^2=-\ell^2\,,
\ee
embedded in a space with ambient metric given by
\be
\textrm{d}s^2=-\textrm{d}T_1^2-\textrm{d}T_2^2+\textrm{d}X_1^2+ \cdots + \textrm{d}X_d^2\,.
\label{eq-metric-emb}
\ee
In what follows, we set $\ell=1$ for simplicity.

In global coordinates, one chooses the following parametrization 
\begin{align}
T_1=\sqrt{\rho^2+1}\cos \tau\,,\,\,\, T_2=\sqrt{\rho^2+1} \sin \tau\,,\,\,\, X_1^2+\cdots + X_d^2=\rho^2\,, 
\end{align}
in terms of which the metric becomes
\be
\textrm{d}s^2= - (1+\rho^2) \textrm{d}\tau^2 + \frac{\textrm{d}\rho^2}{1+\rho^2}+\rho^2 \textrm{d}\Omega_{d-1}^2\,,
\ee
where $\tau \in (-\infty,\infty),\,\, \rho \in [0,\infty)$. The AdS boundary is located at $\rho = \infty$ and the dual field theory description is given in terms of a CFT in $\mathbb{R} \times \mathbb{S}^{d-1}$. In particular, the pure global AdS geometry describes the vacuum state $| 0 \rangle_{\text{global}}$ of such a CFT.
 
Given a constant-time slice of the geometry, we can divide the spatial boundary into two hemispheres $B_L$ and $B_R$. The vacuum state of the CFT defined on the full sphere can be written as a TFD state constructed out of the states of the CFTs defined on the hemispheres. Finally, a conformal transformations can be used to map the domain of dependence of $B_L$ and $B_R$ to $\mathbb{R} \times \mathbb{H}_{d-1}$.
This implies that vacuum state $| 0 \rangle_{\text{global}}$ can be written as a TFD state of two hyperbolic space CFTs \cite{Czech:2012be}
\be
| 0 \rangle_{\text{global}} = Z^{-1/2} \sum_n e^{-\pi E_n}  | E_n \rangle_L  \otimes |E_n\rangle_R\,,
\ee
where $E_L$ and $E_R$ label the energy eigenstates of CFT$_L$ and CFT$_R$.

The dual gravitational picture for a single CFT in hyperbolic space is given in terms of Rindler coordinates in AdS, which can be obtained by parametrizing the embedding coordinates as follows
\begin{align}
\begin{split}
T_1&=\sqrt{r^2-1} \sinh t\,,~~~~~T_2= r \cosh \chi\,,~~~~X_d= \sqrt{r^2-1} \cosh t\,,~~~~X_1^2+\cdots +X_{d-1}^2 = r^2 \sinh^2 \chi\,,
\label{eq-emb-rindler}
\end{split}
\end{align}
where $t \in (-\infty, \infty)$, and $r, \chi \in [0,\infty)$. In terms of these coordinates, the metric becomes
\be
\textrm{d}s^2=-\left(r^2-1\right)\textrm{d}t^2+\frac{\textrm{d}r^2}{r^2-1}+r^2 \textrm{d}\mathbb{H}_{d-1}^2\,,
\label{eq-metric-Rindler}
\ee
where $\textrm{d}\mathbb{H}_{d-1}^2=\textrm{d}\chi^2+\sinh^2\chi\, \textrm{d} \Omega_{d-2}^2$ denotes the unity metric in a $(d-1)-$dimensional hyperbolic space, $\mathbb{H}_{d-1}$, with $\textrm{d} \Omega_{d-2}^2$ being the unity metric on the $(d-2)-$sphere. The AdS boundary is located at $r=\infty$, and the geometry has a horizon at $r=1$, which leads to a non-zero Hawking temperature given by $T=\frac{1}{\beta}=\frac{1}{2 \pi}$. The coordinates $(t,r,{\bf x})$, where ${\bf x} \in \mathbb{H}_{d-1}$, describe an accelerating observer in AdS, and only cover one of the two Rindler wedges in AdS, which are shown in light gray in Figure \ref{fig-rindlerwedge}. 

\begin{figure}
\centering
\includegraphics[width=6cm]{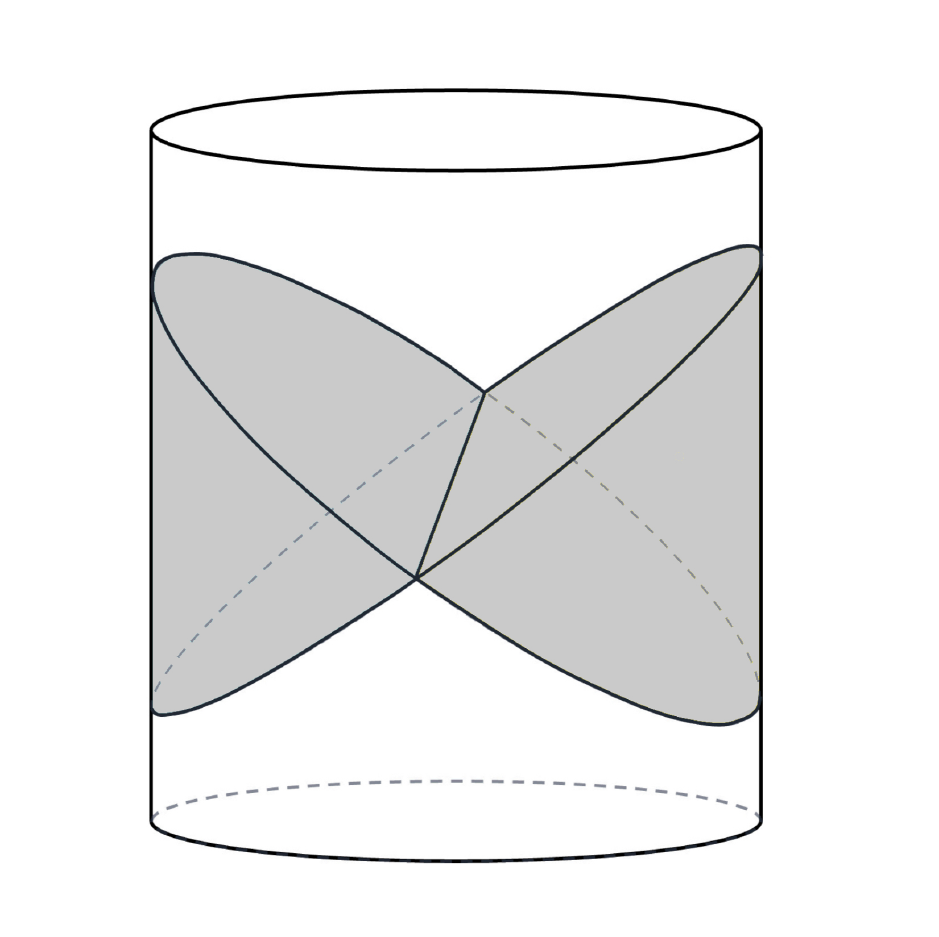}
\caption{Rindler wedges (shown in gray) of global AdS (solid cylinder)}
\label{fig-rindlerwedge}
\end{figure}

The maximally extended geometry of this hyperbolic foliation of AdS can be obtained by using the following embedding coordinates involving Kruskal-Szekeres coordinates $(U,V)$:
\begin{align}
    \begin{split}
T_1&=\frac{U+V}{1+U V} \,,~~~~T_2= \frac{1-U V}{1+U V} \cosh \chi\,,~~~~X_d= \frac{U-V}{1+U V}\,,~~~~X_1^2+\cdots +X_{d-1}^2 = \left( \frac{1-U V}{1+U V} \right)^2 \sinh^2 \chi\,,
\label{eq-emb-kruskal}        
    \end{split}
\end{align}
Substituting the above coordinates into the embedding space metric (\ref{eq-metric-emb}), one obtains
\be \label{eq-metric-kruskal}
\textrm{d} s^2 = -\frac{4 \textrm{d}U \,\textrm{d}V}{(1+U V)^2}+\left( \frac{1-U V}{1+ U V} \right)^2 \textrm{d}\mathbb{H}_{d-1}^2\,.
\ee
In Kruskal-Szekeres coordinates $(U,V,{\bf x})$, the AdS boundary is located at $UV=-1$, and there is a coordinate singularity at $UV=1$. The geometry contains two horizons, located at $U=0$ and $V=0$. The corresponding Penrose diagram is shown in Figure~\ref{fig:Rindler}.

\subsection{Scalar Bulk-to-Bulk Propagator in AdS-Rindler} 
\label{sec:KcompBulkAdS}

We consider a minimally-coupled scalar field $\phi$ of mass $m$ propagating in the background~\eqref{eq-metric-Rindler} with action given by
\begin{equation}
\label{eq:saction}
S_{\textrm{scalar}}=-\frac{1}{2}\int \textrm{d}^{d+1}x \sqrt{-g} \left(g^{\mu \nu} \partial_{\mu} \phi \partial_{\nu} \phi+m^2 \phi^2 \right)\,.
\end{equation}
The corresponding equation of motion is
\begin{equation} 
\label{eq:eom-general}
\left(\Box_{\textrm{AdS-R}}-m^{2}\right)\phi=\frac{1}{\sqrt{-g}}\partial_{\mu} (\sqrt{-g}g^{\mu \nu} \partial_{\nu} \phi)-m^2 \phi=0\,,
\end{equation}
where the mass of the field is given by $m^{2}=\Delta(\Delta-d)$ {and $\Delta$ is the conformal dimension of the dual operator at the boundary.}

\subsection{Exact bulk-to-bulk propagator}
\label{subsec:BulktoBulkProp}
We are interested in the Lanczos coefficients and Krylov complexity associated with bulk massive scalar fields $\phi$ obtained by considering the finite-$r$ autocorrelation function $C(t,r)$ obtained from bulk-to-bulk propagators. In particular, we are interested in understanding how the CFT result is deformed by the finite scale set by $r$.

Our approach with the bulk propagator is similar to the computation for holographic CFTs in~\cite{Dymarsky:2021bjq} since the bulk-to-bulk propagator~\eqref{eq:bbp} is a solution of bulk equations of motion~\eqref{eq:eom-general}
\begin{equation}
    \label{eq:bbpeom}
    \left(\Box_{\textrm{AdS-R}}-m^{2}\right)G_{\Delta}(t,r,\mathbf{x},t',r',\mathbf{x}')=\frac{\delta(t-t')\delta(r-r')\delta^{(d-1)}(\bf{x-x'})}{\sqrt{-g}}~.
\end{equation}
However, the result in~\cite{Dymarsky:2021bjq} is for $r\to\infty$ as the usual computation of the boundary two-point function in AdS/CFT.

The bulk-to-bulk propagator between two points $P=(t,\mathbf{x},r)$ and $P'=(t',\mathbf{x}',r')$ in AdS-Rindler coordinates is given by~\cite{Ahn:2020csv}
\begin{equation}
    \label{eq:bbp}
    G_{\Delta}(P;P')=\langle \phi(t,{\bf x},r) \phi(t',{\bf x'},r') \rangle=c_{\Delta} \cdot\xi^{\Delta}\,\cdot{}_2F_1\left( \frac{\Delta}{2},\frac{\Delta+1}{2};\Delta+1-\frac{d}{2}; \xi^2\right)\,,
\end{equation}
where $\Delta$ is the conformal dimension of the scalar operator, $c_{\Delta}$ is a normalization constant that does not depend on the coordinates, ${}_{2}F_{1}(a,b;c;z)$ is the ordinary hypergeometric function and where $\xi$ is the chordal distance between $P$ and $P'$ given by
\begin{equation}
    \label{eq:xi}
    \begin{split}
    \xi^{-1}&:=\cosh \left(\textrm{dist}(P,P')\right)= - \sqrt{r^2-1} \sqrt{r'^2-1} \cosh(t-t')+r r' \cosh \left(\textrm{dist}({\bf x,x'})\right)\,.
    \end{split}
\end{equation}
Here, $\textrm{dist}(P,P')$ is the geodesic distance between $P$ and $P'$ and $\textrm{dist}({\bf x,x'})$ is the geodesic distance between $\bf x$ and $\bf x'$ in $\mathbb{H}_{d-1}$. 

From bulk-to-bulk correlation functions, one can obtain boundary-boundary correlation functions using the extrapolated dictionary~\cite{Harlow:2011ke}
\begin{equation}
    \langle \mathcal{O}(t_1,{\bf x_1})...\mathcal{O}(t_n,{\bf x_n})\rangle = \lim_{r \rightarrow \infty} r^{n \Delta} \langle \phi(t_1,{\bf x_1},r)...\phi(t_n,{\bf x_n},r)\rangle\,.
\end{equation}
In particular, for $n=2$, the formula above implies that the bulk-to-bulk propagator reduces to a boundary-to-boundary two-point function as one moves the radial location of the bulk point to the asymptotic boundary.

We begin by analyzing bulk-to-bulk propagators between points of the form $P = (t, {\bf x}, r)$ and $P' = (i \epsilon, {\bf x}', r)$, for generic values of the parameter $\epsilon$. We derive the corresponding Lanczos coefficients and Krylov complexity in two limiting regimes: the near-boundary case ($r \rightarrow \infty$) and the near-horizon case ($r \rightarrow 1$). Next, we explore Krylov complexity numerically for generic values of $r$. In this case, we adopt the canonical choice of the Wightman inner product for operator Krylov complexity, setting $\epsilon = \beta/2$, which corresponds to two-sided correlators. To probe the relation between Krylov complexity and the momentum of an infalling particle, we also consider the regime of small $\epsilon$, where this connection becomes more transparent.

As mentioned above, we first choose the points $P$ and $P'$ to be of the form 
\begin{align}
\label{eq:TwoPoints}
r=r', \;\;\; t'=0+i\epsilon, \;\;\; \textrm{dist}({\bf x,x'})=0~.
\end{align}
In this case, the chordal distance~\eqref{eq:xi} between the two points reduces to
\begin{align}
\label{eq:xi2}
\xi^{-1}=-(r^2-1)\cosh(t-i \epsilon)+r^2~.
\end{align}
Starting from the bulk-to-bulk propagator~\eqref{eq:bbp}, we define an autocorrelation function for points $P$ and $P'$~\eqref{eq:TwoPoints} separated by a chordal distance~\eqref{eq:xi2} as follows
\begin{align}
\label{eq:autocorrelation function main}
C(t,r):=N(\Delta, d, r)\, \xi^{\Delta} \,{}_2F_1\left( \frac{\Delta}{2},\frac{\Delta+1}{2};\Delta+1-\frac{d}{2}; \xi^2\right)~,
\end{align}
where $N(\Delta, d, r)$ is an $r$-dependent normalization constant determined by the normalization condition $C(0,r)=1$.
\subsection{Near boundary case \texorpdfstring{$r \rightarrow \infty$}{}}
In the limit, $r \rightarrow \infty$, the inverse cordal distance takes the form
\begin{equation}
    \xi^{-1}=-(r^2-1)\cosh(t-i \epsilon)+r^2 \approx r^2 \left(1-\cosh(t-i \epsilon) \right) =-2 r^2 \sinh^2\left(\frac{t-i \epsilon}{2}\right) 
\end{equation}
Then, one can obtain a boundary-boundary two-point function by taking the limit
\begin{equation} \label{eq-NearBdryAutoCorrelation}
   \lim_{r \rightarrow \infty} C(t,r) =  \frac{\sin(\epsilon/2)^{2\Delta}}{\left[ - \sinh^2 \left( \frac{t-i \epsilon}{2}\right) \right]^{\Delta}}
\end{equation}
where we define $N(\Delta, d, r)$ such that that $\lim_{r \rightarrow \infty}\frac{N(\Delta,d,r)}{(2r^2)^{\Delta}}$ $=\sin(\epsilon/2)^{2\Delta}$ and where we used that $_2F_1(a,b$ $;c;0)$ $=1$. The inverse of the chordal distance $\xi^{-1}$~\eqref{eq:xi2} diverges in the limit $r\to\infty$, which corresponds to the divergence of the geodesic length at the AdS boundary. The $r$-dependent normalization constant $N(\Delta, d, r)$ may therefore be interpreted as the introduction of a regularized geodesic length in the geodesic approximation of the two-point function (see~\cite{Aparicio:2011zy} for example).

\subsection{Near horizon case \texorpdfstring{$r \rightarrow 1$}{}}
In this case, the factor of $\xi^\Delta$ does not diverge, but there is a divergence coming from the hypergeometric function. We first use the identity:
\begin{align} \label{eq-hypergeometricIdentity}
{}_2F_1\left(a,b;c;z\right)=&\frac{\Gamma(c)\Gamma(c-a-b)}{\Gamma(c-a)\Gamma(c-b)}{}_2F_1\left(a,b;a+b-c+1;1-z\right)\notag\\
&+\frac{\Gamma(c)\Gamma(a+b-c)}{\Gamma(a)\Gamma(b)}(1-z)^{c-a-b}{}_2F_1\left(c-a,c-b;c-a-b+1;1-z\right),
\end{align}
with $z=\xi^2$, $a=\Delta/2$, $b=(\Delta+1)/2$ and $c=\Delta+1-d/2$. We can see that the second term in the expansion diverges as $z\rightarrow 1$ because of the term $(1-z)^{c-a-b}=(1-\xi^2)^{-(d-1)/2}$ for $d>1$, because in this case the cordal distance behaves as
\begin{equation}
    \xi^2=1+8\, \delta \,r \sinh^2\left( \frac{t-i \epsilon}{2}\right)
\end{equation}
where we wrote $r=1+\delta r$ and expanded Eq.~\eqref{eq:xi2} squared for small values of $\delta r$. Therefore, one obtains
\begin{equation} \label{eq-AutoCorrelationHorizon}
    \lim_{\delta r \rightarrow 0} C(t,1+\delta r)= \frac{\sin(\epsilon/2)^{d-1}}{\left[- \sinh^2 \left(\frac{t-i\epsilon}{2} \right) \right]^{\frac{d-1}{2}}}
\end{equation}
where we define $N(\Delta, d, r)$ such that 
\begin{equation}
    \lim_{\delta r \rightarrow 0} = \frac{\Gamma(\Delta+1-d/2)\Gamma(\frac{d-1}{2})}{\Gamma(\frac{\Delta}{2}) \Gamma(\frac{\Delta+1}{2})}\frac{N(\Delta,d,1+\delta r)}{\delta r^{\frac{d-1}{2}}}=\sin(\epsilon/2)^{d-1}\,.
\end{equation}
\subsection*{Lanczos coefficients and Krylov complexity for general $\epsilon$}
For an autocorrelation function of the form 
\begin{equation}\label{eq:toda epsilon auto}
    C(t)=\left[ \frac{\sin (\alpha \epsilon)}{ \sin \left(\alpha( t-i \epsilon)\right) }\right]^{2\eta}
\end{equation}
The corresponding Lanczos functions obtained via the recursion method (see Appendix~\ref{app:app1} for details) or the Toda hierarchy technique\footnote{See Appendix A of \cite{Kundu:2023hbk} for a concise review of the Toda hierarchy technique.} are given by\footnote{{{Given that the correlation function \eqref{eq:toda epsilon auto} is not an even function in time, it follows that $a_n\neq0$ in the Lanczos algorithm reviewed in App. \ref{app:app1}.}}}
\begin{equation} \label{eq:LanczosGeneralEpsilon}
    a_n=-2 \alpha (n+\eta)\,\cot \left(\alpha \epsilon \right)\,,~~~~~~~~~~b_n^2=\frac{\alpha^2 n(n-1+2\eta)}{\sin^2\left(\alpha \epsilon\right)}
\end{equation}
and the Krylov complexity is given by~\cite{Kundu:2023hbk, Caputa:2023vyr}\footnote{Note that the cutoff $\epsilon$ introduced here differs from the one in \cite{Caputa:2023vyr} by a minus sign and a factor of two, namely, $\epsilon_\text{here}=-2 \epsilon_\text{there}$.} 
\begin{equation} \label{eq:KrylovGeneralEpsilon}
    K_{\epsilon}(t)=2\eta \, \frac{\sinh^2(\alpha t)}{\sin^2(\alpha \epsilon)}~.
\end{equation}
To apply the above formula for our autocorrelation function near the boundary or near the horizon, we just need to set $\alpha=1/2$, and $\eta=\Delta$ for the near boundary case and $\eta=(d-1)/2$ for the near horizon case. We then obtain:
\begin{equation} 
\label{eq:KCLimitsEpsilon}
  K_{\epsilon}(t)= 
\begin{cases}
2\Delta\, \frac{\sinh^2(t/2)}{\sin^2(\epsilon/2)}\,, \,\,\text{for}\,\,r\rightarrow \infty~,\\
(d-1)\, \frac{\sinh^2(t/2)}{\sin^2(\epsilon/2)}\,,\,\,\text{for}\,\, r\to1~,
\end{cases}  
\end{equation}

\subsection{Pole structure in autocorrelation function}

The pole structure of the autocorrelation function $C(t,r)$~\eqref{eq:autocorrelation function main} near $t = i\epsilon$ can be categorized into two distinct cases: the near-boundary behavior ($r \to \infty$) and the finite-$r$ behavior.

We begin with the near-boundary case. Expanding ~\eqref{eq-NearBdryAutoCorrelation} around $t = i\epsilon + \delta t$, we obtain
\begin{align}
\lim_{r\to\infty} C(i\epsilon + \delta t, r) \approx \sin(\epsilon/2)^{2\Delta} \left(i \frac{\delta t}{2} \right)^{-2\Delta},
\end{align}
indicating a pole of order $2\Delta$ as $\delta t \to 0$.

In the case of finite $r$, the analysis is slightly more subtle. The function $\xi$, defined in~\eqref{eq:xi2}, approaches unity at $t = i\epsilon$, with the following expansion:
\begin{align}
\xi^{-1}(t = i\epsilon + \delta t) \sim 1 - \frac{1}{2}(r^2 - 1)\delta t^2.
\end{align}
At this point, the prefactor $\xi^{\Delta}$ in~\eqref{eq:autocorrelation function main} remains regular, but the hypergeometric function develops a divergence as $\xi^2 \to 1$. This divergence becomes manifest when applying the identity~\eqref{eq-hypergeometricIdentity}, in which the second term is singular at $z = 1$.

Consequently, the behavior of $C(t,r)$ near $t = i\epsilon$ for finite $r$ is governed by the leading divergence of the hypergeometric function:
\begin{align}
C(i\epsilon+\delta t,r)\approx N(\Delta, d, r)\frac{\Gamma(\Delta+1-\frac{d}{2})\Gamma(\frac{d-1}{2})}{\Gamma(\frac{\Delta}{2})\Gamma(\frac{\Delta+1}{2})}\frac{1}{\left((1-r^2)\delta t^2\right)^{(d-1)/2}},\label{polestructurefiniter}
\end{align}
exhibiting a pole of order $d - 1$. 

In summary, the behavior of the autocorrelation function near the pole at $t = i\epsilon$ is
\begin{equation}
C(i\epsilon + \delta t, r) \sim
\begin{cases}
\delta t^{-2\Delta} & \text{for } r \to \infty\,, \\
\delta t^{-(d-1)} & \text{for finite } r\,.
\end{cases}
\end{equation}

\subsection{Two-sided correlators }\label{sec:eps_pi_twosidedcorr}
Now we consider the case in which $\epsilon = \pi$, which corresponds to a two-sided correlator, where $P$ and $P'$ are located in opposite Rindler wedges. See Fig.~\ref{fig:Rindler} for details.
\begin{figure}
    \centering
    \includegraphics[width=7cm]{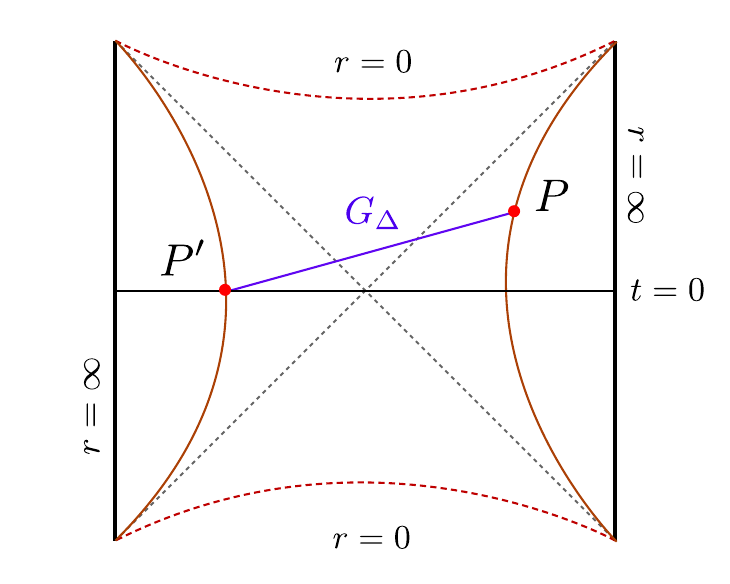}
    \caption{Penrose diagram of the Rindler-AdS geometry representing the location of the two points $P=(t,r,\mathbf{x})$ and $P'=(i \pi,r,\mathbf{x}')$ such that $\textrm{dist}({\bf x,x'})=0$.}
    \label{fig:Rindler}
\end{figure}
In this case, the chordal distance~\eqref{eq:xi} between the two points reduces to
\begin{align}
\label{eq:xi2_eps_pi}
\xi^{-1}=(r^2-1)\cosh(t)+r^2~.
\end{align}

Starting from the bulk-to-bulk propagator~\eqref{eq:bbp}, we define an autocorrelation function for points $P$ and $P'$~\eqref{eq:TwoPoints} separated by a chordal distance~\eqref{eq:xi2} as follows
\begin{align}
\label{eq:afbbp_eps_pi}
C(t,r):=N(\Delta, d, r)\cdot\xi^{\Delta}\cdot\,{}_2F_1\left( \frac{\Delta}{2},\frac{\Delta+1}{2};\Delta+1-\frac{d}{2}; \xi^2\right)~,
\end{align}
where $N(\Delta, d, r)$ is an $r$-dependent normalization constant determined by the normalization condition $C(0,r)=1$. 

The near-boundary and near-horizon limit for \eqref{eq:afbbp_eps_pi} are given by\footnote{Note there is no $\epsilon$ dependence in these limits.}
\begin{equation} \label{eq:AutoCorrelationFunctionBDRYandHORIZON}
   C(t,r)= \begin{cases}
        \cosh(t/2)^{2\Delta}\,,~~\text{for}~~r \rightarrow \infty\,,\\
        \cosh(t/2)^{d-1}\,,~~\text{for}~~r \rightarrow 1\,.
    \end{cases}
\end{equation}
These limits can be obtained from \eqref{eq-NearBdryAutoCorrelation} and ~\eqref{eq-AutoCorrelationHorizon} by setting $\epsilon=\pi$.
Thus,~\eqref{eq:afbbp_eps_pi} with finite $r$ can be interpreted as a deformed autocorrelation function from the usual CFT autocorrelation function, as expected. Note that this expression is valid for CFTs in $(d-1)$-dimensional hyperbolic space $\mathbb{H}_{d-1}$, and it should be contrasted with the expression derived for $d$-dimensional free scalar fields in flat space in~\cite{Dymarsky:2021bjq}. However, the location of the poles at $t=\pm i\beta/2$ in the Wightman two-point correlation of the form $\langle \mathcal{O}(t+i \beta/2) \mathcal{O}(0)\rangle$ is universal for any quantum field theory.

In contrast, \cite{Dymarsky:2021bjq} observed that the same characteristics of exponential growth in Krylov complexity for integrable CFTs can be reproduced from thermal Wightman two-point functions due to the universal pole structure for QFTs. In comparison, our study involves not only to the universal properties for local CFTs but also including non-local ones.

\subsubsection{Lanczos coefficients and Krylov complexity}\label{ssec:Lanczos and complexity}

The Lanczos coefficients for an autocorrelation function of the form \eqref{eq:AutoCorrelationFunctionBDRYandHORIZON} are given by \eqref{eq:LanczosGeneralEpsilon} for $\epsilon =\pi$, $\alpha=1/2$ and $\eta=\Delta$ for the near-boudary case and $\eta=(d-1)/2$ for the near-horizon case:
\begin{equation} 
\label{eq:bnLimits}
  b_{n}= 
\begin{cases}
\frac{1}{2}\sqrt{n(n-1+2\Delta)}\,, \,\,\text{for}\,\,r\rightarrow \infty~,\\
\frac{1}{2}\sqrt{n(n-1+d-1)}\,,\,\,\text{for}\,\, r\to1~.
\end{cases}  
\end{equation}
Thus, we expect to see a transition of $b_n$ from $\frac{1}{2}\sqrt{n(n-1+2\Delta)}$ to $\frac{1}{2}\sqrt{n(n-1+(d-1))}$ by varying $r$ from the near-boundary ($r\to\infty$) to the near-horizon ($r\to1$) limits. Since the order of the pole for finite $r$ is $d-1$, $b_n$ for finite $r$ would approach $\frac{1}{2}\sqrt{n(n-1+(d-1))}$ as $n$ increases. The corresponding Krylov complexity $K(t)$ (obtained from \eqref{eq:KrylovGeneralEpsilon} for $\epsilon=\pi$) is given by
\begin{equation} 
\label{eq:KCLimits}
  K(t)= 
\begin{cases}
2\Delta\sinh^2(t/2)\,, \,\,\text{for}\,\,r\rightarrow \infty~,\\
(d-1)\sinh^2(t/2)\,,\,\,\text{for}\,\, r\to1~,
\end{cases}  
\end{equation}
The results for $r \rightarrow \infty$ can also be derived from the boundary perspective in terms of a CFT in hyperbolic space. We review such a derivation in the Appendix \ref{app:KcompHyperbolic}.

For finite $r$, we expect $K(t)$ to interpolate between the two cases in \eqref{eq:KCLimits} as $r$ varies from the near-boundary to the near-horizon limits. To illustrate the behavior of the Lanczos coefficients and Krylov complexity obtained for finite $r$, there are two interesting regimes that we can consider in the parameter space $\lbrace \Delta,d\rbrace$ that are consistent with unitarity ($\Delta \geq (d-2)/2$) and the Breitenlohner--Freedman (BF) ($m^2\geq -d^2/4$) bounds. From~\eqref{eq:KCLimits}, it can be seen that these regimes are: $2\Delta \in (d-2,d-1]$ and $2\Delta \in (d-1,\infty)$.\footnote{Recall that $_{2}F_{1}(a,b;c;z)$ does not exist when $c=0,-1,-2,\ldots$. Since $c=(2\Delta+2-d)/2$, then the autocorrelation function is not defined when $2\Delta=d-2$.} In Appendix~\ref{app:figsbnKtfiniter} we present a detailed analysis of the Lanczos coefficients and their associated Krylov complexity in both regimes.

\begin{figure}
         \centering
         \includegraphics[width=0.5\textwidth]{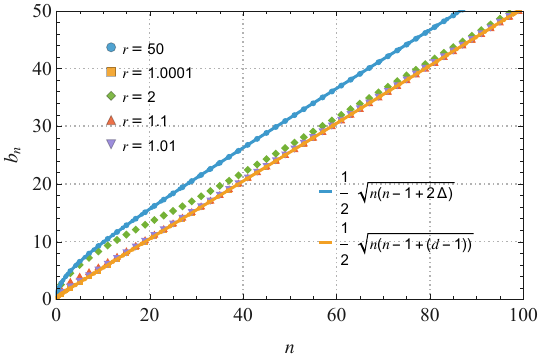}
       \caption{Lanczos coefficient $b_n$ for $C(t,r)$ (\ref{eq:afbbp_eps_pi}) with $\Delta=15$ and $d=4$ for different values of $r=50,2,1.1,1.01,1.0001$. The blue and orange solid curves are the boundary $r\rightarrow \infty$ and horizon $r\rightarrow 1$ Lanczos coefficients respectively in (\ref{eq:bnLimits}). }\label{fig:bnDelta15d4BulkPropagatorHugo}
\end{figure}

\begin{figure}
         \centering
         \includegraphics[width=0.5\textwidth]{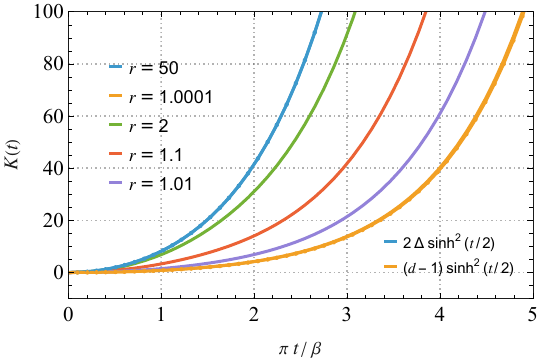}
       \caption{Krylov complexity $K(t)$ for $C(t,r)$ (\ref{eq:afbbp_eps_pi}) with $\Delta=15$ and $d=4$ for different values of $r=50,2,1.1,1.01,1.0001$. The blue and orange solid curves are the boundary $r\rightarrow \infty$ and horizon $r\rightarrow 1$ Krylov complexities respectively in~\eqref{eq:KCLimits}.
       }\label{fig:KCDelta15d4BulkPropagatorHugo}
\end{figure}

\begin{figure}[ht]
        \centering
    \begin{tabular}{cc}
         \includegraphics[width=0.45\textwidth]{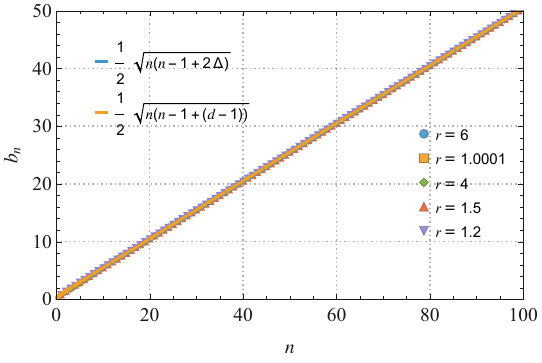} &
         \includegraphics[width=0.45\textwidth]{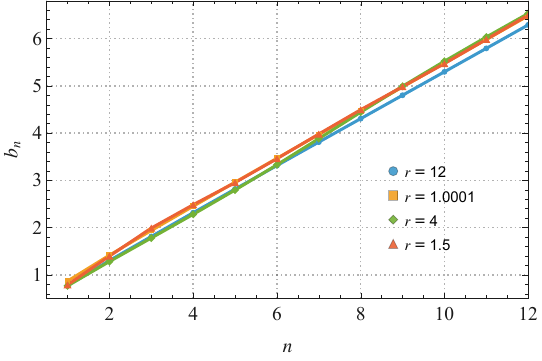}
         \end{tabular}
       \caption{Lanczos coefficient $b_n$ for $C(t,r)$ (\ref{eq:afbbp_eps_pi}) with $\Delta=1.2$ and $d=4$ for different values of $r$. The blue and orange solid curves on the left-hand side plot are the boundary $r\rightarrow \infty$ and horizon $r\rightarrow 1$ Lanczos coefficients respectively in (\ref{eq:bnLimits}). The left-hand side plot shows the behaviour of $b_{n}$ for $n\in [0,100]$ with $r=12,8,1.5,1.2,1.0001$, while the right-hand side plot shows the behaviour for $n\in [0,15]$ with $r=12,4,1.5,1.0001$.}\label{fig:bnDelta1p2d4BulkPropagatorHugo}
\end{figure}

\begin{figure}
         \centering
         \centering
    \begin{tabular}{cc}
         \includegraphics[width=0.45\textwidth]{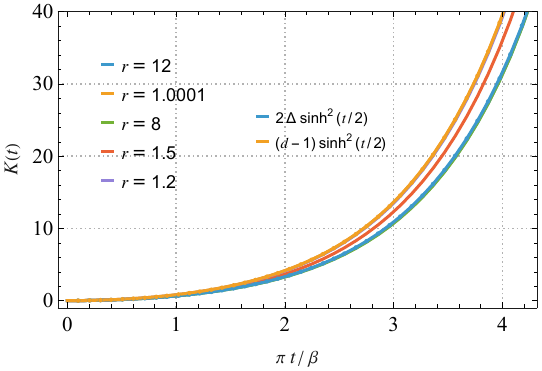} &
         \includegraphics[width=0.45\textwidth]{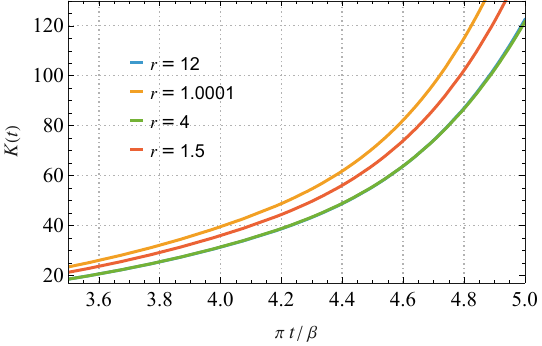}
         \end{tabular}
       \caption{Krylov complexity $K(t)$ for $C(t,r)$ (\ref{eq:afbbp_eps_pi}) with $\Delta=1.2$ and $d=4$ for different values of $r$. The blue and orange solid curves on the left-hand side plot are the boundary $r\rightarrow \infty$ and horizon $r\rightarrow 1$ Krylov complexities respectively in~\eqref{eq:KCLimits}. The left-hand side plot shows the behaviour of $K(t)$ for $t\in [0,4.1]$ with $r=12,8,1.5,1.2,1.0001$ while the right-hand side plot shows the behaviour for $t\in [3.5,5]$ with $r=12,4,1.5,1.0001$.}\label{fig:KCDelta1p2d4BulkPropagatorHugo}
\end{figure}

Figures~\ref{fig:bnDelta15d4BulkPropagatorHugo} and~\ref{fig:bnDelta1p2d4BulkPropagatorHugo} are plots of the Lanczos coefficients $b_n$ computed from $C(t,r)$ (\ref{eq:autocorrelation function main}) with $\Delta=15$ and $d=4$ and $\Delta=1.2$ and $d=4$, respectively, where each point represents $b_n$ for different $r$, and two solid curves that represent $b_n$ in the two limits (\ref{eq:bnLimits}). Meanwhile, figures~\ref{fig:KCDelta15d4BulkPropagatorHugo} and~\ref{fig:KCDelta1p2d4BulkPropagatorHugo} are plots of Krylov complexity $K(t)$ for different $r$ and the same choice of parameters, where we also plot two solid curves corresponding to the horizon and boundary limits (\ref{eq:KCLimits}). These figures show the transition of $b_n$ and $K(t)$ at finite $r$. The first set of plots corresponds to a positive squared mass of the scalar field $m^2>0$, while the second corresponds to a negative squared mass $m^2<0$.

As seen in Figures~\ref{fig:bnDelta15d4BulkPropagatorHugo} and~\ref{fig:bnDelta1p2d4BulkPropagatorHugo}, in both parameter regimes the Lanczos coefficients $b_n$ at large $n$ tend to the line corresponding to the near-horizon limit. This is consistent with the pole structure (\ref{polestructurefiniter}) for any finite $r$. In the case $\Delta=15$ and $d=4$, corresponding to $m^2>0$, the sequence corresponding to the horizon ($r\rightarrow 1$) autocorrelation function lies beneath the one corresponding to the boundary $r\rightarrow \infty$. The Lanczos sequences for intermediate $r\in (1,\infty)$ approach the lower sequence, displaying an intermediate $n$ regime where their transient growth is \emph{sub-linear}. Interestingly, as can be seen from~\ref{fig:bnDelta1p2d4BulkPropagatorHugo}, in the case $\Delta=1.2$ and $d=4$, corresponding to $m^2<0$, the horizon Lanczos sequence lies above the boundary one, with  intermediate-$r$ sequences approaching the horizon one for a finite window in $n$, signaling a transient \emph{super-linear} growth. For more numerical evidence and a detailed analysis, see Ap.~\ref{app:figsbnKtfiniter}. 

Figures~\ref{fig:KCDelta15d4BulkPropagatorHugo} and~\ref{fig:KCDelta1p2d4BulkPropagatorHugo} show that in both parameter regimes $\lbrace \Delta,d\rbrace$ the finite-$r$ Krylov complexities lie between the boundary and horizon $K(t)$ for all times $t$. In Ap.~\ref{app:figsbnKtfiniter} we show that the finite-$r$ Krylov complexities have an approximate $\sinh^2$ behaviour for finite time-window $\Delta t$ of the form
\begin{align}
    \label{eq:KCsinhSq}
    K(t)\approx \tilde{K}_{0}(r)\sinh^{2}(\tilde{\lambda}_{K}(r)t/2)\,\, \textrm{,}  \quad\textrm{for finite }r~,
\end{align}
where $\tilde{K}_{0}(r)$ and $\tilde{\lambda}_{K}(r)$ are approximately constants for a given time window $\Delta t$. Figures~\ref{fig:PlotTableK0rLambdatDelta15d4} and~\ref{fig:PlotTableK0rLambdatDelta1p2d4} in App.~\ref{app:figsbnKtfiniter} show the behaviour of these functions for $t \in [0,4]$. For both parameter regimes, the horizon and boundary behaviour of these functions is consistent with the expectation~\eqref{eq:KCLimits}. However, we remark that the true Krylov complexity obtained for finite $r$ has a more complicated functional dependence on $r$ and $t$. 

If $b_n$ is a smooth function of $n$, it is known that the asymptotic behavior of $b_n$ for large $n$ is determined by the (high-frequency) exponential tail of the power spectrum $\rho(\omega)$ at high frequency~\cite{Parker:2018yvk}.\footnote{The power spectrum is the Fourier transform of the autocorrelation function: $\rho(\omega)=\int \textrm{d}t e^{it\omega}C(t)$.} Thus, the asymptotic $n\rightarrow \infty$ behavior of $b_n$ and the late-time behavior $t\rightarrow \infty$ of Krylov complexity $K(t)$ is determined by the behaviour of their high-energy modes. In the discussion section~\ref{sec:dis}, we comment on these findings from the perspective of holographic renormalization group (RG) flow (Sec.~\ref{sec:disRGusual}) as well as from T\texorpdfstring{$\overline{\text{T}}$}{} deformed CFTs (Sec.~\ref{sec:disTTbar}).  

\subsection{Interpretation of the near horizon limit}
\label{subsec:ConfCoupScalar}

Let us discuss the interpretation of the factor $d-1$ in the second case of (\ref{eq:KCLimits}). Compared with the CFT result, the scaling dimension $\Delta$ of $(d-1)\sinh^2(t/2)$ can be regarded as $\Delta=(d-1)/2$. In AdS/CFT, scalar mass $m$ and $\Delta$ are related via $\Delta_\pm=\frac{d}{2}\pm\sqrt{\frac{d^2}{4}+m^2}$, where we set the AdS radius as one. If the scalar mass is given by\footnote{Note that while the conformally coupled scalar is tachyonic $\forall d>1$, it still obeys the Breitenlohner-Freedman (BF) stability condition \cite{Breitenlohner:1982jf}, which in our notation is expressed as $ m^2 \geq -\frac{d^2}{4} $.}
\begin{align}
m^2=\frac{1-d^2}{4},\label{conformally coupled scalar}
\end{align}
we obtain $\Delta_-=(d-1)/2$. (\ref{conformally coupled scalar}) is known as the mass of a conformally coupled scalar in AdS$_{d+1}$ (see, e.g., \cite{Dorn:2003au}). More explicitly,
\begin{align}
\frac{1}{2}\xi R\phi^2=\frac{1}{2}\left(\frac{1-d^2}{4}\right)\phi^2, \;\;\; \xi:=\frac{d-1}{4d},
\end{align}
where $R=-d(d+1)$ in AdS$_{d+1}$.

\section{Proper momentum/Krylov complexity entry in the holographic dictionary}\label{sec:proper momentum}
Based on the previous result on the Krylov complexity for the thermal correlator in Rindler-AdS \eqref{eq:KCLimits}, in this section, we consider the corresponding rate of growth of complexity:
\begin{equation} 
\label{eq:dKCLdtimits}
  \dot{K}(t)= 
\begin{cases}
\frac{\Delta}{\sin^2(\epsilon/2)} \sinh(t)\,, \,\,\text{for}\,\,r\rightarrow \infty~,\\
\frac{d-1}{2 \sin^2(\epsilon/2)}\sinh(t)\,,\,\,\text{for}\,\, r\to1~,
\end{cases}  
\end{equation}
and we compare it to the radial momentum of a particle falling into Rindler-AdS$_{d+1}$ spacetime.

\subsection{Proper radial momentum in Rindler-AdS\texorpdfstring{$_{d+1}$}{} space}\label{ssecproper radial momentum in RAdS}

The notion of infalling radial particle momentum is inherently coordinate-dependent. In Ref.~\cite{Caputa:2024sux} it was proposed that the \emph{proper} radial momentum of a particle falling in the bulk is related to the rate-change of the Krylov complexity of perturbed thermal states in AdS. As long as the radial coordinate in the chosen coordinate system computes the proper distance from the location of the particle to the black hole horizon (or the center of global AdS or the Poincaré horizon), the holographic identification should hold. On the ohter hand, Ref.~\cite{Li:2025fqz} proposes a different construction of this identification in AdS$_{3}$/CFT$_{2}$, where Krylov state complexity of boundary CFT states is interpreted as the energy measured by a bulk observer, and its rate of change corresponds to the observer’s radial momentum. Because the measured particle energy depends on the coordinate choice (associated with a different symmetry subalgebra on the boundary), both quantities reflect this choice inherently.

Since our computation of the non-local CFT primary operators via the bulk propagator was done in a coordinate  covering one Rindler wedge in AdS, (see Eq.~\eqref{eq-metric-Rindler}) we will work in a frame where the Rindler-AdS$_{d+1}$ metric takes the form 
\begin{equation}
    \rmd s^2=\rmd\rho^2-\sinh^2\rho\rmd t^2+\cosh^2\rho \, \rmd \mathbb{H}_{d-1}^2~,
\end{equation}
where $r=\cosh\rho$ in \eqref{eq-metric-Rindler}. From the perspective of the boundary CFT, this choice of coordinates corresponds to an initial state of the form $\vert\psi_{0}\rangle=\mathcal{O}(x_{0})\vert \textrm{TFD}\rangle $, where $\mathcal{O}(x_{0})$ is a local primary field on the boundary of the Rindler wedge inserted at $(t,x_{0})\in\mathbb{R}\times \mathbb{H}_{d-1}$ on the timeslice $t=0$.
The proper radial momentum of the point particle with mass $m_p$, which we consider to move solely in the radial direction from the asymptotic boundary at $\rho\rightarrow\infty$, can be derived from its action:\footnote{Note that we do not require the mass of the point particle to be identified with the scalar field for our derivations, in contrast with \cite{Caputa:2024sux}, unless the particle is located at the asymptotic boundary. However, as remarked in Sec. \ref{ssec:particle energy} the holographic dictionary relating the probe particle with an CFT operator insertion is no longer valid beyond this limiting case.}
\begin{equation}
    \int\rmd t~\mathcal{L}\equiv-m_p\int\rmd t\sqrt{\sinh^2\rho(t)-\qty(\dv{\rho}{t})^2}~.
\end{equation}
From the equations of motion with $\rho(0)=\rho_0$ and $\dot{\rho}(0)=v_0$ we find:
\begin{equation}\label{eq:rt solut}
    \coth(\rho(t))=\cosh(t)\coth(\rho_0)-v_0\csch^2(\rho_0)\sinh(t)
    \end{equation}
    and moreover, we have that the radial momentum for the particle is
\begin{equation}\label{eq:radial momentum}
  P_\rho=\dv{\mathcal{L}}{\dot{\rho}(t)}= m_p\frac{v_0\cosh(t)-2\sinh2\rho_0\sinh(t)}{\sqrt{\sinh^2\rho_0-v_0^2}}~,
\end{equation}
(where we selected the sign that represents a geodesic where a particle starting at $\rho=\rho_0$ with $v_0=0$ falls into the Rindler-AdS$_{d+1}$ interior). We note that in this case, the proper radial momentum ~\eqref{eq:radial momentum} coincides with the measured radial momentum, as shown in ~\cite{Li:2025fqz}.

{We emphasize that the numerical results for finite-$r$ suggest that the $\sinh^2(t)$ behavior for Krylov complexity is a good approximation, but it cannot be strictly true for all $t$.} The reason is that Krylov complexity grows exponentially at finite times with an effective time dependent exponent, except when the radial location is at the black hole horizon or at the asymptotic boundary. In contrast, the Lyapunov exponent, which is evaluated at $t\rightarrow\infty$ is independent on the radial location. Our results thus suggest that the duality between the particle's proper momentum and the Krylov operator complexity differ for finite bulk radial locations. This is perhaps not surprising since the bulk dictionary relation the particle's energy and the operator's conformal dimension does not hold at arbitrary finite bulk radial locations (see Sec. \ref{ssecproper radial momentum in RAdS}), and since the two-point function used to evaluate Krylov complexity is no longer conformally invariant.

Moreover, given that the proposal in \cite{Caputa:2024sux} states that the proper momentum of the probe particle in AdS space is dual to the Krylov complexity, we will set $v_0=0$ (since the particle does not have velocity according to its own frame). 
{Another way to see that we should impose $v_0=0$ for the identification of $P_\rho$ and $\dot{K}(t)$, is that Krylov complexity for a Hermitian system is generally an even function of $t$ due to the time-reversal symmetry of the autocorrelation function $C^*(t)=C(-t)$ for the usual Hermitian inner product. Thus, the time derivative of Krylov complexity $\dot{K}(t)$ is an odd function of $t$. When $v_0\ne0$, the time-reversal symmetry is broken, and $P_\rho$ (\ref{eq:radial momentum}) is not an odd function of $t$.
}
After plugging (\ref{eq:rt solut}) in (\ref{eq:radial momentum}) we recover that
\begin{equation}\label{eq:radial momentum eval}
    P_\rho=-m_p\cosh(\rho_0)\sinh(t)~.
\end{equation}
Thus, the time dependence agrees with the momentum and the rate of growth of the Krylov complexity  (\ref{eq:dKCLdtimits}) for both cases. Below, we will deduce the relation between the particle's mass and the radial location $\rho_0$.

\subsection{On the particle's and the CFT energy}\label{ssec:particle energy}
In order to gain full information about properties that the particle has in the correspondence above, we also want to express the energy of the particle in terms of CFT quantities. This can be evaluated from
\begin{equation}
    E_\text{particle}= \dot{\rho} \frac{\partial \mathcal{L}}{\partial \dot{\rho}}-\mathcal{L}= m_p \frac{\sinh^2 \rho}{\sqrt{\sinh^2 \rho-\dot{\rho}^2}}\,.
\end{equation}
We are particularly interested in the energy at $\rho_0$ with $\dot{\rho}=0$; where we obtain
\begin{equation}\label{eq:energy particle}
    E_\text{particle}= m_p \sinh \rho_0\,.
\end{equation}
This should match the CFT energy associated to an operator of the form $\mathcal{O}_{\epsilon}=e^{-\frac{\epsilon}{2} H}\mathcal{O}(t)\,e^{\frac{\epsilon}{2} H}$. Motivated by (7) in \cite{Caputa:2024sux}, a definition of the CFT energy for general $\epsilon$ is 
\begin{align}\label{eq:ECFT new}
E_\text{CFT}\coloneqq \bra{\psi(t)}H\ket{\psi(t)}=\bra{\psi(0)}H\ket{\psi(0)}=a_0=\frac{2\pi \Delta}{\beta \tan\left(\frac{\pi \epsilon}{\beta}\right)},
\end{align}
where $\ket{\psi(t)}$ is a locally excited state as defined in App.~\ref{app:KcompHyperbolic}.
When $\epsilon\ll\beta$, the CFT energy becomes $E_\text{CFT}\approx {2\Delta}/{\epsilon}$ which is consistent with (7) in \cite{Caputa:2024sux}. 
In order for the above quantity to be a meaningful measure of the CFT energy we study the energy fluctuations caused by the CFT primary in the CFT spectrum as a point particle. Since
\begin{align}
    \hat{H}\ket{K_0}&=b_1\ket{K_1}+a_0\ket{K_0}~,\\
    \bra{\psi(0)}\hat{H}^2\ket{\psi(0)}&=b_1^2+a_0^2=\qty(\frac{2\pi\Delta}{\beta\tan\qty(\frac{\pi\epsilon}{\beta})})^2+\qty(\frac{\pi\sqrt{2\Delta}}{\beta\sin\qty(\frac{\pi\epsilon}{\beta})})^2
\end{align}
We can then recover the standard deviation
\begin{equation}
    \delta E_{\textrm{CFT}}=\sqrt{\bra{\psi(0)}\hat{H}^2\ket{\psi(0)}-(E_{\textrm{CFT}})^2}=b_1=\frac{\pi\sqrt{2\Delta}}{\beta\sin\qty(\frac{\pi\epsilon}{\beta})}~.
\end{equation}
Thus, in order to describe a localized particle at a given radial location with a well-defined energy, we require that
\begin{equation}
    \frac{\delta E_{\textrm{CFT}}}{E_{\textrm{CFT}}}=\frac{b_1}{a_0}=\frac{\pi\sqrt{2\Delta}}{\beta\sin\qty(\frac{\pi\epsilon}{\beta})}\frac{\beta\tan\qty(\frac{\pi\epsilon}{\beta})}{2\pi\Delta}=\frac{1}{\sqrt{2\Delta}\cos(\frac{\pi\epsilon}{\beta})}\ll1
\end{equation}
which is satisfied when $\Delta\gg1$, as in the original analysis in \cite{Caputa:2024sux} (which also assumes $\epsilon\rightarrow0$). Under this consideration, one may impose the equality $E_\text{particle}=E_\text{CFT}$ to derive part of the holographic dictionary relating the particle's parameters and those of the CFT:\footnote{There are other ways of defining the CFT energy, see e.g. \cite{Berenstein:2019tcs}, which include additional terms with respect to (\ref{eq:dictionaryParameters}).} 
\begin{equation} \label{eq:dictionaryParameters}
    m_p \sinh \rho_0 = \frac{2\Delta}{\epsilon}~.
\end{equation}
Eq.\eqref{eq:dictionaryParameters} establishes the dictionary between $\rho_0$ and the conformal dimension $\Delta$ of the CFT operator and the parameter $\epsilon$.\footnote{Particularly for $\rho_0 \gg 1$, we have $\cosh \rho_0 \approx \sinh \rho_0 \approx e^{\rho_0}/2$, so that (\ref{eq:dictionaryParameters}) together with (\ref{eq:radial momentum eval}) returns one of the main results reported in \cite{Caputa:2024sux}
 \begin{equation}
     P_{\rho} \approx- \frac{2\Delta}{\epsilon}  \sinh t~.
 \end{equation}}
We stress that in order to have a concrete identification of dictionary above, we indeed require a small cutoff parameter. Nevertheless, our numerical studies relied on other values of the cutoff to investigate how smeared boundary reconstruction works under more general conditions.

\subsection{Holographic dictionary}\label{ssec:holo dictionary}
According to the proposal \cite{Caputa:2024sux}, there should be a match between the Krylov complexity for locally excited states due to primary CFT$_2$ operator insertions, ${K}(t)$; and the radial momentum of a particle falling into an asymptotically AdS$_3$ spacetime, denoted by $P_\rho$; i.e.
\begin{equation}\label{HolographicProposal}
    \dot{K}(t)=-\frac{2}{\epsilon}P_\rho~,
\end{equation}
where $\epsilon$ is a regulator parameter associated with the conformal boundary.

We have found that the time dependence of the Krylov complexity of primary CFT operators computed holographically with the bulk propagator in Rindler-AdS$_{d+1}$ matches the proper radial momentum of an infalling particle in the same background, showing that the relation proposed in \cite{Caputa:2024sux} is valid in higher dimensions. Moreover, we show that a similar relation is valid in the near-horizon limit for bulk-to-bulk propagators.
For $\epsilon <<1$, Eq. \eqref{eq:dKCLdtimits} takes the form 
\begin{equation} \label{eq:KdotwithEta}
    \dot{K}(t)=\frac{4 \eta}{\epsilon^2} \sinh t\,,~~~~\eta = \begin{cases} \Delta\,,~~\text{for}~r \rightarrow \infty\,.\\
    \frac{d-1}{2}\,,~~\text{for}~r \rightarrow 1\,.
        \end{cases}
\end{equation}
In the same limit ($\epsilon <<1$), Eq. \eqref{eq:dictionaryParameters} is valid and together with \eqref{eq:radial momentum eval} implies the relations
\begin{equation}\label{eq:K dot final}
    \dot{K}(t)=
    \begin{cases}
        - \frac{2}{\epsilon} \, P_\rho~,&\text{for }\rho_0\rightarrow\infty\\
        -\frac{1}{\epsilon^2} \frac{2(d-1)}{m_p}\, P_\rho~,&\text{for }\rho_0\rightarrow0~,
    \end{cases}
\end{equation}
The first equation in \eqref{eq:K dot final} reproduces Eq.(27) in \cite{Caputa:2024sux} for boundary-boundary two-point functions, while the second equation shows that a similar relation is valid for bulk-to-bulk two-point functions in the near-horizon limit. Therefore, our results extend to those in \cite{Caputa:2024sux} by generalizing the result to higher dimensions (albeit in the s-wave sector of the dual CFT$_{d}$), as well as when the particle is located near the horizon. We stress that for the identification to work, we needed the initial condition $\dot{\rho}(0)=0$ when we place the probe particle at either the asymptotic boundary and in the near-horizon region.

As an additional comparison, the rate of growth of operator Krylov complexity was matched with the radial momentum of \emph{massless} particles in \cite{Fan:2024iop} for planar AdS$_{d+1}$ black hole backgrounds. Our results instead consider massive particles, and we are extending the cases considered in the literature \cite{Caputa:2025dep,Caputa:2024sux} to the near-horizon region.

\paragraph{Intermediate locations inside the bulk}
So far, we have considered the rate of change of Krylov operator complexity near the asymptotic AdS boundary and near the horizon. For intermediate bulk regions, we can arrive at similar conclusions, as follows. First, we know that Krylov complexity at early times behaves as \cite{Aguilar-Gutierrez:2023nyk}
\begin{equation}\label{eq:early Kt}
    K(t)=b_1^2 t^2+\mathcal{O}(t^4)~.
\end{equation}
In order to recover a match of $\dot{K}(t)$ with the proper momentum in (\ref{eq:radial momentum}) as in (\ref{HolographicProposal}) we see that we must impose $v_0=0$ for any value of $\rho_0$, which indicates the probe particle is at rest in its proper frame for the relation to dictionary entry to work. On the other hand, for the large-time behavior, we need to study the asymptotic behavior of $b_n$ for $n\gg1$ \cite{Parker:2018yvk}. We notice from the pole structure for general finite bulk radial location in \eqref{polestructurefiniter} that the autocorrelation function contains a pole of order $d-1$ at $t= i\epsilon$. We can then deduce from the pole structure (see App. A.2 in \cite{Parker:2018yvk}) the asymptotic behavior of the Lanczos coefficients being $b_n \sim n$. This indicates that at very late times the Krylov complexity asymptotes to $K(t)\propto\rme^t$, which agrees with the large-time dependence in (\ref{eq:radial momentum eval}). Indeed, for large times, the above considerations and Eq.~\eqref{eq:KdotwithEta} suggest that the following relation should hold: $\dot K(t) = \tilde K_0(\rho_0) e^t/\epsilon^2$, where $\tilde K_0(\rho_0)$ varies from $2\Delta$ to $d-1$ as $\rho_0$ ranges from $\infty$ to $0$. In this case, one can derive the following relation:
\begin{equation}
\frac{\dot{K}}{P_\rho}=-\frac{2}{\epsilon^2} \frac{\tilde K_0(\rho_0)}{m_p} \cosh \rho_0,,~~~(t \rightarrow \infty)
\end{equation}
which is expected to hold at late times for arbitrary values of $\rho_0$.
Thus, one can match the rate of growth of Krylov complexity with the proper radial momentum at arbitrary locations in the bulk in the early and late times. Meanwhile, to study intermediate time regimes we employ numerical methods. Our results from Sec. \ref{app:figsbnKtfiniter} indicates that at intermediate times and generic radial bulk locations, the relation $\dot{K}\propto\sinh(t)$ does not strictly hold. Nevertheless, it is still a good approximation to the evolution of Krylov complexity, as substantiated in Sec. \ref{app:figsbnKtfiniter}. The reasons that this relation is only approximate have been explained in the paragraph after \eqref{eq:radial momentum}, and they are further is discussed in the next section.

\section{Discussion and Future Directions}\label{sec:dis}

In this work, we have investigated the evolution of Krylov complexity of holographic CFTs from the bulk-to-bulk propagator in Rindler-AdS space for local operators in the bulk, corresponding to non-local observables in the boundary theory. In particular, the near horizon limit of our results show that the propagator corresponding to a probe minimally coupled scalar field in the near the horizon region has a very similar structure to a conformally coupled scalar field in the asymptotic boundary. In order to match the rate of growth of Krylov complexity with the proper radial momentum, we had to impose that the initial velocity of the particle in the proper frame vanishes, just as if it were located on the asymptotic boundary, as per the original proposal~\cite{Caputa:2024sux}, while the specific coefficients involved in the holographic dictionary are altered. We found that the bulk dual of the rate of growth of Krylov complexity for a TFD state perturbed by a non-local operator still is described by the proper momentum of a probe particle in the bulk. However, there is no exact matching, and the correspondence relation is only approximate at finite radial distance since conformal symmetries in the corresponding observables are broken. It is interesting to note, however, that one recovers exact matching for bulk fields in the near-horizon region of the geometry, suggesting an emergent conformal symmetry for the autocorrelation function in this region.

{{\paragraph{Comparizon with spread complexity}{To conclude the summary, we stress the similarities and differences between our study with the original work \cite{Caputa:2024sux} establishing a match between Krylov spread complexity of excited states (due to operator insertions on the vacuum) holographic CFTs with the proper radial momentum of a bulk probe particle (generated by the operator insertion in the boundary theory). Our work finds this same type of match for Krylov complexity for the CFT operators instead of spread complexity. In both cases Krylov complexity is proportional to the proper radial momentum when the particle is placed at asymptotic boundary. Thus, Krylov operator complexity in \eqref{eq:K dot final} reproduces exactly Krylov state complexity \cite{Caputa:2024sux}) in this system. The reason for this has been proven in more general systems in \cite{Sanchez-Garrido:2024pcy} (also discussed in \cite{Ambrosini:2024sre} App F.2) which shows that the Lanczos algorithm for the thermofield double state with an operator insertion as reference state evolving unitarily can be isometrically mapped to the Lanczos algorithm for Krylov complexity of the operator insertion, and then Krylov spread/operator complexity under these conditions are exactly equal to each other. Our work extended the previous discussion to higher dimensions and for smearing CFT operators.
Based on the proof by \cite{Sanchez-Garrido:2024pcy}, spread complexity would also only match the proper radial momentum approximately when conformal symmetries are broken, as our evaluations from Krylov operator complexity indicate.}}}

We now elaborate on the interpretation of our results, its connection with other parts of the literature, and related future directions. 

\subsection{Analogous interpretation of our results with holographic RG flow}
\label{sec:disRGusual}

Here we comment on qualitative similarity between our results for Lanczos coefficients and Krylov complexity for two-sided correlators from perspective of holographic RG flow~\cite{deHaro:2000vlm}. However, we stress that our results are not intended to reflect a RG flow, since we work with free scalar field correlation function is AdS-Rindler; while to have a UV-IR flow, one would need to introduce a relevant operator insertion that would backreact on the black hole background. As shown in Sec.~\ref{ssec:Lanczos and complexity}, depending on the relation between the conformal (scaling) dimension $\Delta$ of the dual operator and the spatial dimension $d$, the Lanczos sequence for the horizon $r\rightarrow 1$ lies above ($2\Delta<(d-1)$) or below ($2\Delta>(d-1)$) the boundary sequence $r\rightarrow \infty$, with intermediate finite-$r$ sequences lying between these two for all $n$ and $r$. Moreover, for sufficiently large $n$, any finite-$r$ sequence approaches the \emph{horizon} sequence, as can be seen in Figs.~\ref{fig:bnDelta15d4BulkPropagatorHugo} and~\ref{fig:bnDelta1p2d4BulkPropagatorHugo}. In general, the asymptotic $n\rightarrow \infty$ behavior of $b_n$ and the late-time behavior $t\rightarrow \infty$ of Krylov complexity $K(t)$ in QFTs are determined by the behaviour of the high frequency regime in the power spectrum, typically corresponding to the UV limit of the theory undergoing a RG flow. Here we offer an analogous interpretation of this result from the perspective of holographic RG flow (see~\cite{Skenderis:2002wp} and references therein); which, again, does not strictly speaking apply to the setting we have studied.

In essence, holographic RG flow (see \cite{Fukuma:2002sb} for a modern review) is based on the statement that the radial coordinate $r$ in AdS/CFT can be identified with a renormalization scale $\mu$ of the boundary field theory. In this case, the UV theory, corresponding to $\mu\rightarrow \infty$ is typically identified with the boundary CFT at $r\rightarrow \infty$ \cite{Fukuma:2002sb}. At the same time, there exists an effective IR theory at the horizon $r=1$ (in units where the AdS radius is $\ell=1$), where the renormalization scale in this case is $\mu \sim r-1\rightarrow 0$. As we move from the boundary deeper into the bulk, the effective scalar field theory at finite $r$ ``flows'' from the UV theory at the boundary to the IR theory at the horizon. In other words, the scalar field theory at finite $r$, QFT($r$), corresponds to a would-be ``effective'' theory in the analogous holographic RG flow. Keep in mind that throughout this argument we keep the UV (CFT) theory and boundary conditions for all fields fixed at $r\rightarrow \infty$.

From Eq.~\eqref{polestructurefiniter} it can be seen that for any finite $r$, the autocorrelation function $C(t,r)$ has a pole in $\delta t$ of the order $d-1$ (corresponding to the horizon pole-order), which implies that the high-frequency modes of the power spectrum $\rho(\omega,r)$ corresponding to $C(t,r)$ will be dominated by the high-frequency modes of the horizon theory (see Sec.~\ref{subsec:ConfCoupScalar}). This is consistent with the behaviour of the Lanczos coefficients for finite $r$ in both parameter regimes: $2\Delta >(d-1)$ and $2\Delta < (d-1)$. In both cases, the  finite-$r$ Lanczos sequences tend to the horizon sequence, regardless of the relation between $2\Delta$ and $d-1$. This implies that the late-time behaviour of the Krylov complexity will be dominated by the high-frequency modes of the horizon theory. From this perspective, it is the effective IR theory at $r\rightarrow 1$ that dominates the late-time behaviour of Krylov complexity.

It is interesting to contrast this behaviour with the case of a massive scalar field in flat space~\cite{Avdoshkin:2022xuw, Camargo:2022rnt}. In that case, the mass $m$ of the field acts as an IR scale. If we imagine that in that case we also have an ``RG flow" of theories, where the renormalization scale $\mu$ is set by the mass $m$,  QFT($m$), then the UV theory corresponds to the massless - conformal - limit $m\rightarrow 0$, QFT$(m\rightarrow 0)=$ CFT, and the IR would correspond to an infinite mass $m\rightarrow \infty$, QFT$(m\rightarrow \infty)=\overline{\textrm{QFT}}$. Consider each $m\geq0$ corresponding to a different theory QFT($m$). In that case, if we look at the power spectrum of the constant $m$ theory, then the late-time behaviour of Krylov complexity and large-$n$ behaviour of Lanczos coefficients (dominated by high-frequency modes $\vert \omega \vert \gg m$ in the power spectrum $\rho (\omega)$) {are closer to} the ones expected from the UV theory, the CFT with $m\rightarrow 0$.\footnote{{To be precise, the nonzero mass causes non-smoothness and staggering in the Lanczos coefficients \cite{Avdoshkin:2022xuw, Camargo:2022rnt}.}} This is simply because the mass $m$, being an IR scale, does not affect the high-energy/short-distance behaviour of the QFT$(m)$ theory for any $m\geq 0$. Indeed, for any $m\geq 0$ there is a sufficiently large $\vert \omega \vert$ such that the high-frequency modes of the power spectrum are insensitive to it. So, in the end, any theory with finite $m$ will ``behave'' like the UV theory (the massless CFT) at sufficiently late times. In this sense, the mass of the field $m$ acts as an ``inverse'' high-energy scale $m\sim 1/\Lambda_{\textrm{UV}}$.

We remind the reader that the above discussion relating the results with a holographic RG flow is intended only as analogy, since one would need to deform the boundary theory to produce those effects. Another subtlety in this discussion is that our autocorrelation function $C(t,r)$ describes correlations between local bulk fields in AdS, which are non-local from the perspective of boundary operators (see App.~\ref{app:appHKLL}). Close to the boundary, the effective theory is well-approximated by the boundary CFT behaviour, where the dual operator to the field is effectively local. As the scalar field probes deeper into the bulk, the autocorrelation function describes the correlations between increasingly non-local CFT operators, which are smeared in an increasingly large boundary subregion. In the limit where the scalar fields are taken to the horizon, the boundary operators are effectively smeared over the whole boundary subregion. Moreover, the pole structure of the autocorrelation function for $\delta t$ and finite $r$ (Eq.~\eqref{polestructurefiniter}) implies that for any finite $r>1$, the evolution of the autocorrelation function between the smeared boundary operators will flow towards the late-time behaviour at the horizon at sufficiently large times.

\subsection{Analogous interpretation of our results with T\texorpdfstring{$\overline{\text{T}}$}{} deformed CFTs}
\label{sec:disTTbar}

One is tempted to compare the finite-$r$ autocorrelation function~\eqref{eq:autocorrelation function main} with the two-point function of primary operators in a T\texorpdfstring{$\overline{\text{T}}$}{}-deformed CFT living at a finite radial surface, \emph{\`{a} la} finite cutoff holography \cite{McGough:2016lol}. This analogy could then be na\"{i}vely used to find the Krylov complexity in a holographic T\texorpdfstring{$\overline{\text{T}}$}{}-deformed CFT (the reader is referred to \cite{Jiang:2019epa,He:2025ppz} for reviews on T\texorpdfstring{$\overline{\text{T}}$}{}-deformations). Indeed, the bulk-to-bulk propagator~\eqref{eq:BulkCorrelator} shares similarities with the finite cutoff AdS correlation function, where pure gravity with the finite radial cutoff is holographically dual to T\texorpdfstring{$\overline{\text{T}}$}{}-deformed CFT$_2$ and T$^2$-deformed CFTs in higher dimensions \cite{Hartman:2018tkw}.

In the context of holographic T$\overline{\text{T}}$ deformations, the asymptotic boundary is IR (since T$\overline{\text{T}}$ is a irrelevant deformation). For instance, we treat $1/r^2$ and $\lambda$ similarly in eqs.~(\ref{eq:AutocPertr}) and (\ref{eq:AutocPertlambda}). Thus, $r\to\infty$ corresponds to $\lambda\to0$, which is IR because the T$\overline{\text{T}}$ deformation is an irrelevant deformation. If we regard that the asymptotic boundary is IR and the horizon is UV, our results in Sec.~\ref{ssec:Lanczos and complexity} imply that the large-$n$ behavior of the Lanczos coefficient and the late-time behavior of the Krylov complexity for finite-$r$ is determined by the information at UV (the horizon $r\to1$). Note that there is a caveat regarding the energy spectrum in the finite cutoff holography: the energy spectrum becomes complex if the radial coordinate is inside the horizon. A similar issue may appear in our computation of the Krylov complexity if we use the bulk propagator with $r<1$.

It is interesting and important to deeply understand the difference between the two interpretations, the analogous holographic RG flow and the T$\overline{\text{T}}$ deformation, from the universal behavior of the Lanczos coefficient and the Krylov complexity.

However, finding the Krylov complexity in a boundary theory dual to finite cutoff AdS space would require integrating out the bulk degrees of freedom corresponding to high-energy modes in the boundary, while still imposing Dirichlet boundary conditions on the finite timelike boundary cutoff~\cite{Kraus:2018xrn,Hartman:2018tkw}. Moreover, the bulk side would preferably be pure gravity in finite cutoff holography, since otherwise one needs to introduce matter contributions in the stress tensor by hand. This implies that the finite cutoff correlator does \emph{not} coincide with the bulk-to-bulk propagator (\ref{eq:bbp}), though their structure is very similar. Formally, the perturbative expansions of the bulk and T\texorpdfstring{$\overline{\text{T}}$}{}-deformed CFT autocorrelation functions with perturbation parameters $1/r^{2}$ and $\lambda$ can be written as
\begin{align}
    \label{eq:AutocPertr}
    C(t,r)=C^{(0)}_{r}(t)+\frac{1}{r^{2}}\,C^{(1)}_{r}(t)+\left(\frac{1}{r^{2}}\right)^{2}\,C^{(2)}_{r}(t)+\cdots\,
\end{align}
\begin{align}
    \label{eq:AutocPertlambda}
    C^{\lambda}(t)=C^{(0)}_{\lambda}(t)+\lambda \,C^{(1)}_{\lambda}(t)+\lambda^{2} \,C^{(2)}_{\lambda}(t)+\cdots\,
\end{align}
where $C^{(0)}_{r}(t):=\lim_{r\rightarrow \infty} C(t,r)$ and $C^{(0)}_{\lambda}(t)=\lim_{\lambda\rightarrow 0} C^{\lambda}(t)$ are the unperturbed autocorrelation functions and where the ellipsis denotes higher order terms in $1/r^{2}$ and $\lambda$ respectively.

To understand this relation in more detail, we perform a perturbative analysis of the finite $r$ autocorrelation function (\ref{eq:bbp}) in Appendix~\ref{app:figs}, and its corresponding Lanczos coefficients and Krylov operator complexity. We compare the results with an analogous perturbative expansion for the thermal two-point correlation function in T\texorpdfstring{$\overline{\text{T}}$}{}-deformed CFT$_2$, as done in~\cite{Chattopadhyay:2024pdj}. We find certain puzzling features, including a decrease in the Lanczos coefficients, reaching even negative values. In contrast, these effects are not present in our study with the exact bulk-to-bulk autocorrelation function (\ref{eq:autocorrelation function main}) as the input in the Lanczos algorithm for Krylov operator complexity. The same problem has been observed in studying Krylov operator complexity for T$\overline{\text{T}}$-deformed CFTs \cite{Chattopadhyay:2024pdj}. In those cases, one relies on a perturbative expansion of the thermal two-point function for primary CFT operators, in terms of the T$\overline{\text{T}}$ deformation parameter. This suggests that the results in \cite{Chattopadhyay:2024pdj} would benefit from revision with a non-perturbative completion of the correlation functions used in the evaluation of Krylov complexity. There is no known non-perturbative thermal two-point correlation function for T$\overline{\text{T}}$-deformed CFTs; which is the reason that \cite{Chattopadhyay:2024pdj} relied in a perturbative expansion. It is therefore an important future development to find the exact thermal two-point function and confirm our observations. To accomplish this task, one might study the little string sector in the TvT transformations \cite{Apolo:2019zai} to obtain the momentum space two-point correlation function of the vacuum theory holographically and consider an appropriate analytic continuation for the thermal one.\footnote{We thank Monica Guica for the suggestion.} This might settle one of the possible interpretations from the findings in \cite{Chattopadhyay:2024pdj}, that the Krylov operator complexity (under the assumption of the universal operator growth hypothesis \cite{Parker:2018yvk}) could indicate a possible violation in the chaos bound \cite{Maldacena:2015waa} in the presence of a finite radial cutoff in the bulk. {It should be remarked, nonetheless, that there is a non-perturbative expression for the T$\overline{\text{T}}$-deformed correlator in momentum space (see e.g. \cite{Asrat:2017tzd}). An alternative approach would consist in transforming it to position space, although one still finds a divergence in $\epsilon$ similarly to~\eqref{eq:ttbarautoc}. To the best of our knowledge, there is no consensus on how to properly renormalize it.}

\subsection{The switchback effect in Krylov complexity}
As pointed out in \cite{Ambrosini:2024sre}, there is a relation between the Krylov operator complexity and the switchback effect. This is defined in circuit complexity due to the insertion of precursor operators in a quantum circuit \cite{Susskind:2014jwa}. The switchback effect is one of the defining properties of holographic complexity \cite{Belin:2021bga,Belin:2022xmt}. In the bulk description, the switchback effect is caused by multiple shockwaves \cite{Shenker:2013pqa,Shenker:2013yza} backreacting in the bulk due to the insertion of heavy or light operators in the boundary CFT \cite{Afkhami-Jeddi:2017rmx} which should be inserted in the boundary at different time steps. In contrast, our study focused on Krylov operator complexity of a single probe field in AdS$_{d+1}$ space (which therefore does not generate shockwave backreaction). 

To properly study the Krylov complexity of multiple precursor operators acting on an AdS shockwave geometry, we are studying the Krylov complexity of a set of primary CFT operators (with appropriate smearing) acting on the TFD state \cite{Afkhami-Jeddi:2017rmx}. We propose building a CFT state with an arbitrary number of precursor operator insertions in the two-sided boundaries and evaluate the corresponding two-sided two-point correlation function with a probe bulk matter field. Similar to our study, the two-sided two-point function can be related to a bulk geodesic length in an AdS, albeit with multiple shockwaves \cite{Shenker:2013yza} due to the precursor operator insertion. This can allow us to directly relate the Complexity=Volume conjecture \cite{Susskind:2014rva,Stanford:2014jda,Susskind:2014jwa} with Krylov complexity as recently argued in \cite{Heller:2024ldz} (see also e.g. \cite{Das:2024tnw,Jian:2020qpp}).

\subsection{Comparison of Krylov operator complexity for holographic CFTs with the double-scaled SYK model}
Exciting developments have taken place in the holographic dictionary of Krylov complexity in terms of a Lorentzian wormhole in the double-scaled SYK (DSSYK) model (the reader is referred to \cite{Berkooz:2024lgq} for a modern review); see \cite{Ambrosini:2024sre,Rabinovici:2023yex,Heller:2024ldz,Aguilar-Gutierrez:2025pqp}. We briefly compare with our results in arbitrary dimensional CFTs to uncover new lessons in Krylov operator complexity. We now focus on the semiclassical limit of Krylov complexity for matter chord operators, $\mathcal{O}_\Delta$, where $\Delta$ is the total conformal weight, and the inner product is defined with respect to a TFD state at infinite temperature (see more details therein \cite{Ambrosini:2024sre}; and a discussion of how to incorporate finite temperature effects in the corresponding Lanczos algorithm in \cite{Aguilar-Gutierrez:2025pqp}). The authors uncover the following relation
\begin{equation}\label{eq:DSSYK Krylov growth}
K(t)=\frac{2}{\abs{\log q}}\log(1+(1-q^\Delta)\sinh^2(Jt))~,
\end{equation}
where $q\in[0,~1]$ is a parameter controlling quantum effects, where the semiclassical limit corresponds to $\abs{\log(q)}\rightarrow0$, and $J$ is a constant which measures the strength of the (Gaussian distributed) coupling constant in the DSSYK model, which plays the role of the parameter $\beta/2\pi$ in our analysis (which we have set $1$).

The leading order contribution in Krylov complexity in the semiclassical limit appears when $q^\Delta\rightarrow1-\Delta \abs{\log q}$ (which corresponds to light fields in the bulk), and for $Jt\sim\mathcal{O}(1)$. (\ref{eq:DSSYK Krylov growth}) becomes
\begin{equation}
    K(t)=2\Delta\sinh^2(Jt)~.
\end{equation}
We see that Krylov complexity takes exactly the form that we derived in (\ref{eq:KCLimits}) when $J=1$ (as we have taken $\beta/2\pi=1$), and the operators are inserted at the asymptotic boundary (i.e. $r\rightarrow\infty$). We, therefore, expect that Krylov operator complexity will have the same universal type of behavior in the higher dimensional CFTs probed in our study with respect to this model (or a finite-dimensional completion \cite{Balasubramanian:2024lqk,Miyaji:2025ucp}).

\subsection{First principle derivation for Krylov complexity growth/proper radial momentum in the holographic dictionary}
In principle there are many other bulk observables that may have the same time dependence as the proper radial momentum of a probe particle, which also depends on the chosen foliation, so the proposed proposed entry in the holographic dictionary relating the proper radial momentum of a particle with the rate of growth of Krylov complexity for states and operators needs further development to. There are other holographic examples where the is indeed satisfied \cite{Caputa:2024sux,Caputa:2025dep,Li:2025fqz,Aguilar-Gutierrez:2025pqp,Fan:2024iop,He:2024pox}. Our work provides new evidence for this relation both at the asymptotic boundary and at the near horizon region; and we show evidence that it is only approximately satisfied at other locations for local free scalar bulk operators. However a first principles holographic derivation of this relation is out of scope in this work; but we hope it sparks further developments towards establishing this entry in the holographic dictionary. A productive testground to explore this connection would be in lower dimensional quantum gravity models such as sine dilaton gravity theory and the DSSYK model, an effort started in \cite{Aguilar-Gutierrez:2025pqp}.

\subsection{Autocorrelation function from the bulk inner product}
In the usual Lanczos algorithm for operator growth,~\cite{Parker:2018yvk}, one constructs a Hilbert space for the operators in the algebra of the theory, through the Choi–Jamiołkowski isomorphism \cite{Jamiolkowski:1972pzh,Choi:1975nug} (see App. \ref{app:app1}). In this process we map an operator to state in the Hilbert space which is defined based on the inner we associate it with. In this context, the inner product between two states in the resulting Hilbert space corresponds to an autocorrelation function. For instance, the Wightman inner product (\ref{eq:CFTInnerProductGeneral}) is a natural choice to evaluate autocorrelation function of QFT in a finite temperature ensemble, in the sense that it simplifies several evaluations. More generally, there are a formal set of axioms (\ref{eq:innerprodpropCFT}) that the inner product must satisfy in order for the Hilbert space to be well defined (see App. \ref{app:appHKLL} for a more detailed discussion). The notion of autocorrelation that we used (\ref{eq:autocorrelation function main}) needs to obey the same type of axioms in order to properly define the Hilbert space as well as the corresponding Krylov space by the Lanczos algorithm (such as in the moment method, App. \ref{app:app1}). In App. \ref{app:HKLL}, we have confirmed that two-point correlation functions obtained by the HKLL method applied for local bulk operators indeed satisfies the expected axioms (\ref{eq:innerprodpropCFT}) in the asymptotic boundary limit. However, the technical methods are quite intricate to apply in more general situations, such as when we are faced with finite bulk radial location dependence. {In particular, to the best of our knowledge there is no closed-form expression for the smearing function in Rindler-AdS$_{d+1}$ due to an exponentially increasing factor. Ref.~\cite{Sugishita:2022ldv} discusses this point from the perspective of wavepacket transmission in Rindler-AdS.} For this reason we have not been able to show that (\ref{eq:autocorrelation function main}) is a well defined inner product at any radial location in the bulk. It is a future important development to treat this problem rigorously, to show that the method of smeared bulk reconstruction which we have applied in this work is formally well behaved at any bulk location. This would be particularly important in order to provide an interpretation of our results from the perspective of the bulk Hilbert space. A useful next step in this direction might be to perform some of the numerical integrals in App. \ref{app:HKLL} to verify that the finite $r$ satisfies the properties of inner products (\ref{eq:innerprodpropCFT2}).

\section*{Acknowledgments}
We thank Luis Apolo, David Berenstein, Monica Guica, Kuntal Pal, and Yichao Fu for useful discussions. We also thank Pawel Caputa and Seiji Terashima for discussions and comments on the draft. We are grateful to the organizers of ``Quantum Extreme Universe: Matter, Information and Gravity 2024'' in OIST, which allowed part of this collaboration. The authors also thank the Yukawa Institute for Theoretical Physics at Kyoto University, where this work was completed during the YITP-I-25-01 on "Black Hole, Quantum Chaos and Quantum Information". SEAG thanks Don Marolf in UC Santa Barbara for hospitality during the development of this work, and the QISS consortium for travel support.
SEAG is supported by the Okinawa Institute of Science and Technology Graduate University. 
This work was supported by the Basic Science Research Program through the National Research Foundation of Korea (NRF) funded by the Ministry of Science, ICT $\&$ Future Planning (NRF-2021R1A2C1006791),  by the Ministry of Education (NRF-2020R1I1A2054376) and the AI-based GIST Research Scientist Project grant funded by the GIST in 2025.
This work was also supported by Creation of the Quantum Information Science R$\&$D Ecosystem (Grant No. 2022M3H3A106307411) through the National Research Foundation of Korea (NRF) funded by the Korean government (Ministry of Science and ICT). H.~A. Camargo, and V.~Jahnke were supported by the Basic Science Research Program through the National Research Foundation of Korea (NRF) funded by the Ministry of Education (NRF-2022R1I1A1A01070589, and RS-2023-00248186). This project/publication was also made possible through the support of the ID\#62312 grant from the John Templeton Foundation, as part of the ‘The Quantum Information Structure of Spacetime’ Project (QISS), as well as Grant ID\# 62423 from the John Templeton Foundation. The opinions expressed in this project/publication are those of the author(s) and do not necessarily reflect the views of the John Templeton Foundation. 

\appendix

\section{Krylov complexity for CFTs in hyperbolic space}
\label{app:KcompHyperbolic}
In this Appendix, we review the derivation of Krylov complexity for CFTs in hyperbolic space~\cite{Camargo:2022rnt} and show that the result matches the corresponding calculation on the gravity side when one moves the operators to the asymptotic boundary of the spacetime. 

Using a conformal map from $\mathbb{R}^d$ to $\mathbb{S}^{1}\times \mathbb{H}^{d-1}$, we can compute the Krylov complexity in CFTs whose spatial geometry is a hyperbolic space $\mathbb{H}^{d-1}$. 
Due to the conformal map, the curvature scale of $\mathbb{H}^{d-1}$ is proportional to $(2\pi/\beta)^2$, where $\beta$ is the period of Euclidean time.
The scalar conformal two-point function on $\mathbb{S}^{1}\times \mathbb{H}^{d-1}$  is given by \cite{Haehl:2019eae}
\begin{align}
\langle \mathcal{O}_{\Delta}(\tau_{1},\mathbf{x}_{1})\mathcal{O}_{\Delta}(\tau_{2},\mathbf{x}_{2})\rangle = \frac{\mathcal{N}}{\left(-2\cos\left(\frac{2\pi}{\beta}(\tau_1-\tau_2)\right)+2\cosh \left[\mathbf{d}(1,2)\right]\right)^{\Delta}}~,\label{etfsh}
\end{align}
where $\Delta$ is a scaling dimension, $\tau_i$ is the Euclidean time, $\mathbf{d}(1,2)$ is the spatial distance between $\mathbf{x}_{1}$ and $\mathbf{x}_{2}$ in $\mathbb{H}^{d-1}$, and $\mathcal{N}$ is a normalization factor. Substituting $\tau_1-\tau_2=\rmi t+\epsilon+\frac{\beta}{4}$ and $\mathbf{d}(1,2)=0$ for the points (\ref{eq:TwoPoints}) to the Euclidean two-point function (\ref{etfsh}), we obtain the normalized thermal Wightman two-point function
\begin{align}
C(t)=\left(\frac{\sinh \left(\frac{\pi i\epsilon}{\beta}\right)}{\sinh \left(\frac{\pi(t-\rmi\epsilon)}{\beta}\right)}\right)^{2\Delta}\label{wtfh}
\end{align}
where we choose $\mathcal{N}$ such that $C(t=0)=1$. Note that (\ref{wtfh}) agrees with eq.~(\ref{eq:AutoCorrelationFunctionBDRYandHORIZON}) when $\beta=2\pi$ and $r \rightarrow \infty$.
The Lanczos coefficients and Krylov complexity corresponding to the autocorrelation \eqref{wtfh} are given by~\cite{Caputa:2023vyr}
\begin{align}
 a_n=\frac{2\pi (n+\Delta)}{\beta \tan\left(\frac{\pi \epsilon}{\beta}\right)},\;\;\; b_n=\frac{\pi\sqrt{n(n-1+2\Delta)}}{\beta \sin\left(\frac{\pi \epsilon}{\beta}\right)}\,,~~~~K(t)=2\Delta\frac{\sinh^2\left(\frac{\pi t}{\beta}\right)}{\sin^2\left(\frac{\pi \epsilon}{\beta}\right)}.\label{kotcft}
 \end{align}
The above results can also be interpreted as the Krylov state complexity for locally perturbed thermal states considered in \cite{Caputa:2024sux}. The authors of \cite{Caputa:2024sux} considered the time evolution of locally excited states by primary operators in holographic CFTs as
\begin{align}
|\psi(t)\rangle=\mathcal{N}e^{-iHt}e^{-\frac{ \epsilon}{2}H}\mathcal{O}(x_0)|\psi_\beta\rangle,
\end{align}
where $|\psi_\beta\rangle$ is the thermofield-double state. Its amplitude $S(t)$ is given by 
\begin{align}
S(t)=\langle\psi(t)|\psi(0)\rangle=\left(\frac{\sinh \left(\frac{\pi(t-i\epsilon)}{\beta}\right)}{\sinh \left(\frac{\pi i\epsilon}{\beta}\right)}\right)^{-2\Delta},
\end{align}
which is identical to the autocorrelation \eqref{wtfh}, leading to the same Lanczos coefficients and Krylov complexity given in \eqref{kotcft}.

In particular, for $\epsilon=\beta/2=\pi$, we can identify $S(t)$ with Wightman two-point function \eqref{eq:AutoCorrelationFunctionBDRYandHORIZON} for a thermal CFT, obtaining
\begin{align}
S(t)=\frac{1}{ (\cosh (t/2))^{2\Delta}}, \;\;\; a_n=0, \;\;\; b_n=\frac{1}{2}\sqrt{n(n-1+2\Delta)}, \;\;\; K(t)=2\Delta\sinh^2(t/2),
\end{align}
which is consistent with our Eqs.\eqref{eq:AutoCorrelationFunctionBDRYandHORIZON}, (\ref{eq:bnLimits}), and (\ref{eq:KCLimits}) in the limit $r \to \infty$. This result also aligns with \cite{Caputa:2021sib}; see their Eq.(30).

\section{The Inner Product in Rindler-AdS and the HKLL Formula} 
\label{app:appHKLL}
In this Appendix, we review the Hamilton--Kabat--Lifschytz--Lowenstein (HKLL) formula for Rindler-AdS$_{3}$. The reader may refer to~\cite{DeJonckheere:2017qkk,Kajuri:2020vxf} for more details.

\subsection{The HKLL Formula }
\label{app:HKLL}

In AdS/CFT, the HKLL formula~\cite{Hamilton:2005ju,Hamilton:2006az}  provides a way of expressing a local bulk field $\phi(y,t,\mathbf{x})$ in AdS$_{d+1}$, where $(y,t,\mathbf{x})$ are local bulk coordinates, in terms of CFT$_{d}$ operators in the boundary, with local coordinates $(t,\mathbf{x})$. To be precise, a massive scalar bulk field satisfying the free equations of motion $(\Box_{\textrm{AdS}}-m^{2})\phi(y,t,\mathbf{x})=0$ can be written as
\begin{equation}
    \label{eq:HKLLFieldExpansion}
    \phi(y,t,\mathbf{x})=\int \textrm{d}t'\,\int \textrm{d}\mathbf{x}'\,\mathcal{K}(y,t,\mathbf{x};t',\mathbf{x}')\,\mathcal{O}_{\Delta}(t',\mathbf{x}')~,
\end{equation}
where $\mathcal{K}(y,t,\mathbf{x};t',\mathbf{x}')$ is called the \emph{smearing function} that has support on operators $\mathcal{O}_{\Delta}(t',\mathbf{x}')$ that are spacelike-separated from $\phi(y,t,\mathbf{x})$. This allows for the computation of bulk correlators from their CFT counterpart
\begin{equation}
\begin{split}
    \label{eq:HKLLCorrleator}
    &\langle \phi_{1}(y_{1},t_{1},\mathbf{x}_{1})\cdots\phi_{N}(y_{N},,t_{N},\mathbf{x}_{N})\rangle_{\textrm{AdS}}=\\
    &\int \textrm{d}t_{1}'\,\ldots\textrm{d}t_{N}'\,\int \textrm{d}\mathbf{x}_{1}'\,\ldots\textrm{d}\mathbf{x}_{N}'\,\mathcal{K}_{1}(y_{1},t_{1},\mathbf{x}_{1};t_{1}',\mathbf{x}'_{1})\,\cdots \mathcal{K}_{N}(y_{N},t_{N},\mathbf{x}_{N};t_{N}',\mathbf{x}'_{N})\\
    &\times\langle\mathcal{O}_{\Delta_{1}}(t_{1}',\mathbf{x}'_{1})\cdots\mathcal{O}_{\Delta_{N}}(t_{N}',\mathbf{x}'_{N})\rangle_{\textrm{CFT}}~.
    \end{split}
\end{equation}

\subsection{The HKLL Formula in Rindler-AdS}
\label{app:HKLLRindler}
We are interested in the HKLL formula for Rindler-AdS$_{d+1}$. For simplicity and concreteness, we focus on Rindler-AdS$_{3}$. In local coordinates $(r,t,\chi)$, the right Rindler wedge of AdS has a metric given by
\begin{equation}
    \label{eq:RindlerAdS3}
    \textrm{d}s^{2}= \frac{\ell^{2}}{r^{2}-\ell^{2}}\textrm{d}r^{2}-\frac{r^{2}-\ell^{2}}{\ell^{2}}\textrm{d}t^{2}+\frac{r^{2}}{\ell^{2}}\textrm{d}\chi^{2}~,
\end{equation}
where $\ell$ is the AdS radius. In these coordinates, where $-\infty < t,\chi<+\infty$ and $0<\ell\leq r<\infty$, the Rindler-AdS horizon is located at $r=\ell$ and the conformal boundary at $r\rightarrow \infty$. It is more convenient to write the metric in a slightly different form. Following~\cite{Sugishita:2022ldv}, we define a radial coordinate $\mathfrak{r}=\sqrt{r^{2}-\ell^{2}}$ such that $0\leq\mathfrak{r}<\infty$, for which the metric~\eqref{eq:RindlerAdS3} takes the form
\begin{equation}
    \label{eq:RindlerAdS3Xi}
    \textrm{d}s^{2}= \frac{1}{1+\left(\mathfrak{r}/\ell\right)^{2}}\textrm{d}\mathfrak{r}^{2}-(\mathfrak{r}/\ell)^{2}\textrm{d}t^{2}+\left(1+\left(\mathfrak{r}/\ell\right)^{2}\right)\textrm{d}\chi^{2}~.
\end{equation}
In these coordinates, the Rindler-AdS horizon is located at $\mathfrak{r}=0$ and the asymptotic boundary $\mathbb{R}_{t}^{1}\times \mathbb{H}^{1}$ (conformal to $\mathbb{R}^{1,1}$) is located at $\mathfrak{r}\rightarrow \infty$. In addition, the left Rindler-AdS wedge can be obtained by a transformation $t\mapsto t-i \pi$.

In this case, the equation of motion for a massive scalar field $\Box_{\textrm{R-AdS}}\,\phi(\mathfrak{r},t,\chi)$$=m^{2}\phi(\mathfrak{r},t,\chi)$ is given by
\begin{equation}
    \label{eq:ScalarEoM}
    \left(g^{tt}\,\partial^{2}_{t}+\frac{1}{\sqrt{-g}}\partial_{\mathfrak{r}}\left(\sqrt{-g}\,g^{\mathfrak{r}\mathfrak{r}}\,\partial_{\mathfrak{r}}\right)+g^{\chi\chi}\,\partial_{\chi}^{2}-m^{2}\right)\phi(\mathfrak{r},t,\chi)=0~,
\end{equation}
with $\sqrt{-g}=\mathfrak{r}/\ell$. Setting $\ell=1$, the general solution to this equation on the right Rindler-AdS wedge can be written as
\begin{equation}
    \label{eq:SolPhiRightW}
    \phi(\mathfrak{r},t,\chi)= \int_{-\infty}^{\infty} \textrm{d}\lambda\int_{0}^{\infty} \textrm{d}\omega\,\frac{1}{\sqrt{2\pi}}\tilde{\psi}_{\omega,\lambda}(\mathfrak{r})\left(a_{\omega,\lambda}\,e^{-i\,\omega\,t+i\,\lambda\,\chi}+a^{\dagger}_{\omega,\lambda}\,e^{i\,\omega\,t-i\,\lambda\,\chi}\right)~,
\end{equation}
where $\lbrace a_{\omega,\lambda},a^{\dagger}_{\omega,\lambda}\rbrace $ are ladder operators satisfying $[a_{\omega,\lambda},a_{\omega',\lambda'}^{\dagger}]=\delta(\omega-\omega')\delta(\lambda-\lambda')$ and where
\begin{equation}
\begin{split}
    \label{eq:PsiTilde}
    \tilde{\psi}_{\omega,\lambda}(\mathfrak{r})=&\frac{N_{\omega,\lambda}}{\Gamma(\Delta)}\mathfrak{r}^{i\,\omega}(1+\mathfrak{r}^{2})^{-\frac{i\,\omega}{2}-\frac{\Delta}{2}}\times\phantom{\,}_{2}F_{1}\left(\frac{i\,\omega-i\,\lambda+\Delta}{2},\frac{i\,\omega+i\,\lambda+\Delta}{2};\Delta;\frac{1}{1+\mathfrak{r}^{2}}\right)~,
    \end{split}
\end{equation}
where $\Delta:=1+\sqrt{m^{2}+1}$ and where the normalization constant $N_{\omega,\lambda}$ is fixed such that
\begin{equation}
    \label{eq:NormPsiTildeCond}
    \int_{0}^{\infty}\textrm{d}\mathfrak{r}\,\frac{1}{\mathfrak{r}}\tilde{\psi}_{\omega,\lambda}(\mathfrak{r})\tilde{\psi}_{\omega',\lambda}(\mathfrak{r})=\frac{1}{2\omega}\delta(\omega-\omega')~,
\end{equation}
and is given by
\begin{equation}
    \label{eq:NormPsiTilde}
    N_{\omega,\lambda}=\frac{\vert \Gamma\left(\frac{i\omega-i\lambda+\Delta}2{}\right)\vert \vert \Gamma\left(\frac{i\omega+i\lambda+\Delta}{2}\right)\vert}{\sqrt{4\pi\omega}\vert\Gamma(i\omega)\vert}~.
\end{equation}
In~\cite{Sugishita:2022ldv} it is argued that the implementation of the Banks--Douglas--Horowitz--Martinec (BDHM) formula~\cite{Banks:1998dd} fails for Rindler-AdS due to the presence of tachyonic (negative mass) modes ($\omega^{2}<\lambda^{2}$) that break the causality and unitarity of the CFT. This suggests that the part of the bulk local field which is reconstructable in the right Rindler wedge is given by
\begin{equation}
    \label{eq:ReconstField}
    \tilde{\phi}(\mathfrak{r},t,\chi)=\int_{-\infty}^{\infty}\frac{\textrm{d}t'}{2\pi}\int_{-\infty}^{\infty}\frac{\textrm{d}\chi'}{2\pi}\,\mathcal{K}(\mathfrak{r},t,\chi;t',\chi')\,\mathcal{O}^{\textrm{CFT},\textrm{flat}}_{\Delta}(t',\chi')~,
\end{equation}
where the smearing function $\mathcal{K}(\mathfrak{r},t,\chi;t',\chi')$ is given by
\begin{equation}
    \label{eq:SmearingAdSRindler}
    \mathcal{K}(\mathfrak{r},t,\chi;t',\chi'):=\int_{-\infty}^{\infty}\textrm{d}\lambda\int_{\vert \lambda\vert}^{\infty}\textrm{d}\omega\left(e^{-i\omega(t-t')+i\lambda(\chi-\chi')}+e^{i\omega(t-t')-i\lambda(\chi-\chi')}\right)f_{\omega,\lambda}(\mathfrak{r})~,
\end{equation}
and where $f_{\omega,\lambda}(\mathfrak{r})=\Gamma(\Delta)\,\tilde{\psi}_{\omega,\lambda}(\mathfrak{r})/N_{\omega,\lambda}$ is given by
\begin{equation}
    \label{eq:fFunctionSmear}
    f_{\omega,\lambda}(\mathfrak{r})=\mathfrak{r}^{i\omega}(1+\mathfrak{r}^{2})^{-\frac{i\omega}{2}-\frac{\Delta}{2}}\phantom{\,}_{2}F_{1}\left(\frac{i\,\omega-i\,\lambda+\Delta}{2},\frac{i\,\omega+i\,\lambda+\Delta}{2};\Delta;\frac{1}{1+\mathfrak{r}^{2}}\right)~.
\end{equation}
Note that the smearing function $\mathcal{K}$ depends only on differences $t-t'$ and $\chi-\chi'$: $\mathcal{K}(\mathfrak{r},t-t',\chi-\chi')$ and that it is, in general, complex. These facts will be relevant in the following section.

The operator $\mathcal{O}_{\Delta}^{\textrm{CFT},\textrm{flat}}(t',\chi')$ in~\eqref{eq:ReconstField} is a large-$N$ primary field evaluated at a point $(t',\chi')$ on $\mathbb{R}^{1,1}$ that can be expressed as
\begin{equation}
    \label{eq:CFTPrimary}
    \mathcal{O}_{\Delta}^{\textrm{CFT},\textrm{flat}}(t',\chi')= \int_{-\infty}^{\infty}\textrm{d}\lambda\int_{\vert \lambda\vert}^{\infty}\textrm{d}\omega\frac{N_{\omega,\lambda}^{\textrm{CFT}}}{\sqrt{2\pi}\Gamma(\Delta)}\left(a_{\omega,\lambda}^{\textrm{CFT}}e^{-i\omega t'+i\lambda \chi'}+\left(a_{\omega,\lambda}^{\textrm{CFT}}\right)^{\dagger}e^{i\omega t'-i\lambda \chi'}\right)~,
\end{equation}
where $\lbrace a^{\textrm{CFT}}_{\omega,\lambda},(a^{\textrm{CFT}}_{\omega,\lambda})^{\dagger}\rbrace $ are normalized ladder operators and where
\begin{equation}
    \label{eq:NormCFT}
    N_{\omega,\lambda}^{\textrm{CFT}}=\begin{cases} \left(\frac{\omega^{2}-\lambda^{2}}{4}\right)^{\frac{\Delta-1}{2}}&\textrm{for}\quad \omega^{2}\geq \lambda^{2}~,\\
    0 &\textrm{for}\quad \omega^{2}< \lambda^{2}~,
    \end{cases}
\end{equation}
is the normalization in large-$N$ CFTs on Minkowski space $\mathbb{R}^{1,1}$. The relation between $\tilde{\phi}(\mathfrak{r},t,\chi)$ and $\mathcal{O}_{\Delta}^{\textrm{CFT},\textrm{flat}}(t,\chi)$ is given by
\begin{equation}
    \label{eq:LimitFieldtoOperator}
    \lim_{\mathfrak{r}\rightarrow \infty}\mathfrak{r}^{\Delta}\tilde{\phi}(\mathfrak{r},t,\chi)=\mathcal{O}_{\Delta}^{\textrm{CFT},\textrm{flat}}(t,\chi)~.
\end{equation}
The primary field $\mathcal{O}_{\Delta}^{\textrm{CFT},\textrm{flat}}(t,\chi)$ at a point $(t,\chi)$ on Minkowski $\mathbb{R}^{1,1}$ can also be obtained by a conformal transformation of the primary field on the cylinder $\mathcal{O}_{\Delta}^{\textrm{CFT}}(\tau,\theta)$ according to $\mathcal{O}_{\Delta}^{\textrm{CFT},\textrm{flat}}(t,\chi)=e^{\Delta \Phi(t,\chi)}\mathcal{O}_{\Delta}^{\textrm{CFT}}(\tau(t,\chi),\theta(t,\chi))$ where
\begin{equation}
    \label{eq:ConformalFactor}
    e^{\Phi(t,\chi)}=\frac{1}{\sqrt{\cosh^{2}(\chi)+\sinh^{2}(t)}}~,
\end{equation}
is the conformal factor taking the asymptotic metric $-\textrm{d}\tau^{2}+\textrm{d}\theta^{2}$ on the asymptotic boundary of the global patch of AdS (a cylinder $\mathbb{R}_{\tau}^{1}\times \mathbb{S}^{1}$ with $-\pi/2\leq \theta \leq \pi/2$) to the metric $e^{2\Phi}(-\textrm{d}t^{2}+\textrm{d}\chi^{2})$ on $\mathbb{R}_{t}^{1}\times\mathbb{H}^{1}$ via
\begin{equation}
    \label{eq:GlobaltoRindler}
    \tan(\tau)=\frac{\sinh(t)}{\cosh(\chi)}~,\quad\tan(\theta)=\frac{\sinh(\chi)}{\cosh(t)}~.
\end{equation}
In terms of null coordinates $u=t-\chi$, $v=t+\chi$, $e^{\Phi(u,v)}=(\cosh(u)\cosh(v))^{-1/2}$.

\subsection{Bulk Inner Product}

With these ingredients, we can tackle the question of defining an inner product in the bulk of Rindler-AdS from the perspective of the boundary inner product of operators. Here, we distinguish the coordinates on the right Rindler wedge $(\mathfrak{r}_{R},t_{R},\chi_{R})$ from the left Rindler wedge $(\mathfrak{r}_{L},t_{L},\chi_{L})$. Recall that the left wedge can be obtained from the right one by a shift in imaginary time $t_{L}=t_{R}-i\pi$. In the following, we will drop the superscripts ``CFT,flat'' from the CFT operators and the tilde from the scalar fields to simplify the notation.

Following the HKLL formula~\eqref{eq:ReconstField}, the bulk correlator of the reconstructible part of two free massive scalar fields in Rindler-AdS$_{3}$ can be formally written as
\begin{equation}
    \label{eq:BulkCorrelator}
    \begin{split}
&\langle\phi_{1}^{\dagger}(\mathfrak{r}_{1},t_{1},\chi_{1})\phi_{2}(\mathfrak{r}_{2},t_{2},\chi_{2})\rangle_{\textrm{R-AdS}}=\\
    &\frac{1}{(2\pi)^{4}}\int_{-\infty}^{\infty}\textrm{d}t_{1}'\,\textrm{d}\chi_{1}'\,\textrm{d}t_{2}'\,\textrm{d}\chi_{2}'\Bigg[\mathcal{K}_{1}^{\ast}(\mathfrak{r}_{1},t_{1}-t_{1}',\chi_{1}-\chi_{1}')\mathcal{K}_{2}(\mathfrak{r}_{2},t_{2}-t_{2}',\chi_{2}-\chi_{2}')\times\\
    & \times\langle\mathcal{O}^{\dagger}_{\Delta_{1}}(t_{1}',\chi_{1}')\mathcal{O}_{\Delta_{2}}(t_{2}',\chi_{2}')\rangle_{\textrm{CFT}}\Bigg]~,
    \end{split}
\end{equation}
with $\Delta_{1,2}=1+\sqrt{m_{1,2}^{2}+1}$.
The boundary correlator $\langle\mathcal{O}^{\dagger}_{\Delta_{1}}(t_{1}',\chi_{1}')\mathcal{O}_{\Delta_{2}}(t_{2}',\chi_{2}')\rangle_{\beta}$, which in $\mathbb{R}_{t}^{1}\times \mathbb{H}^{1}\cong \mathbb{R}^{1,1}$ at finite temperature $\beta^{-1}=1/(2\pi)$ is given by
\begin{equation}
    \label{eq:BdryProp}
    \langle\mathcal{O}^{\dagger}_{\Delta_{1}}(t_{1}',\chi_{1}')\mathcal{O}_{\Delta_{2}}(t_{2}',\chi_{2}')\rangle_{\beta=2\pi}=\frac{\delta_{\Delta_{1},\Delta_{2}}c_{\Delta_1}}{\left(-\cosh(t_{1}'-t_{2}')+\cosh(\chi_{1}'-\chi_{2}')\right)^{\Delta_1}}~,
\end{equation}
where $c_{\Delta}$ is a normalization constant. {Performing an appropriate analytic continuation, (\ref{eq:BdryProp})} can be used to define an inner product for scalar primary fields in the CFT using the relation
\begin{equation}
    \label{eq:CFTCorrelatorAndInnerProduct}
    \begin{split}
    \left(\mathcal{O}_{\Delta_{1}}(t_{1}',\chi_{1}')\vert\mathcal{O}_{\Delta_{2}}(t_{2}',\chi_{2}')\right)_{\textrm{CFT}}&:=\langle\mathcal{O}^{\dagger}_{\Delta_{1}}(t_{1}'-i\pi,\chi_{1}')\mathcal{O}_{\Delta_{2}}(t_{2}',\chi_{2}')\rangle_{\beta=2\pi}=\\
    &=\frac{\delta_{\Delta_{1},\Delta_{2}}c_{\Delta_1}}{\left(\cosh(t_{1}'-t_{2}')+\cosh(\chi_{1}'-\chi_{2}')\right)^{\Delta_1}} ~.
    \end{split}
\end{equation}
This inner product can be seen as a special case of the more general expression for the Wightman (thermal) inner product for CFT$_{2}$ fields in $\mathbb{R}^{1}_{t}\times \mathbb{H}^{1}\cong \mathbb{R}^{1,1}$ $A(t_{A},\chi_{A}),$ $B(t_{B},\chi_{B})$ with $\beta=2\pi$ given formally by
\begin{equation}
    \label{eq:CFTInnerProductGeneral}
    \begin{split}
&\left(A(t_{A},\chi_{A})\vert B(t_{B},\chi_{B})\right)^{W}:=\langle A^{\dagger}(t_{A}-\frac{i\beta}{2},\chi_{A})B(t_{B},\chi_{B})\rangle_{\beta}\equiv\\
&\frac{\textrm{tr}\left(e^{-\beta H}A^{\dagger}(t_{A}-\frac{i\beta}{2},\chi_{A})B(t_{B},\chi_{B})\right)}{Z(\beta)}=\frac{\textrm{tr}\left(e^{-\frac{\beta H}{2}}A^{\dagger}(t_{A},\chi_{A})e^{-\frac{\beta H}{2}}B(t_{B},\chi_{B})\right)}{\textrm{tr}\left(e^{-\beta H}\right)}~,
\end{split}
\end{equation}
As a consequence, we propose a formula for the inner product of reconstructible scalar fields $\phi_{1},\phi_{2}$ in Rindler-AdS$_{3}$ using the HKLL formula~\eqref{eq:HKLLCorrleator} as follows:
\begin{equation}
    \label{eq:BulkInnProdfromHKLL1}
    (\phi_{1}(\mathfrak{r}_{1}, t_{1},\chi_{1})\vert\phi_{2}(\mathfrak{r}_{2},t_{2},\chi_{2}))_{\textrm{R-AdS}}:=\langle\phi_{1}^{\dagger}(\mathfrak{r}_{1},t_{1}-i\pi,\chi_{1})\phi_{1}(\mathfrak{r}_{2},t_{2},\chi_{2})\rangle_{\textrm{R-AdS}}~.
\end{equation}
To be precise:
\begin{equation}
    \label{eq:BulkInnProdfromHKLL}
    \begin{split}
    &(\phi_{1}(\mathfrak{r}_{1}, t_{1},\chi_{1})\vert\phi_{2}(\mathfrak{r}_{2},t_{2},\chi_{2}))_{\textrm{R-AdS}}\equiv\\
    &\frac{1}{(2\pi)^{4}}\int_{-\infty}^{\infty}\textrm{d}t_{1}'\,\textrm{d}\chi_{1}'\,\textrm{d}t_{2}'\,\textrm{d}\chi_{2}'\Bigg[\mathcal{K}_{1}^{\ast}(\mathfrak{r}_{1},t_{1}-t_{1}',\chi_{1}-\chi_{1}')\,\mathcal{K}_{2}(\mathfrak{r}_{2},t_{2}-t_{2}',\chi_{2}-\chi_{2}')\times\\
    & (\mathcal{O}_{\Delta_{1}}(t_{1}',\chi_{1}')\vert\mathcal{O}_{\Delta_{2}}(t_{2}',\chi_{2}'))_{\textrm{CFT}}\Bigg]~.
    \end{split}
\end{equation}
Note that we also perform the analytic continuation to imaginary time $t_{1}\rightarrow t_{1}-i\pi,t_{1}'\rightarrow t_{1}'-i\pi$ in the smearing function of the first field $\mathcal{K}_{1}$, which amounts to placing the first field on the left Rindler wedge. However, because this function depends only on the time difference $t_{1}-t_{1}'$, the imaginary time rotation by $-i\pi$ leaves the function unchanged. Note as well that the CFT inner product~\eqref{eq:CFTCorrelatorAndInnerProduct} only depends on the differences $t_{1}'-t_{2}'$ and $\chi_{1}'-\chi_{2}'$ and not on their independent values. This also means that the inner product is invariant under replacements $t_{1}'\leftrightarrow t_{2}'$ $\chi_{1}'\leftrightarrow \chi_{2}'$.

The first question is whether~\eqref{eq:BulkInnProdfromHKLL} is a well-defined inner product. It is known that the CFT inner product~\eqref{eq:CFTCorrelatorAndInnerProduct} satisfies the necessary and sufficient conditions for a well-defined inner product, namely (see e.g. \cite{bogachev2017topological}):
\begin{subequations} \label{eq:innerprodpropCFT}
\begin{equation}
    \label{eq:innerprodpropCFT1}
    (\mathcal{O}_{\Delta_{1}}(t_{1}',\chi_{1}')\vert \mathcal{O}_{\Delta_{2}}(t_{2}',\chi_{2}'))_{\textrm{CFT}}=(\mathcal{O}_{\Delta_{2}}(t_{2}',\chi_{2}')\vert \mathcal{O}_{\Delta_{1}}(t_{1}',\chi_{1}'))_{\textrm{CFT}}^{\ast}~,
\end{equation}
\begin{equation}
    \label{eq:innerprodpropCFT2}
    \begin{split}
    &(a_{1}\,\mathcal{O}_{\Delta_{1}}(t_{1}',\chi_{1}')+a_{2}\,\mathcal{O}_{\Delta_{2}}(t_{2}',\chi_{2}')\vert \mathcal{O}_{\Delta_{3}}(t_{3}',\chi_{3}'))_{\textrm{CFT}}=\\
    &a_{1}^{\ast}\,(\mathcal{O}_{\Delta_{1}}(t_{1}',\chi_{1}')\vert \mathcal{O}_{\Delta_{3}}(t_{3}',\chi_{3}'))_{\textrm{CFT}}+a_{2}^{\ast}\,(\mathcal{O}_{\Delta_{2}}(t_{2}',\chi_{2}')\vert \mathcal{O}_{\Delta_{3}}(t_{3}',\chi_{3}'))_{\textrm{CFT}}~,
    \end{split}
\end{equation}
\begin{equation}
    \label{eq:innerprodpropCFT3}
    (\mathcal{O}_{\Delta_{1}}(t_{1}',\chi_{1}')\vert \mathcal{O}_{\Delta_{1}}(t_{1}',\chi_{1}'))_{\textrm{CFT}}\geq 0~.
\end{equation}
\end{subequations}
The first three properties are indeed simple to prove. The last one is in contrast more complicated since $\mathcal{K}$ is in general complex so it might generically not be satisfied. 

What we would like to show explicitly is that by substituting the expression for the CFT inner product~\eqref{eq:CFTCorrelatorAndInnerProduct} into the expression for the bulk inner product~\eqref{eq:BulkInnProdfromHKLL} and after evaluating the smearing function appropriately~\eqref{eq:SmearingAdSRindler} that we recover the autocorrelation function~\eqref{eq:afbbp_eps_pi} for two-sided correlators used in Sec.~\ref{sec:eps_pi_twosidedcorr} of the main text. However, performing this task for arbitrary bulk points is in general difficult, due to the distributional nature and mathematical structure of the smearing function $\mathcal{K}$. Nevertheless, in the important case where bulk reconstruction is performed by an observer at the asymptotic boundary $\mathfrak{r}\rightarrow\infty$, we can do the derivation explicitly and recover the usual CFT result.

Taking $\mathfrak{r}\rightarrow\infty$ in (\ref{eq:fFunctionSmear}), we have
\begin{equation}
    f_{\omega,\lambda}(\mathfrak{r}\rightarrow\infty)=r^{-\Delta}~,
\end{equation}
since ${}_1F_2(a,b;c;1)=1$. This means that $f_{\omega,\lambda}$ is just an overall scaling parameter that we should regularize. 

Next, to evaluate $\mathcal{K}$ in (\ref{eq:SmearingAdSRindler}), we note that $f_{\omega,\lambda}=$constant implies
\begin{equation}\label{eq:new K}
    \mathcal{K}=2\int_{-\infty}^{\infty}\rmd\lambda\qty(\int_{0}^{\infty}-\int_0^{\abs{\lambda}})\rmd\omega~(\cos[\lambda(\chi-\chi')]\cos[\omega(t-t')]+\sin[\lambda(\chi-\chi')]\sin[\omega(t-t')])~.
\end{equation}
We can use the integral representation found \cite{friedman1990principles}
\begin{equation}
    \delta(x-x')=\frac{1}{\pi}\int_{0}^{\infty}\cos{(k(x-x'))}\rmd k~,
\end{equation}
so that $\mathcal{K}$ in (\ref{eq:new K})
\begin{equation}
    \mathcal{K}={4\pi^2}\delta(\chi-\chi')\delta(t-t')+{\rm others}
\end{equation}
where the other terms seem to depend on the way that we evaluate them in the complex plane by the Cauchy principal value. Explicitly:
\begin{align}
    &\begin{aligned}
    \int_{-\infty}^{\infty}\rmd\lambda\int_0^{\abs{\lambda}}\rmd\omega &\qty(\cos[\lambda(\chi-\chi')]\cos[\omega(t-t')]+\sin[\lambda(\chi-\chi')]\sin[\omega(t-t')])\\
    &=2\int_{0}^{\infty}\rmd\lambda\frac{\sin[\lambda(t-t')]}{t-t'}~,\end{aligned}\label{eq:First int}\\
    &\int_{-\infty}^{\infty}\rmd\lambda\int_{-\infty}^{\infty}\rmd\omega \sin[\lambda(\chi-\chi')]\sin[\omega(t-t')]=0~.\label{eq:second int}
\end{align}
Note that (\ref{eq:First int}) is not well defined, so we should regularize it by the Cauchy principal value; while (\ref{eq:second int}) is consequence of the odd integrand in the $\lambda$ integral. So that (\ref{eq:ReconstField}) becomes
\begin{equation}
    \tilde{\phi}(r_B,t,\chi)=r_B^{-\Delta} \mathcal{O}_\Delta^{\rm CFT,flat}(t,\chi)~,
\end{equation}
where $r_B\rightarrow\infty$ is a regularization radial parameter. It follows that at least in this limit the bulk inner product (\ref{eq:BulkCorrelator}) is well defined; although, this is not useful for the case of interest in the main text for arbitrary points inside the bulk.

However, in general, we do not have a way to analytically evaluate the bulk operators when the reconstruction is done inside the bulk. One might attempt to perform the numerical integration above and compare with CFT correlators. We leave a more involved exploration in this direction for future directions.

\section{Moment method}\label{app:app1}
Starting from the Heisenberg picture, we use the Choi-Jamiokowski isomorphism \cite{Jamiolkowski:1972pzh,Choi:1975nug} which maps an operator $\hat{O}$ acting on states in Hilbert space $\mathcal{H}$ to a state $\ket{O}$ in a double-copy Hilbert space $\mathcal{H}\otimes\mathcal{H}$, where $\ket{O}$ can be expanded in a complete basis of states $\qty{\ket{\chi_n}}\in\mathcal{H}$ as
\begin{equation}\label{eq:Op as state Krylov}
\begin{aligned}
  |O)&\equiv\sum_{m,\,n}O_{nm}\ket{\chi_m,~\chi_n}~,
\end{aligned}
\end{equation}
where $O_{nm}\equiv\bra{\chi_m}\hat{O}\ket{\chi_n}$. In the following, we will consider the Frobenius product for defining the inner product of these states as:
\begin{equation}\label{eq:inner prod}
    (X|Y)=\frac{1}{D_{\mathcal{H}}}\tr(X^\dagger Y)~,
\end{equation}
where $D_{\mathcal{H}}$ refers to the Hilbert space dimension.\footnote{{{For thermal ensembles instead, one can define the operator scalar product \cite{Parker:2018yvk}
\begin{align}\label{eq:inner H}
(A|B)_\beta^g:=\frac{1}{Z}\int_0^\beta g(\lambda)\text{Tr}[e^{-(\beta-\lambda)H}A^\dagger e^{-\lambda H}B] d\lambda~,
\end{align}
where $Z:=\text{Tr}[e^{-\beta H}]$, and $g(\lambda)$ is a function that satisfies
\begin{align}
g(\lambda)\ge0,\;\;\;g(\beta-\lambda)=g(\lambda),\;\;\;\int_0^\beta g(\lambda) d\lambda=1~.
\end{align}
The corresponding autocorrelation function in terms of \eqref{eq:inner H} is
\begin{align}\label{eq:new correlator}
C_\beta^g(t):=(\mathcal{O}|\mathcal{O}(t))_\beta^g=\frac{1}{Z(\beta)}\int_0^\beta g(\lambda)\text{Tr}[\rme^{-\beta H}\mathcal{O} \mathcal{O}(t+\rmi\lambda)] d\lambda~,
\end{align}
where we assume $\mathcal{O}^\dagger=\mathcal{O}$.
One can also choose $g(\lambda)$ as \cite{Parker:2018yvk}
\begin{align}
g(\lambda)=\frac{1}{2}[\delta(\lambda-\epsilon)+\delta(\beta-\lambda-\epsilon)],
\end{align}
with $0<\epsilon\ll1$
so that the autocorrelation function \eqref{eq:new correlator} becomes an even function through the sum of two correlation functions with $t+\rmi\epsilon$ and $t-\rmi\epsilon$; explicitly:
\begin{equation}
    C_\beta^g(t)=\frac{1}{2Z(\beta)}\qty(\Tr\qty[\rme^{-\beta H}\mathcal{O}\mathcal{O}(t+\rmi\epsilon)]+\Tr\qty[\rme^{-\beta H}\mathcal{O}(t-\rmi\epsilon)\mathcal{O}])~.
\end{equation}}
The autocorrelation function \eqref{eq:toda epsilon auto} is not an even function in time since it corresponds to $g(z)=\delta(\beta-\lambda+\epsilon)$ in \eqref{eq:new correlator}.} Other choices of inner products inherently related to finite temperature ensembles can be found in \cite{Parker:2018yvk,Barbon:2019wsy}.}

We can represent the evolution of the operator through the Heisenberg equation as
\begin{align}\label{eq: Heisenberg time evol}
\partial_t|O(t))&=\rmi\mathcal{L}|O(t))~,
\end{align}
where $\mathcal{L}$ is called the Liouvillian super-operator,
\begin{equation}
    \mathcal{L}=\qty[H,~\cdot~],\quad O(t)=\rme^{\rmi\mathcal{L}t}O~.
\end{equation}
We can then solve (\ref{eq: Heisenberg time evol}) in terms of a Krylov basis, $\qty{|O_n)}$,
\begin{equation}\label{eq:amplitudes}
\begin{aligned}
    |O (t))&=\sum_{n=0}^{\mathcal{K}-1} i^n \varphi_n(t) |O_n)~,\\
    \varphi_n(t)&=(O_n|\rme^{\rmi \mathcal{L}t}|O_n)~,\quad (O_m|O_n)=\delta_{mn}~.
\end{aligned}
\end{equation}
The Lanczos algorithm for determining the Krylov basis, $\qty{|O_n)}$, takes the form \cite{Kundu:2023hbk}
\begin{align}
    &|A_{n+1})=\mathcal{L}|O_{n})-b_{n}|O_{n-1})-a_n|O_n)~,\quad a_n:=(O_n|\mathcal{L}|O_n)~,\label{eq:Lanczos op}\\
    &|O_{n})=b_n^{-1}|A_{n})~,\quad b_{n\geq1}=(A_{n}|A_{n})^{1/2}~,~b_0=1~.\label{eq:K amplitudes}
\end{align}
There are further implications if we assume that $O(t)$ is a Hermitian operator. In this case, the correlation function is an even function in $t$ that can be expanded as a Taylor series as
\begin{equation}\label{eq:2pnt correlator Krylov}
    \varphi_0(t)=(O(t)|O(0))=\sum_nm_{2n}\frac{(-1)^{n}t^{2n}}{(2n)!}~,
\end{equation}
where $m_{2n}$ are referred to as the ``moments". The Lanczos coefficients $b_n$ can be then determined from the moments using an equivalent algorithm to (\ref{eq:K amplitudes}) \cite{Parker:2018yvk,viswanath1994recursion,Bhattacharjee:2022ave}
\begin{align}\label{eq:Alt Lanczos}
    b_n=\sqrt{Q_{2n}^{(n)}}~,\quad Q_{2k}^{(m)}=\frac{Q_{2k}^{(m-1)}}{b_{m-1}^2}-\frac{Q_{2k-2}^{(m-2)}}{b_{m-2}^2}~,
\end{align}
where $Q_{2k}^{(0)}=m_{2k}$, and $Q_{2k}^{(-1)}=0$.

The other amplitudes can be determined through the Lanczos algorithm (\ref{eq:K amplitudes}) and the Heisenberg equation (\ref{eq: Heisenberg time evol}), resulting in the recursion relation:
\begin{equation}\label{eq:sch eq K operator}
    \partial_t\varphi_n(t)=b_n\varphi_{n-1}(t)-b_{n+1}\varphi_{n+1}(t)~.
\end{equation}
Krylov complexity is then defined as
    \begin{equation}\label{eq:Krylov complexity}
        K(t)\equiv\sum_{n=0}^{\mathcal{K}-1}n|\varphi_n(t)|^2\,.
    \end{equation}
The above definition was originally motivated \cite{Parker:2018yvk} to describe the size of the operator under Hamiltonian evolution, as it measures the mean width of a wavepacket in the Krylov space.

{\section{Analytic Domains and Krylov Growth for Smeared Operators}}
{In this appendix, we discuss the conditions under which the Krylov
complexity of non-local operators exhibits the same exponential behavior
as that of local operators.
As emphasized in \cite{Parker:2018yvk}, the
Krylov exponent associated with an autocorrelation function is fixed by
the region in the complex $t$–plane where the correlator is analytic,
which is typically a horizontal strip. The width of this strip then
determines the Krylov exponent}.

{In the particular case where the smearing kernels arise from HKLL
reconstruction, the resulting smeared autocorrelators provide the
boundary dual of the bulk calculation in the main text for local bulk
fields.}

{\subsection{Analytic domain of Wightman functions involving local operators}}

{We begin by showing that a positive-frequency Wightman two-point function of the form,\footnote{{Positive frequency in this context means that the two-point function in momentum space only picks up contributions by integrating around the negative frequency pole.}}}
\begin{equation}
    {G^{+}(t,x)
    = \langle \mathcal{O}(t,x)\mathcal{O}(0,0) \rangle_\beta
    = \frac{\mathrm{Tr}\!\left[e^{-\beta H}\,\mathcal{O}(t,x)\mathcal{O}(0,0)\right]}
           {\mathrm{Tr}\!\left[e^{-\beta H}\right]} }\,,
\end{equation}
{is analytic for $-\beta < \Im \,t < 0$. We follow the argument in footnote~11 of \cite{Ohya:2016gto}.}

{Expanding in an energy eigenbasis $H |n\rangle = E_n |n\rangle$, we obtain}
\begin{equation}
    {G^{+}(t,x)
    = \frac{1}{Z} \sum_{n,m}
      \langle n|\mathcal{O}(0,x)|m\rangle\,
      \langle m|\mathcal{O}(0,0)|n\rangle\,
      \rme^{E_n (\rmi t - \beta)} e^{- \rmi E_m t}}\,.
\end{equation}
{Writing $t = \Re\,t + \rmi \Im\,t$, this becomes}
\begin{equation} \label{eq:Gsum}
   { G^{+}(t,x)
    = \frac{1}{Z} \sum_{n,m}
      \langle n|\mathcal{O}(0,x)|m\rangle\,
      \langle m|\mathcal{O}(0,0)|n\rangle\,
      \rme^{\rmi E_n \Re\,t} e^{-\rmi E_m \Re\,t}
      \rme^{-(\beta + \Im\,t) E_n} \rme^{E_m\Im\,t}\,.}
\end{equation}
{Restricting to $H$ with positive spectrum, so that $E_n, E_m \geq 0$ for all $n,m$, we see that the sum \eqref{eq:Gsum} converges only if}
{\[
\beta + \Im\,t > 0,
\qquad
\Im\,t < 0,
\]}
{which together imply $-\beta < \Im\,t < 0$. This defines the strip in the complex $t$-plane inside which $G^{+}(t,x)$ is analytic.  
Similarly, one finds that the negative-frequency Wightman function}
{\[
G^{-}(t,x) = \langle \mathcal{O}(0,0)\mathcal{O}(t,x) \rangle_\beta
\]}
{is analytic in the strip $0 < \Im\,t < \beta$.}

{We will now use this result to show that Wightman functions involving smeared operators remain analytic in the same region of the complex plane, provided the smearing kernels $K$ are analytic everywhere.}

{\subsection{Analytic domain of Wightman functions involving non-local operators}}

{We now consider a non-local operator obtained by smearing a local field with a kernel $K$,}
\begin{equation}
{\mathcal{O}_{K}(t,x)
=
(K * \mathcal{O})(t,x)
\equiv
\int dt'\, d\!x'\; K(t-t',x-x')\,\mathcal{O}(t',x')\,.}
\end{equation}
{The corresponding thermal Wightman function is}
\begin{equation}
{G_{K}(t,x)
=
\langle \mathcal{O}_{K}(t,x)\,\mathcal{O}_{K}(0,0)\rangle_{\beta}
=
K * G^{+} * K \, ,}
\end{equation}
{where the stars denote convolution in $(t,x)$. To determine the analytic structure of the smeared correlator it is more transparent to work in frequency space. Let $\widetilde G^{+}(\omega,k)$ and $\widetilde K(\omega,k)$ denote the Fourier transforms of $G^{+}$ and $K$. Since convolutions become products, we have }
\begin{equation}
{\widetilde G_{K}^{+}(\omega,k)
= 
\widetilde K(\omega,k)\,
\widetilde K(-\omega,-k)\,
\widetilde G^{+}(\omega,k)\,.}
\label{eq:freq_prod}
\end{equation}
T{hermal Wightman functions have a well–known frequency–space analytic structure: $G^{+}(t,x)$ is analytic in the strip $-\beta < \Im t < 0$, and this is encoded in the fact that $\widetilde G^{+}(\omega,k)$ has poles (the Matsubara poles) at }
{\[
\omega = - i \frac{2\pi n}{\beta}\,,\qquad n=1,2,\dots,
\]}
{with no singularities between the real axis and the line $\Im\omega = - 2\pi/\beta$. Thus the location of the singularities of $\widetilde G^{+}(\omega,k)$ completely fixes the strip of analyticity of $G^{+}(t,x)$.}

{Equation \eqref{eq:freq_prod} shows that the smeared correlator $\widetilde G_{K}^{+}$ can only develop new singularities if $\widetilde K(\omega,k)$ itself has poles or branch cuts. If the smearing kernel is analytic everywhere in the frequency space—for example a Gaussian—then $\widetilde K(\omega,k)$ is analytic for all complex $\omega$ and therefore cannot generate new singularities. In this case, $\widetilde G_{K}^{+}(\omega,k)$ has exactly the same singularities as $\widetilde G^{+}(\omega,k)$, and no new obstructions arise when performing the inverse Fourier transform. Consequently,}
\begin{equation}
{G_{K}^{+}(t,x)\ \text{is analytic in the same strip}\quad
-\beta < \Im t < 0\,.}
\end{equation}
{Thus the analytic domain of the smeared Wightman function is determined entirely by the singularities of $\widetilde K$ and $\widetilde G^{+}$ in the complex $\omega$–plane. If $K$ is analatyic everywhere, the smearing cannot move, remove, or create singularities, and therefore the analyticity strip of the original correlator is preserved.
Since the analytic structure of the autocorrelation function determines the Krylov exponent, one does not expect this exponent to change when the smearing kernel is analytic everywhere; in such cases the large-time behavior should remain unaffected.}

{The situation is more subtle for the HKLL smearing kernel. In the main text we analyze boundary operators smeared over the boundary using HKLL kernels
$K$, which exhibit singularities whenever the bulk point is spacelike-separated from the boundary. Despite this more complicated analytic structure, our numerics indicate that the Krylov exponent remains unchanged for arbitrary bulk radial position 
$r$. This is a nontrivial result, given that the HKLL kernel introduces additional singularities in frequency space that, a priori, could have modified the analyticity domain relevant for the Krylov exponent.}

{\subsection{Numerical exploration of effects of locality on the Lanczos coefficients}}

{In this section, we discuss the effects of non-locality of operators in the growth of Lanczos coefficients/Krylov exponent in the context of lattice discretizations of CFTs. Here, we focus on the case of the transverse-field (XXZ) Ising chain}
\begin{align}
    \label{eq:XXZChain}
    {\hat{H}=-\sum_{i=1}^{N}\left(2J\,\hat{S}^{x}_{i}\hat{S}^{x}_{i+1}+h\,\hat{S}^{z}_{i}\right)~,}
\end{align}
{which in the critical limit $J=h$ and for large $N$ corresponds to a lattice discretization of the free $(1+1)$-dimensional $c=1/2$ CFT. Here $\hat{S}^{\alpha}_{i}$ are spin-$1/2$ operators in the $\alpha$ direction at position $i$ in the chain and are given by}
\begin{align}
    \label{eq:SpinOpLattice}
  {  \hat{S}^{\alpha}_{i}:=\mathbb{I}_{2}^{\otimes (i-1)}\otimes \left(\frac{\hat{\sigma}_{\alpha}}{2}\right)\otimes \mathbb{I}_{2}^{\otimes(N-i)}~,}
\end{align}
{where $\hat{\sigma}_{\alpha}$ are the usual Pauli matrices for $\alpha \in \lbrace x,y,z\rbrace$ and where we use the identification $\hat{S}^{\alpha}_{N+1}=\hat{S}^{\alpha}_{1}$. Thus, the system consists of $N$ spin-$1/2$ degrees of freedom with periodic boundary conditions $N+i\sim i$. The spin-$1/2$ operators $\hat{S}^{\alpha}_{i}$ act locally at each chain site and the Hamiltonian~\eqref{eq:XXZChain} has an interaction term which acts locally on two neighboring sites. As is well-known, this model becomes analytically solvable via the Jordan--Wigner transform~\cite{Jordan:1928wi}. This transformation is based on the introduction of fermionic creation and annihilation operators $\hat{f}^{\dagger}_{i}$ and $\hat{f}_{i}$. Defining the spin-$1/2$ ladder operators $\hat{S}^{\pm}_{i}:=\hat{S}^{x}_{i}\pm i\hat{S}^{y}_{i}$, these are related to the fermionic operators via}
\begin{align}
    \label{eq:JWTrafo}
    \begin{split}
    &{\hat{S}^{+}_{i}=\exp\left(+i\pi\sum_{j=1}^{i-1}\hat{n}_{j}\right)\hat{f}^{\dagger}_{i}~,}\\
    &{\hat{S}^{-}_{i}=\exp\left(-i\pi\sum_{j=1}^{i-1}\hat{n}_{j}\right)\hat{f}_{i}~,}
    \end{split}
\end{align}
{where $\hat{n}_{j}:=\hat{f}^{\dagger}_{j}\hat{f}_{j}$ is the fermionic occupation number at site $j$. Equivalently, the inverse transformation is given by}
\begin{align}
    \label{eq:InvJWTrafoInverse}
    \begin{split}
    &{\hat{f}^{\dagger}_{i}=\left(\prod_{j=1}^{i-1}\hat{S}^{z}_{j}\right) \hat{S}^{+}_{i}~,}\\
    &{\hat{f}_{i}=\left(\prod_{j=1}^{i-1}\hat{S}^{z}_{j}\right) \hat{S}^{-}_{i}~.}
    \end{split}
\end{align}
{The exponential factor appearing in the fermionic operators~\eqref{eq:JWTrafo} represents a string operator that acts non-locally on sites preceding the insertion of the corresponding fermionic operator. Because of this, from the spin perspective, the fermionic operators are non-local. This is represented in Fig.~\ref{fig:JWTrafo}.}

\begin{figure}
    \centering
    \includegraphics[width=0.5\linewidth]{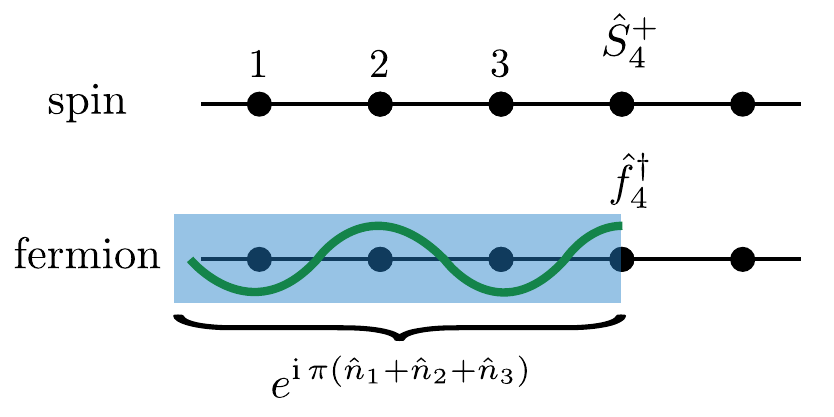}
    \caption{Representation of the Jordan--Wigner transformation on a lattice. The ladder spin operator $\hat{S}^{+}_{4}$ is written as the product of the fermionic creation operator $\hat{f}^{\dagger}_{4}$ and the string (parity) operator $e^{i\pi(\hat{n}_{1}+\hat{n}_{2}+\hat{n}_{3})}$.}
    \label{fig:JWTrafo}
\end{figure}

{The concrete idea that we want to test in this appendix is whether a local spin operator on a fixed site grows faster or slower than a fermionic operator inserted on the same site. To be precise, we want to compare the Lanczos coefficients from two choices of initial operators: $\mathcal{O}_{0}=\hat{S}^{+}_{4},\hat{f}^{\dagger}_{4}$. Fig.~\ref{fig:bnIsingL8} shows the Lanczos coefficients for these initial operators using the Wightmann inner product for $N=8$ and for $\beta=1$ and computed via the moment method described in Appendix~\ref{app:app1}.}

\begin{figure}
    \centering
    \includegraphics[width=0.5\linewidth]{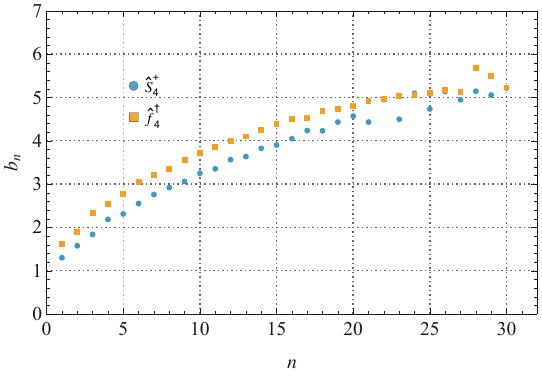}
    \caption{Lanczos coefficients for $N=8$ and $\beta=1$.}
    \label{fig:bnIsingL8}
\end{figure}
 {Although the numerical values of the coefficients computed from $\hat{f}^{\dagger}_{4}$ lie above the ones from $\hat{S}^{+}_{4}$, their growth rate does not appear to vary qualitatively. There are some fluctuations arising from numerical accuracy, but we do not expect this behavior to change significantly for larger $N$. This is consistent with the analysis in the previous subsection for smeared operators. We should remark, nonetheless, that the approach discussed here differs from that one, in that here the non-locality here arises from a non-local contribution (in the form of $\prod_{j=1}^{4}\hat{S}^{z}_{j}$) rather than the smearing of the operator with a smooth function.}}

\section{Lanczos coefficients at a finite radial location from the full propagator}\label{app:figsbnKtfiniter}
In this Appendix, we collect and analyze numerical results for the Lanczos coefficients as well as their associated Krylov complexity obtained from the autocorrelation function~\eqref{eq:autocorrelation function main} arising from the full bulk-to-bulk propagator~\eqref{eq:bbp} for a fixed radial location $r$ at for a finite time window $\Delta t$. We have chosen values for $\Delta$ such that obey the following bounds
\begin{itemize}
    \item Unitarity bound: $\Delta \geq \frac{d-2}{2}$~.
    \item Breitenlohner-Freedman bound $ m^2 \geq -\frac{d^2}{4} $, i.e. $\Delta(\Delta - d) \geq -\frac{d^2}{4}$~.
\end{itemize}
Due to the structure of the Lanczos coefficients at the AdS boundary $r\rightarrow \infty$ and horizon $r\rightarrow 1$, two interesting regimes that we consider are the following: $\Delta \in [\frac{d-2}{2},\frac{d-1}{2}]$ and $\Delta \in [\frac{d-1}{2},\infty)$. In both regimes, both bounds are satisfied. In the first regime, the squared mass of the dual scalar field to the CFT operator is always negative, whereas in the second case it can be positive or negative, depending on whether $\Delta > d$ or $d>\Delta >(d-1)/2$.

\subsection{\texorpdfstring{$2\Delta > (d-1)$}{}}\label{app:figsbnKtfiniterDelta15d4}

\begin{figure}[ht]
    \centering
    \begin{tabular}{cc}
        \includegraphics[width=0.45\textwidth]{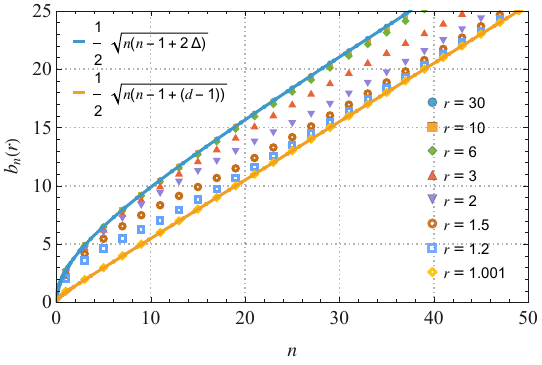} & 
        \includegraphics[width=0.45\textwidth]{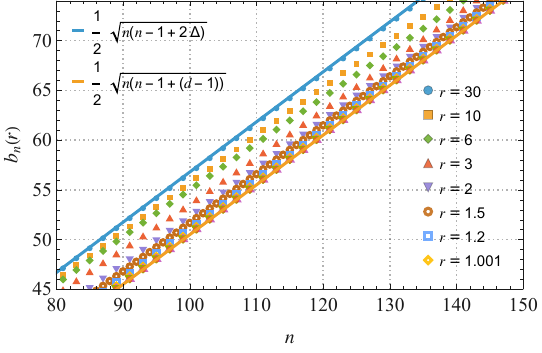} \\
        (a) & (b) \\
        \multicolumn{2}{c}{\centering\includegraphics[width=0.45\textwidth]{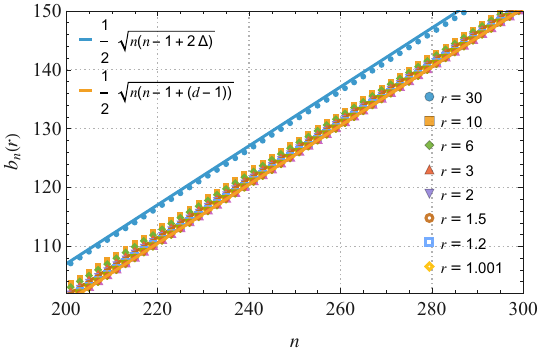}}  \\
       \multicolumn{2}{c}{(c)} \\
    \end{tabular}
     \caption{Behaviour of Lanczos coefficients $b_n$ for $C(t,r)$ (\ref{eq:autocorrelation function main}) with $\Delta=15$ and $d=4$ for different $r$ in the early (a), intermediate (b), and large (c) $n$ regime. Blue and orange solid curves are $b_n=\frac{1}{2}\sqrt{n(n-1+2\Delta)}$ and $b_n=\frac{1}{2}\sqrt{n(n-1+(d-1))}$ respectively.}
\label{fig:bnBulkPropagatorHUGO}
\end{figure}

The first interesting regime is $2\Delta > (d-1)$. We choose $\Delta=15$ and $d=4$. In this case, the squared mass of the scalar field is positive. In Fig.~\ref{fig:bnBulkPropagatorHUGO}, detailed plots of the small, intermediate, and large $n$ behaviour of the Lanczos coefficients are shown for different values of $r$. These are obtained from the moments $m_{2n}$ of the autocorrelation function, as discussed in the previous section.

From these plots, we infer that the Lanczos coefficients have a finite $r$ behavior of the form
\begin{align}
    \label{eq:AppLanczosAnsatz}
    b_{n}(r)=\alpha(r)n^{c(r)}\,,
\end{align}
where $\alpha(r)$ and $c(r)$ are the $r$-dependent power-law coefficient and power-law exponent respectively. Based on Fig.~\ref{fig:bnBulkPropagatorHUGO}, we find that in the limit $n\rightarrow \infty$, the finite-$r$ Lanczos coefficients~\eqref{eq:AppLanczosAnsatz} approach the near-horizon Lanczos coefficients given by $b_{n}\rightarrow n/2+O(1)$ according to Eq.~\eqref{eq:bnLimits}, and in this limit behave independently of $r$ according to: $\lim_{n\rightarrow \infty}\alpha(r)\rightarrow 1/2$ and $\lim_{n\rightarrow \infty}c(r)\rightarrow 1$. However, for finite $n$, they are $r$-dependent. To find their behaviour, we consider the following expression for the power-law exponent $c(r)$ in terms of derivatives of $b_{n}(r)$ with respect to $n$
\begin{align}
    \label{eq:AppExponentLanczos}
    c(r)=1+\frac{\textrm{d}}{\textrm{d}(\log(n))}\log\left(\frac{\textrm{d}(b_{n}(r))}{\textrm{d}n}\right)~.
\end{align}
By discretizing the derivatives in Eq.~\eqref{eq:AppExponentLanczos} and using the numerical data for $b_{n}$, we can obtain the numerical expression for $c(r)$ at finite values of $n$. Similarly, using this expression, we can find $\alpha(r)$ for intermediate values of $n$ by substituting~\eqref{eq:AppExponentLanczos} in Eq.~\eqref{eq:AppLanczosAnsatz}.

\begin{figure}[ht]
    \centering
    \begin{tabular}{cc}
        \includegraphics[width=0.45\textwidth]{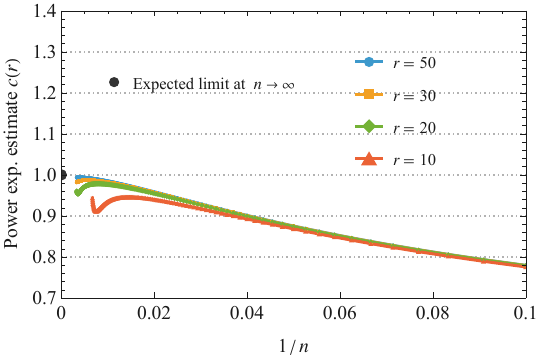} & 
        \includegraphics[width=0.45\textwidth]{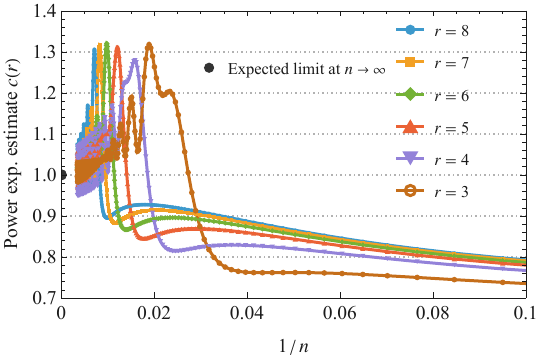} \\
        (a) & (b) \\
        \multicolumn{2}{c}{\centering\includegraphics[width=0.45\textwidth]{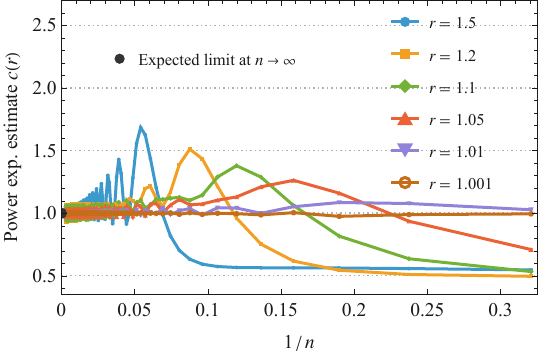}}  \\
       \multicolumn{2}{c}{(c)} \\
    \end{tabular}
     \caption{Power-law exponent $c(r)$~\eqref{eq:AppExponentLanczos} from numerical data of $b_{n}(r)$ for 
 (a) $r=50,30,20,10$, (b) $r=8,7,6,5,4,3$ and (c) $r=1.5,1.2,1.1,1.05,1.001$.}
\label{fig:bnPowerEstimateDelta15d4}
\end{figure}

In Fig.~\ref{fig:bnPowerEstimateDelta15d4}, we show the numerical behavior of the power-law exponent $c(r)$ obtained by discretizing the derivatives in~\eqref{eq:AppExponentLanczos} and for different values of $r$ for finite $1/n$. From these figures it can be seen that the value of $c(r)$ changes for different values of $1/n$, as can be seen from the fact that the Lanczos sequences grow slower than linearly for small $n$, but for sufficiently large $n$ (which depends on the specific value of $r$), these approach the expected behavior at $n\rightarrow \infty$, that is, $c(r)\rightarrow 1$. Moreover, oscillations around the expected value $\lim_{n\rightarrow \infty }c(r)=1$ arise from numerical fluctuations of the discrete derivatives and do not correspond to physical violations of the maximum growth rate of Lanczos coefficients due to locality constraints in many-body systems~\cite{Parker:2018yvk}. To smooth out these numerical oscillations, a \emph{moving average} is implemented in the numerical data for small $r$ for small $1/n$. Importantly, before these oscillations, the value of $c(r)$ for finite $1/n$ is smaller than $1$, implying that in this case and regime, the derivative
\begin{align}
    \label{eq:AppDerivativeLogn}
    \frac{\textrm{d}}{\textrm{d}(\log(n))}\log\left(\frac{\textrm{d}(b_{n}(r))}{\textrm{d}n}\right)\leq0~,
\end{align}
is non-positive, signaling a sub-linear growth of the coefficients for finite $1/n$. While, we currently, do not have a direct confirmation, we expect that these oscillations are the result of staggering, due to the finite mass of the scalar particle (which has been previously studied in flat spacetimes with CFT matter in \cite{Camargo:2022rnt}). We may confirm this observation by testing if they are absent for $m=0$ limiting scale.

\begin{figure}[ht]
    \centering
    \begin{tabular}{cc}
        \includegraphics[width=0.45\textwidth]{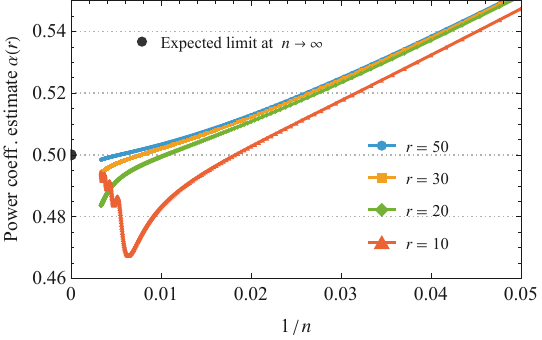} & 
        \includegraphics[width=0.45\textwidth]{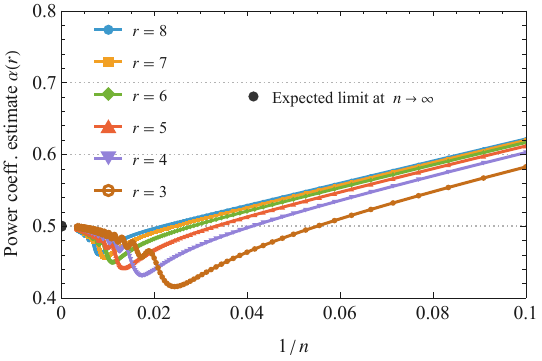} \\
        (a) & (b) \\
        \multicolumn{2}{c}{\centering\includegraphics[width=0.45\textwidth]{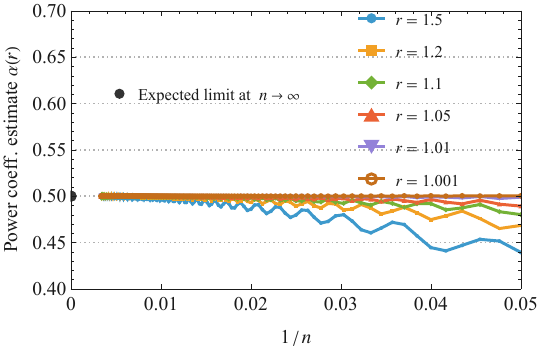}}  \\
       \multicolumn{2}{c}{(c)} \\
    \end{tabular}
     \caption{Power-law coefficient estimate $\alpha(r)$~\eqref{eq:AppLanczosAnsatz} from numerical data of $b_{n}(r)$ for (a) $r=50,30,20,10$, (b) $r=8,7,6,5,4,3$ and (c) $r=1.5,1.2,1.1,1.05,1.001$.}
\label{fig:bnPowerGrowthRateDelta15d4}
\end{figure}

In Fig.~\ref{fig:bnPowerGrowthRateDelta15d4} we show the numerical behavior of the power-law coefficient $\alpha(r)$ for different values of $r$ obtained by discretizing~\eqref{eq:AppLanczosAnsatz}. We include contributions from $c(r)$~\eqref{eq:AppExponentLanczos}. From these plots it can be seen that $\alpha(r)$ has a monotonically-decreasing behaviour for small $n$, starting from initial values larger than the asymptotic $n\rightarrow \infty$ value $1/2$. However, for large $n\gg 1$ ($1/n\ll 1$) and after becoming smaller than $1/2$, there is a turning point after which $\alpha(r)$ increases toward the expected value at $n\rightarrow \infty$, with small oscillations. This is consistent both for small and large values of $r$. Similarly to the power law exponent (\ref{eq:AppExponentLanczos}), these oscillations might signal \emph{staggering} around $\alpha\rightarrow 1/2$.

\begin{figure}[ht]
    \centering
    \begin{tabular}{cc}
        \includegraphics[width=0.45\textwidth]{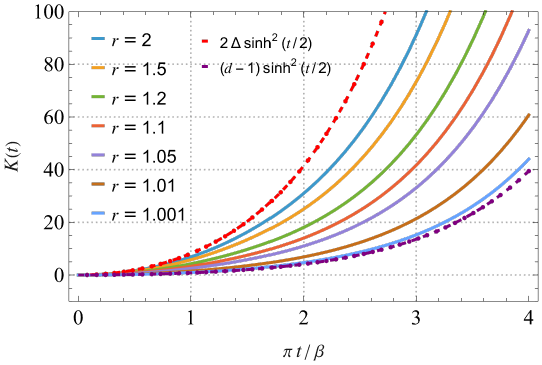} &
        \includegraphics[width=0.45\textwidth]{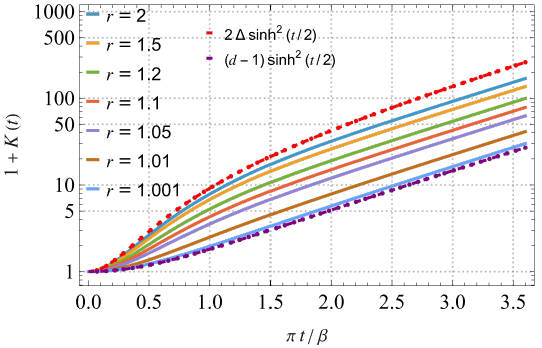} 
    \end{tabular}
     \caption{Plot of Krylov complexity $K(t)$ for different finite values of $r$ and for $\Delta=15$ and $d=4$ using data from the Lanczos coefficients up to $n=400$ in linear (left) and logarithmic (right) scales.}
\label{fig:KrylovPlotDelta14d4}
\end{figure}

From the numerical Lanczos coefficients $b_{n}$, we can obtain the probability amplitudes $\varphi_{n}(t)$ by solving the recursion relation Eq.~\eqref{eq:sch eq K operator} and subsequently the Krylov complexity via Eq.~\eqref{eq:Krylov complexity}. In Fig.~\ref{fig:KrylovPlotDelta14d4} we show the behaviour of Krylov complexity $K(t)$ for different values of $r$. Here, we used the numerical data from the Lanczos coefficients up to $n=400$. The plot on the right side displays the Krylov complexity for the same range as the left one, but in a logarithmic scale. Note that in this case, the initial value of $K(t)$ has been shifted by $1$ for convenience, to improve the presentation of the results.

The plot on the right in Fig.~\ref{fig:KrylovPlotDelta14d4} suggests that there is a regime for large $t$ in which the Krylov complexity $K(t)$ grows exponentially, independently of $r$. In this regime, the behaviour of Krylov complexity for finite $r$ can be written as
\begin{align}
    \label{eq:AppKrylovr}
    K(t,r)\propto K_{0}(r)e^{\lambda_{K}(r)t/2}~,
\end{align}
where $K_{0}(r)=K(t=0,r)\neq0$ and where $\lambda_{K}(r)$ is the Krylov exponent. Similarly to the Lanczos coefficients, the exponent $\lambda_{K}(r)$ can be extracted from the numerical data using the formula
\begin{align}
\label{eq:AppLambdaKDer}
    \lambda_{K}(r)=2\frac{\textrm{d}}{\textrm{d}t}\left(\log\left(\frac{\textrm{d}K(t,r)}{\textrm{d}t}\right)\right)~.
\end{align}
This expression assumes that the exact behaviour of the complexity is of the form~\eqref{eq:AppKrylovr}. In such a case, $\lambda_{K}(r)$ will not depend on $t$.  However, if the behaviour is not of the form~\eqref{eq:AppKrylovr} for all times, then in general ~\eqref{eq:AppLambdaKDer} will be a time-dependent function. For example, the right-hand side of~\eqref{eq:AppLambdaKDer} for the Krylov complexities at the boundary $r\rightarrow \infty$ and near the horizon $r= 1+\delta r$ with $\delta r\ll 1$ takes the form $2\pi\coth(2\pi t/\beta)/\beta$. However, in this case, it is clear that for sufficiently large values of $t$, this approaches the $t\rightarrow \infty$ value $2\pi/\beta$.

\begin{figure}[ht]
    \centering
    \begin{tabular}{cc}
        \includegraphics[width=0.45\textwidth]{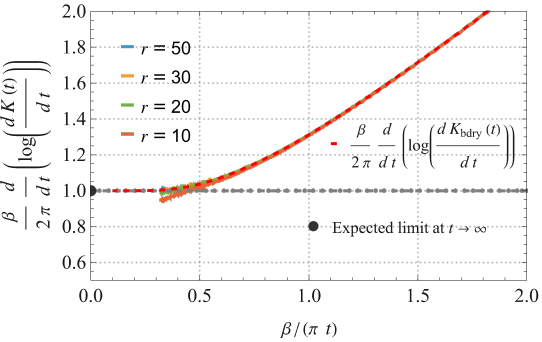} & 
        \includegraphics[width=0.45\textwidth]{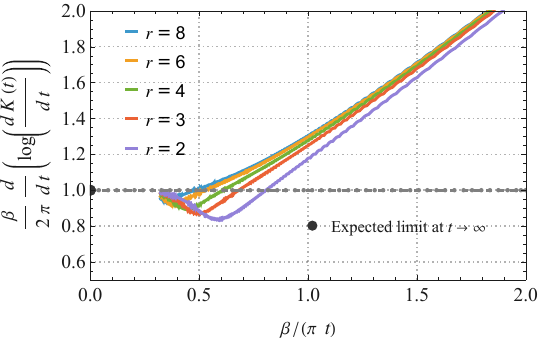} \\
        (a) & (b) \\
        \multicolumn{2}{c}{\centering\includegraphics[width=0.45\textwidth]{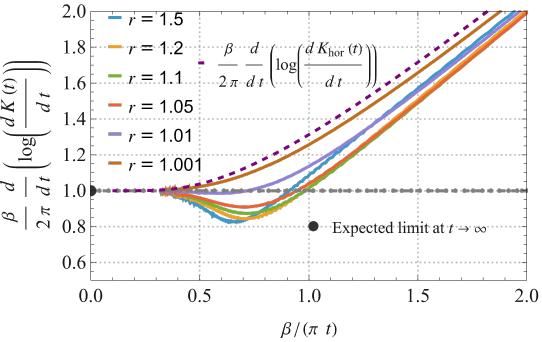}}  \\
       \multicolumn{2}{c}{(c)} \\
    \end{tabular}
     \caption{Plot of the Krylov exponent $\beta \,\lambda_{K}(r)/(2\pi)$ with $\beta=2\pi$ extracted from Eq.~\eqref{eq:AppLambdaKDer} as a function of $1/t$ for different values of $r=50,30,20,10$ (a), $r=8,6,4,3,2$ (b) and $r=1.5,1.2,1.1,1.05,1.01,1.001$ (c), for $\Delta=15$ and $d=4$ using data from the Lanczos coefficients up to $n=300$.}
\label{fig:LambdaKDelta14d4}
\end{figure}

Figure~\ref{fig:LambdaKDelta14d4} shows $\lambda_{K}(r)$ computed using~\eqref{eq:AppLambdaKDer} for different values of $r$. From these plots, it can be seen that the exponent $\lambda_{K}(r)$ decreases monotonically as a function of $\pi t /\beta$ in the regime $ O(1)\lesssim \beta/(\pi t)$. In the figures, this corresponds to viewing the plot from right to left. As $\beta /(\pi t) $ decreases, the data corresponding to different $r$ reach a different minimum at a specific $r$-dependent value of $t$, which is below the expected behaviour at $t\rightarrow \infty$. They subsequently grow again and approach the expected limit $t\rightarrow \infty$. These figures, therefore, imply the following: 1)the Ansatz~\eqref{eq:AppKrylovr} holds only for sufficiently large $t$ for the complexities obtained for finite $r$. For intermediate time-regimes, the specific behaviour of Krylov complexity is not exactly exponential and in general $r$-dependent, 2)the regime for which~\eqref{eq:AppKrylovr} holds, depends on the specific value of $r$, 3)regardless of the value of $r$, the Krylov exponent $\lambda_{K}(r)$ approaches the asymptotic near-horizon and boundary value for sufficiently small $\pi t/\beta$. These observations suggest that in the limit $t\rightarrow \infty$ the Krylov exponent $\lambda_{K}(r)$ becomes insensitive to the specific radial position $r$ where the scalar bulk fields are located. Thus, in this limit the Krylov exponent $\lambda_{K}(r)$ is insensitive to the holographic renormalization group (RG) flow.

\begin{figure}[ht]
    \centering
    \begin{tabular}{cc}
        \includegraphics[width=0.45\textwidth]{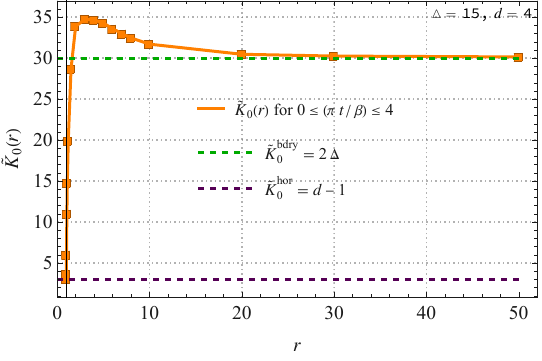} & 
        \includegraphics[width=0.45\textwidth]{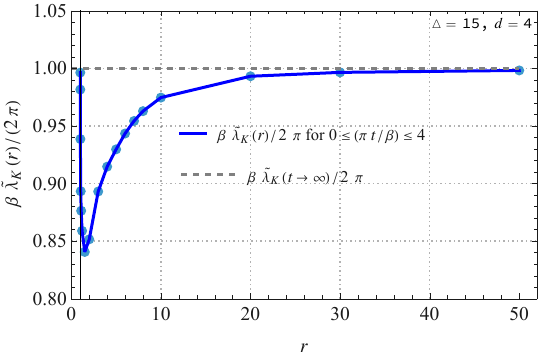} \\
        (a) & (b) \\
    \end{tabular}
     \caption{Numerical plots for $\tilde{K}_{0}(r)$ (a) and $\tilde{\lambda}_{K}(r)$ (b) obtained from $K(t)$ in the time region $0\leq (\pi t / \beta)\leq 4$ for different values of $r$ for $\Delta=15$ and $d=4$. Lines corresponding to the asymptotic values $\tilde{K}_{0}(r\rightarrow \infty):=\tilde{K}^{\textrm{bdry}}_{0}=2\Delta$, $\tilde{K}_{0}(r\rightarrow 1):=\tilde{K}^{\textrm{hor}}_{0}=d-1$ and $\beta \tilde{\lambda}_{K}(t\rightarrow \infty)/2\pi=1$ are shown in the respective plots. The black vertical line corresponds to the location of the horizon at $r=1$.}
\label{fig:PlotTableK0rLambdatDelta15d4}
\end{figure}

\begin{figure}[ht]
    \centering
    \begin{tabular}{cc}
        \includegraphics[width=0.45\textwidth]{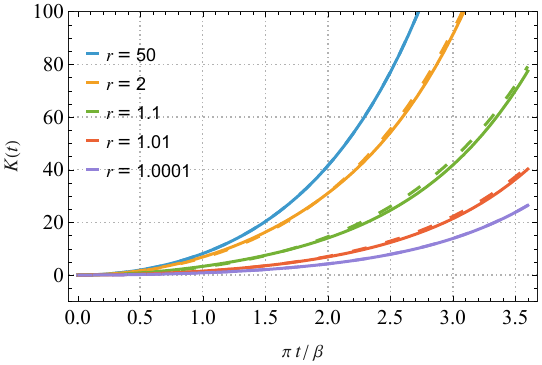} & 
        \includegraphics[width=0.45\textwidth]{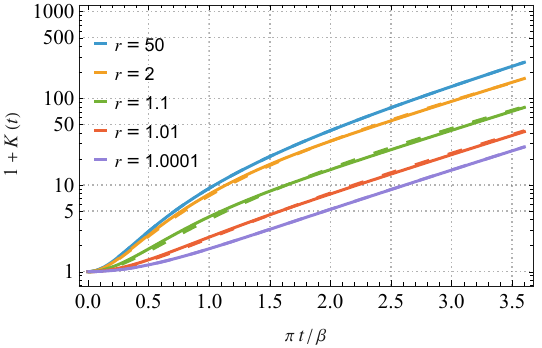} \\
        (a) & (b) \\
    \end{tabular}
     \caption{Plots comparing the numerical results for $K(t)$ (solid) and the fit~\eqref{eq:AppSinhSqAnsatzKR} (dashed) for $r=50,2, 1.1,1.01,1.0001$ for $\Delta=15$ and $d=4$ in normal (a) and logarithmic (b) scales.}
\label{fig:PlotCompSinhSqFitDelta15d4}
\end{figure}

This analysis at finite $t$ can be made more precise by considering a different Ansatz for the behaviour of the Krylov complexity at finite $r$. Suppose that the behaviour is of the form
\begin{align}
    \label{eq:AppSinhSqAnsatzKR}
    K(t,r)=\tilde{K}_{0}(r)\sinh^{2}(\tilde{\lambda}_{K}(r)t/2)~.
\end{align}
One can try to extract $\tilde{\lambda}_{K}(r)$ from~\eqref{eq:AppSinhSqAnsatzKR} in a similar way to what was done for the exponential Ansatz~\eqref{eq:AppKrylovr}. However, given the functional form of~\eqref{eq:AppSinhSqAnsatzKR}, it is not possible to isolate $\tilde{\lambda}_{K}(r)$ completely in terms of derivatives and logarithms of $K(t,r)$. One can, nevertheless, get rid of the pole-order dependence by considering the ratio
\begin{align}
    \label{eq:AppRatioSinhKr}
    \frac{1}{K(t,r)}\frac{\textrm{d}K(t,r)}{\textrm{d}t}=\tilde{\lambda}_{K}(r)\coth(\tilde{\lambda}_{K}(r)t/2)~.
\end{align}
Fitting $\tilde{\lambda}_{K}(r)$ and subsequently $\tilde{K}_{0}(r)$ using the numerical data for $K(t)$ in this way yields Fig.~\ref{fig:PlotTableK0rLambdatDelta15d4}. Moreover, Fig.~\ref{fig:PlotCompSinhSqFitDelta15d4} displays a comparison between the numerical results for $K(t)$ and the fit~\eqref{eq:AppSinhSqAnsatzKR} for representative values of $r$, showing close agreement between them.

From Fig.~\ref{fig:PlotTableK0rLambdatDelta15d4} it can be seen that at finite $t$ both $\tilde{\lambda}_{K}(r)$ and $\tilde{K}_{0}(r)$ have a nonmonotonic behaviour as a function of $r$. It should be noted that plot (a) in Fig.~\ref{fig:PlotTableK0rLambdatDelta15d4} does not contradict~\eqref{polestructurefiniter}, as the latter is the order of the pole with respect to $t=\pm i \pi$, which determines the $t\rightarrow \infty$ behaviour. 

Finally, given that neither $\tilde{\lambda}_{K}(r)$ nor $\tilde{K}_{0}(r)$ vary monotonously as $r$ moves from the boundary ($r\rightarrow \infty$) to the horizon ($r\rightarrow 1$), this implies that neither function can have the interpretation of a \emph{holographic $c$-function} as a function of $r$ at finite $t$. This should be compared with~\cite{Chapman:2024pdw}, where it is found in mass-deformed SYK models, the Krylov exponent behaves monotonically along the renormalization group (RG) flow, acting as a $c$-function. Our interpretation of the results is that the $\sinh^2$ fit is an approximation to the true Krylov complexity and is valid only at finite intervals in $0\leq t \leq 4$. The true Krylov complexity is a more complicated function of $t$ and $r$, where the subleading, finite $r$ effects decay as $t$ increases, leading to the equal Lyapunov exponent for any given radial location when we do the evaluation at $t\rightarrow \infty$, as we discussed in Sec. \ref{ssec:holo dictionary}. Moreover, as seen from Fig.~\ref{fig:LambdaKDelta14d4}, the Krylov exponent extracted from the bulk-to-bulk correlator becomes insensitive to $r$ for sufficiently large $t$, and becomes constant and independent of $r$ in the limit $t\rightarrow \infty$. In this case we find $\beta \partial_{\beta}(\beta \lambda_{K}/2\pi)\geq 0$ and $\beta \partial_{r}(\beta \lambda_{K}(t\rightarrow \infty)/2\pi)= 0$ with $\beta\lambda_{K}^{\textrm{UV}}/2\pi=1=\beta\lambda_{K}^{\textrm{IR}}/2\pi$. An alternative explanation is that the autocorrelation function~\eqref{eq:autocorrelation function main} might not obey the positivity condition (\ref{eq:innerprodpropCFT3}) for finite $r$, which is required for finite-temperature inner product of bulk scalar fields. We have not been able to verify this possibility due to technical limitations in our analytical and numerical analysis. Nevertheless, the fact that \eqref{eq:autocorrelation function main} at any finite radial location display the same early and late time behavior, and that we have not encountered cases where the autocorrelator (seen as a bulk inner product) reaches negative values in our evaluations gives us confidence that it obeys the axions (\ref{eq:innerprodpropCFT2}). However, given that the different plots indicate a varying exponential growth, while, as discussed in Sec. \ref{ssec:holo dictionary}, the proper radial momentum of a probe particle is a $\sinh$ function, our results imply that the proper radial momentum/ rate of growth of Krylov complexity (sec. \ref{sec:proper momentum}) is not strictly obeyed at any bulk radial location. Nevertheless, Fig. \ref{fig:PlotCompSinhSqFitDelta15d4} indicate that the relation (\ref{eq:radial momentum eval}) is approximately satisfied even at generic finite radial locations.

\subsection{\texorpdfstring{$(d-2)<2\Delta<(d-1)$}{}}\label{app:figsbnKtfiniterDelta1p2d4}

\begin{figure}[ht]
    \centering
    \begin{tabular}{cc}
        \includegraphics[width=0.45\textwidth]{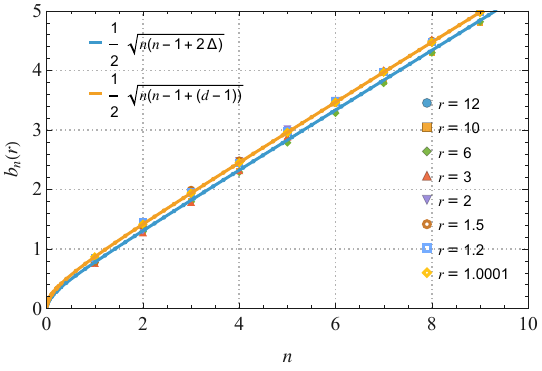} & 
        \includegraphics[width=0.45\textwidth]{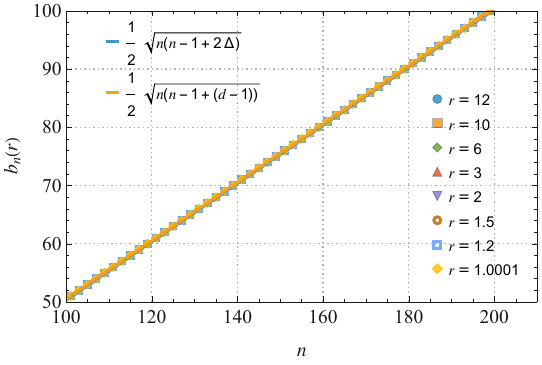} \\
        (a) & (b) \\
    \end{tabular}
     \caption{Behaviour of Lanczos coefficients $b_{n}$ for $\Delta=1.2$ and $d=4$ for different $r$ in the early (a) and large (b) $n$ regimes. Blue and orange solid curves are $b_{n}=(1/2)\sqrt{n(n-1+2\Delta)}$ and $b_{n}=(1/2)\sqrt{n(n-1+(d-1))}$ respectively.}
\label{fig:LanczosDelta1p2d4}
\end{figure}

The second interesting regime is $(d-1)>2\Delta > (d-2)$. We choose $\Delta=1.2$ and $d=4$. In this case, the squared mass of the scalar field is negative. In Fig.~\ref{fig:LanczosDelta1p2d4}, detailed plots of the small and large $n$ behaviour of the Lanczos coefficients are shown for different values of $r$. Similarly to the previous case, these are obtained from the moments $m_{2n}$ of the autocorrelation function.

\begin{figure}[ht]
    \centering
    \begin{tabular}{cc}
        \includegraphics[width=0.45\textwidth]{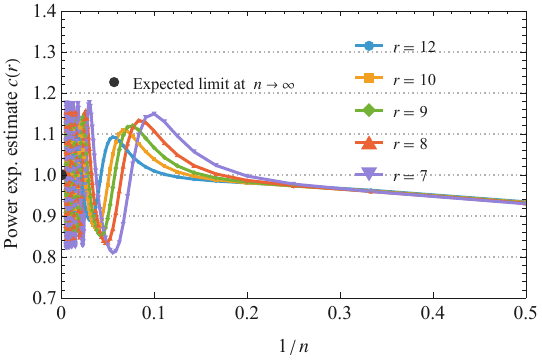} & 
        \includegraphics[width=0.45\textwidth]{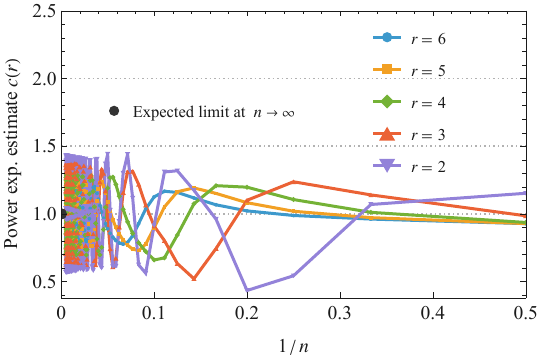} \\
        (a) & (b) \\
        \multicolumn{2}{c}{\centering\includegraphics[width=0.45\textwidth]{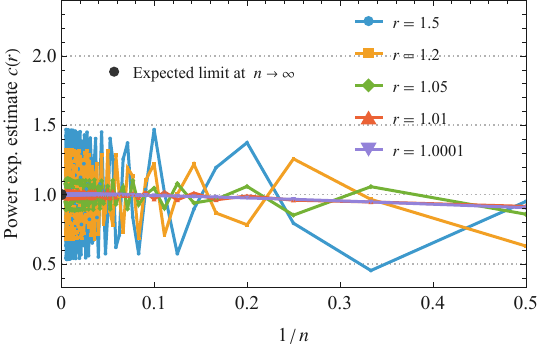}}  \\
       \multicolumn{2}{c}{(c)} \\
    \end{tabular}
     \caption{Power-law exponent estimate $c(r)$~\eqref{eq:AppExponentLanczos} from numerical data of $b_{n}(r)$ for (a) $r=12,10,9,8,7$, (b) $r=6,5,4,3,2$ and (c) $r=1.5,1.2,1.1,1.05,1.01,1.001,1.0001$.}
\label{fig:PowerEstimateDelta1p2d4}
\end{figure}

\begin{figure}[ht]
    \centering
    \begin{tabular}{cc}
        \includegraphics[width=0.45\textwidth]{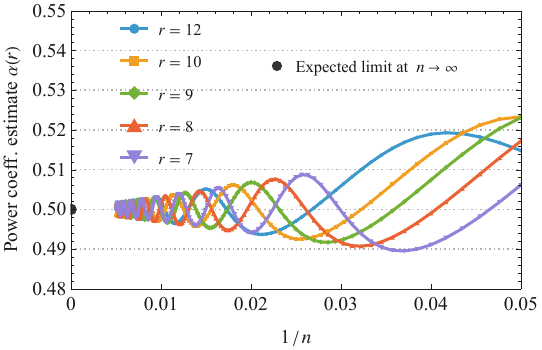} & 
        \includegraphics[width=0.45\textwidth]{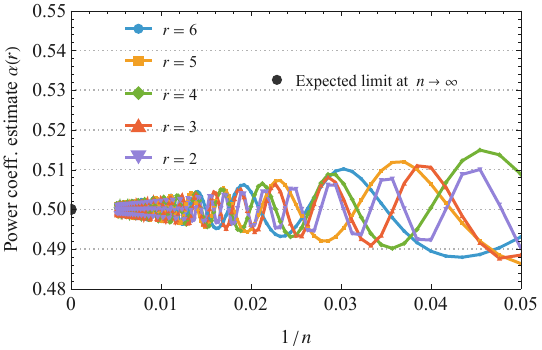} \\
        (a) & (b) \\
        \multicolumn{2}{c}{\centering\includegraphics[width=0.45\textwidth]{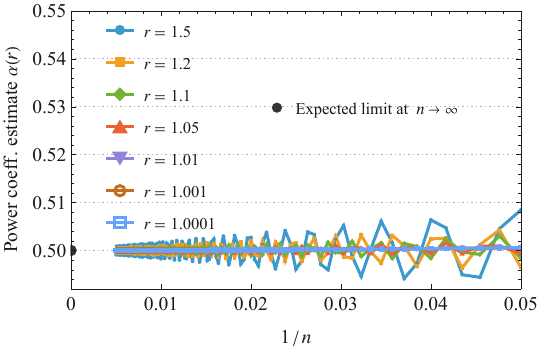}}  \\
       \multicolumn{2}{c}{(c)} \\
    \end{tabular}
     \caption{Power-law coefficient estimate $\alpha(r)$~\eqref{eq:AppLanczosAnsatz} from numerical data of $b_{n}(r)$ for (a) $r=12,10,9,8,7$, (b) $r=6,5,4,3,2$ and (c) $r=1.5,1.2,1.1,1.05,1.01,1.001,1.0001$.}
\label{fig:PowerSlopeDelta1p2d4}
\end{figure}

The first contrast with the case studied in the previous section is that in this case the curve corresponding to the boundary Lanczos coefficients (blue) lies beneath the one for the horizon coefficients (orange). Similarly to the previous section, by assuming a behaviour of the Lanczos coefficients of the form~\eqref{eq:AppLanczosAnsatz} we can find estimates for the power-law coefficient $\alpha(r)$ and exponent $c(r)$. These estimates are found in Figs.~\ref{fig:PowerEstimateDelta1p2d4} and~\ref{fig:PowerSlopeDelta1p2d4}.

These figures show a similar tendency to the previous case. The power-law exponent $c(r)$ is smaller than $1$ for small $n$, but gradually increases before it oscillates around the expected limit at $n\rightarrow \infty$. Similarly to the previous case, these oscillations are a numerical artifact of taking the discrete derivatives of the numerical data. The power-law coefficient $\alpha(r)$ oscillates around the expected limit with the oscillations decreasing as $n$ increases. This behaviour is particularly reminiscent of the small $r$ behaviour for the $\Delta=15$, $d=4$ case shown in plot (c) of Fig.~\ref{fig:bnPowerGrowthRateDelta15d4}.

\begin{figure}[ht]
    \centering
    \begin{tabular}{cc}
        \includegraphics[width=0.45\textwidth]{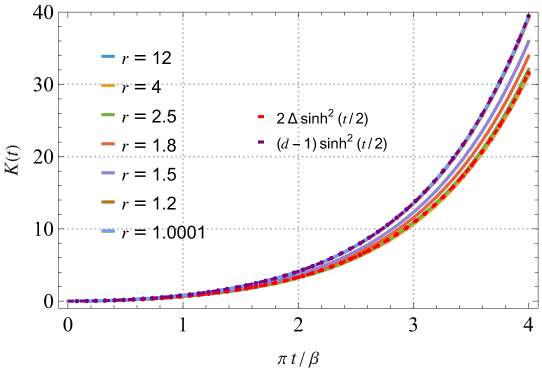} &
        \includegraphics[width=0.45\textwidth]{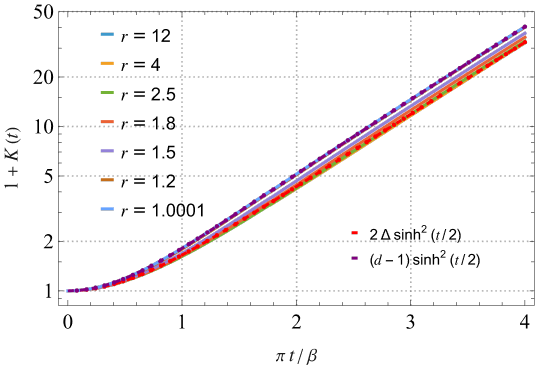} 
    \end{tabular}
     \caption{Plot of Krylov complexity $K(t)$ for different finite values of $r$ and for $\Delta=1.2$ and $d=4$ using data from the Lanczos coefficients up to $n=200$ in linear (left) and logarithmic (right) scales.}
\label{fig:KrylovPlotDelta1p2d4}
\end{figure}

Just as in the previous section, from the numerical Lanczos coefficients $b_{n}$, we can obtain the probability amplitudes $\varphi_{n}(t)$ and subsequently the Krylov complexity $K(t)$. In Fig.~\ref{fig:KrylovPlotDelta1p2d4} we show the behaviour of $K(t)$ for different values of $r$. Here, we used the numerical data from the Lanczos coefficients up to $n=200$. This data is sufficient to find the Krylov complexity for the same range ($0\leq (\pi t /\beta)\leq 4$) as in the previous case. The plot on the right side displays the Krylov complexity for the same range as the left one, but in a logarithmic scale, where the initial value of $K(t)$ has been shifted by $1$.

\begin{figure}[ht]
    \centering
    \begin{tabular}{cc}
        \includegraphics[width=0.45\textwidth]{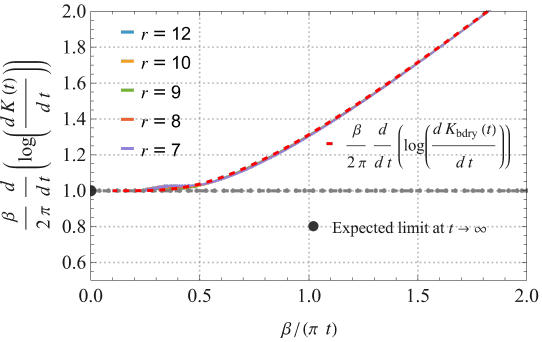} & 
        \includegraphics[width=0.45\textwidth]{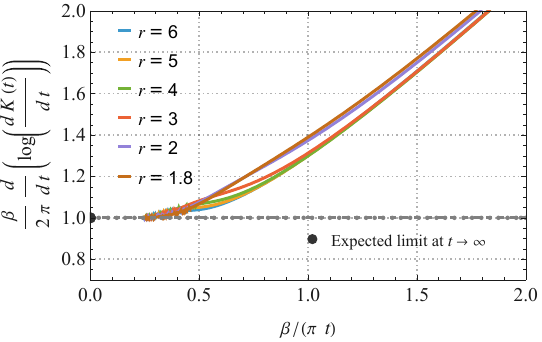} \\
        (a) & (b) \\
        \multicolumn{2}{c}{\centering\includegraphics[width=0.45\textwidth]{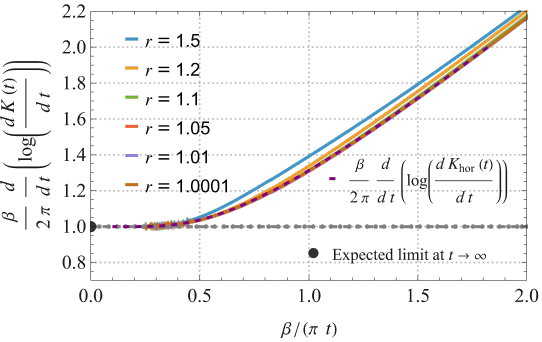}}  \\
       \multicolumn{2}{c}{(c)} \\
    \end{tabular}
     \caption{Plot of the Krylov exponent $\beta \,\lambda_{K}(r)/(2\pi)$ with $\beta=2\pi$ extracted from Eq.~\eqref{eq:AppLambdaKDer} as a function of $1/t$ for different values of $r=12,10,9,8,7$ (a), $r=6,5,4,3,2$ (b) and $r=1.5,1.2,1.1,1.05,1.01,1.0001$ (c), for $\Delta=1.2$ and $d=4$ using data from the Lanczos coefficients up to $n=200$.}
\label{fig:LambdaKDelta1p2d4}
\end{figure}

The most striking difference in these plots with respect to the previous case is that here, the curve of the boundary $r\rightarrow \infty$ Krylov complexity $K(t)=2\Delta \sinh^{2}(t/2)$ lies beneath the curve for the horizon $r\rightarrow 1$ Krylov complexity $K(t)=(d-1) \sinh^{2}(t/2)$, with the curves for the finite-$r$ Krylov complexities lying between them. Following the previous section, we can propose an exponential Ansatz for the late-time behaviour of the Krylov complexity of the form~\eqref{eq:AppKrylovr}, for which we can extract the Krylov exponent using~\eqref{eq:AppLambdaKDer}. Fig.~\ref{fig:LambdaKDelta1p2d4} shows the behaviour of this function for the complexities in Fig.~\ref{fig:KrylovPlotDelta1p2d4} for different values of $r$. In all cases, all estimated Krylov exponents tend to the expected limit at $t\rightarrow \infty$.

\begin{figure}[ht]
    \centering
    \begin{tabular}{cc}
        \includegraphics[width=0.45\textwidth]{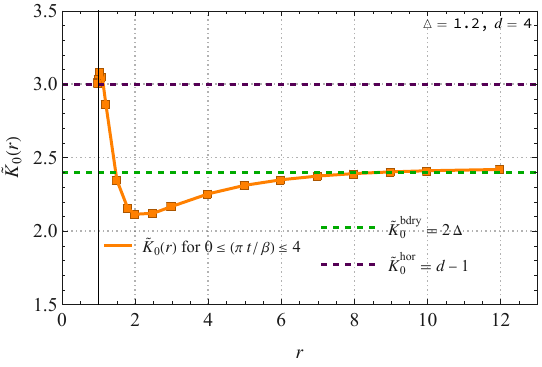} & 
        \includegraphics[width=0.45\textwidth]{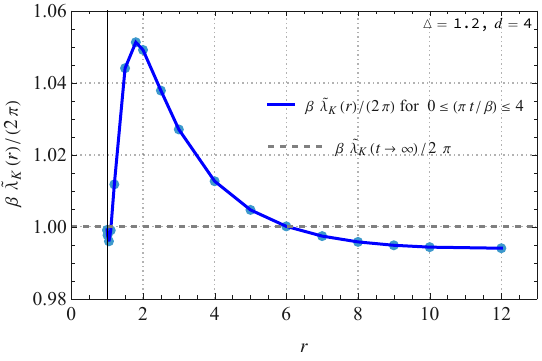} \\
        (a) & (b) \\
    \end{tabular}
     \caption{Numerical plots for $\tilde{K}_{0}(r)$ (a) and $\tilde{\lambda}_{K}(r)$ (b) obtained from $K(t)$ in the time region $0\leq (\pi t / \beta)\leq 4$ for different values of $r$ for $\Delta=1.2$ and $d=4$. Lines corresponding to the asymptotic values $\tilde{K}_{0}(r\rightarrow \infty):=\tilde{K}^{\textrm{bdry}}_{0}=2\Delta$, $\tilde{K}_{0}(r\rightarrow 1):=\tilde{K}^{\textrm{hor}}_{0}=d-1$ and $\beta \tilde{\lambda}_{K}(t\rightarrow \infty)/2\pi=1$ are shown in the respective plots. The black vertical line corresponds to the location of the horizon at $r=1$.}
\label{fig:PlotTableK0rLambdatDelta1p2d4}
\end{figure}

\begin{figure}[ht]
    \centering
    \begin{tabular}{cc}
        \includegraphics[width=0.45\textwidth]{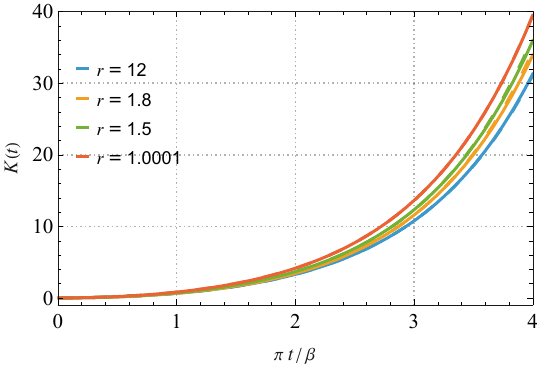} & 
        \includegraphics[width=0.45\textwidth]{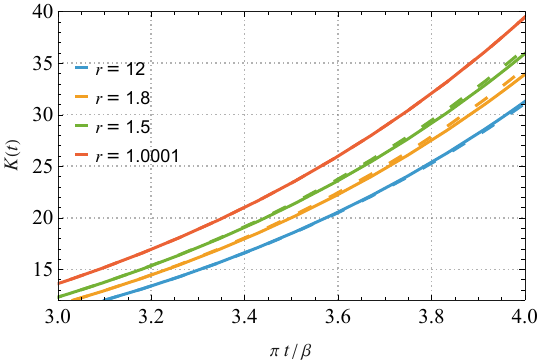} \\
        (a) & (b) \\
    \end{tabular}
     \caption{Plots comparing the numerical results for $K(t)$ (solid) and the fit~\eqref{eq:AppSinhSqAnsatzKR} (dashed) for representative values $r=12,1.8,1.5,1.0001$ for $\Delta=1.2$ and $d=4$ for (a) $0\leq (\pi t/\beta)\leq 4$ and (b) $3\leq (\pi t/\beta)\leq 4$..}
\label{fig:PlotCompSinhSqFitDelta1p2d4}
\end{figure}

\begin{figure}[ht]
    \centering
    \begin{tabular}{cc}
        \includegraphics[width=0.45\textwidth]{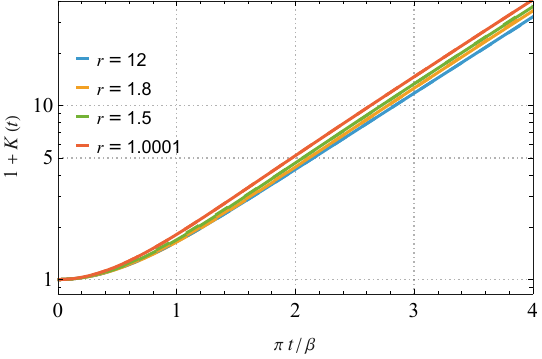} & 
        \includegraphics[width=0.45\textwidth]{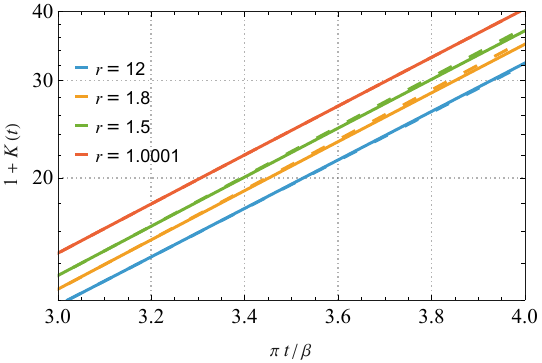} \\
        (a) & (b) \\
    \end{tabular}
     \caption{Plots in logarithmic scale comparing the numerical results for $1+K(t)$ (solid) and the fit~\eqref{eq:AppSinhSqAnsatzKR} (dashed) for representative values $r=12,1.8,1.5,1.0001$ for $\Delta=1.2$ and $d=4$ for (a) $0\leq (\pi t/\beta)\leq 4$ and (b) $3\leq (\pi t/\beta)\leq 4$.}
\label{fig:PlotLogCompSinhSqFitDelta1p2d4}
\end{figure}

We can also use a $\sinh^2$ Ansatz, following eq.~\eqref{eq:AppSinhSqAnsatzKR} for a finite intervale $0\leq t \leq 4$, and extract the behaviour of the functions $\tilde{K}_{0}(r)$ and $\tilde{\lambda}_{K}(t)$. Fig.~\ref{fig:PlotTableK0rLambdatDelta1p2d4} displays their behaviour as a function of $r$. From it can be seen that, similarly to the case discussed in the previous section, at finite $t$ both $\tilde{\lambda}_{K}(r)$ and $\tilde{K}_{0}(r)$ have non-monotonic behaviour as a function of $r$. Figures~\ref{fig:PlotCompSinhSqFitDelta1p2d4} and~\ref{fig:PlotLogCompSinhSqFitDelta1p2d4} display a comparison between the numerical results for $K(t)$ and the fit~\eqref{eq:AppSinhSqAnsatzKR} for representative values of $r$, also showing close agreement between them.

\section{Comparison with T\texorpdfstring{$\overline{\text{T}}$}{}-deformations of holographic CFTs}\label{app:figs}

In this Appendix, we show that the bulk-to-bulk propagator (\ref{eq:BdryProp}) shares similarities with the finite cutoff AdS correlation function, which in pure gravity is holographically dual to T$\overline{\text{T}}$-deformed CFT$_2$ (introduced in \cite{Smirnov:2016lqw}) and $T^2$-deformed CFTs in higher dimensions \cite{Hartman:2018tkw}. Our driving motivation for this comparison is the finite radial cutoff interpretation of the bulk dual to T$\overline{\text{T}}$-deformations \cite{McGough:2016lol,Kraus:2018xrn}. This involves the removal of the asymptotic boundary of the bulk geometry. We emphasize that our approach is different from introducing a cutoff in the geometry; instead, we focus on studying a local bulk field at a fixed AdS radial coordinate $r$. Despite this distinction, the bulk-to-bulk correlation function expanded in powers of $1/r$ exhibits a structure analogous to those found in T$\overline{\text{T}}$-deformations, potentially providing valuable qualitative insights into previous studies on Krylov complexity in T$\overline{\text{T}}$-deformed CFTs. In this appendix, we collect numerical plots of the Lanczos coefficients for the bulk-to-bulk propagator in a perturbative expansion in the radial location, and for the two-point correlation function of T$\overline{\text{T}}$-deformed CFTs.

\subsection{Lanczos coefficients of the perturbed bulk propagator}
For comparison purposes, we will consider a perturbative expansion of (unnormalized) bulk propagator
\begin{equation}\label{eq:full G(t)}
\begin{aligned}
     &G_{\Delta}(P;P')=\xi^{\Delta}{}_2F_1\left( \frac{\Delta}{2},\frac{\Delta+1}{2};\Delta+1-\frac{d}{2}; \xi^2\right)=\\
     &=\left(\frac{1}{r^2 (\cosh t+1)}\right)^{\Delta } \left(1+\frac{\Delta  \cosh
   t}{(\cosh t+1) r^2}+\frac{\frac{\Delta  (\Delta +1) \cosh ^2t}{2 (\cosh
   t+1)^2}+\frac{\Delta  (\Delta +1)}{4 \left(\Delta +1-\frac{d}{2}\right) (\cosh
   t+1)^2}}{r^4}+\mathcal{O}\left(\frac{1}{r^6}\right)\right).
\end{aligned}
\end{equation}
Instead of using the exact bulk propagator, we analyze the Lanczos coefficient from the perturbative expression of the bulk propagator.

Let us perform the approximation to the first subleading term
\begin{align}
G_{p1}(t,r)&:=\left(\frac{1}{r^2 (\cosh t+1)}\right)^{\Delta } \left(1+\frac{\Delta  \cosh
   t}{(\cosh t+1) r^2}\right),\label{eq:Gp1(t)}\\
   C_{p1}(t,r)&:=\frac{G_{p1}(t,r)}{G_{p1}(0,r)}.
\end{align}
We can compute the moment $m_n$ from the autocorrelation function $C_{p1}(t,r)$. We find that $m_n$ of $C_{p1}(t,r)$ can be negative. 

To understand the origin of these results, it is convenient to show explicitly the perturbative autocorrelation function $C_{p1}(t,r)$ is explicitly given by
\begin{align}\label{eq:C1}
    C_{p1}(t,r)&=\frac{1+\frac{\Delta}{r^2}}{1+\frac{\Delta}{2r^2}}\frac{1}{\left(\cosh \frac{t}{2}\right)^{2\Delta}}-\frac{\frac{\Delta}{2r^2}}{1+\frac{\Delta}{2r^2}}\frac{1}{\left(\cosh \frac{t}{2}\right)^{2\Delta+2}}\\
    &=\frac{1}{\left(\cosh \frac{t}{2}\right)^{2\Delta}}+\frac{\Delta}{2r^2}\left(\frac{1}{\left(\cosh \frac{t}{2}\right)^{2\Delta}}-\frac{1}{\left(\cosh \frac{t}{2}\right)^{2\Delta+2}}\right)+\mathcal{O}\left(\frac{1}{r^4}\right),
\end{align}
which has a remarkably similar structure to (3.13) of \cite{Chattopadhyay:2024pdj} (the autocorrelation function of the deformed CFT in a perturbative expansion of the deformation parameter), while there are notable differences including coefficients and logarithmic terms. Our calculations also allow us to obtain the Lanczos coefficient of $b_n$ corresponding to (\ref{eq:C1})
\begin{align}\label{eq:lanczpertbulkprop}
    b_n=\frac{1}{2} \sqrt{n (2 \Delta +n-1)}-\frac{\sqrt{n (2 \Delta +n-1)} (2 \Delta +2 n-1)}{8
   (2 \Delta +1) r^2}+\mathcal{O}\left(\frac{1}{r^4}\right).
\end{align}
The negative sign in the second term is problematic at large $n$.

Next, we consider up to the sub-subleading terms
\begin{align}
G_{p2}(t,r)&:=\left(\frac{1}{r^2 (\cosh t+1)}\right)^{\Delta } \left(1+\frac{\Delta  \cosh
   t}{(\cosh t+1) r^2}+\frac{\frac{\Delta  (\Delta +1) \cosh ^2t}{2 (\cosh
   t+1)^2}+\frac{\Delta  (\Delta +1)}{4 \left(\Delta +1-\frac{d}{2}\right) (\cosh
   t+1)^2}}{r^4}\right)~,\label{eq:Gp2(t)}\\
   C_{p2}(t,r)&:=\frac{G_{p2}(t,r)}{G_{p2}(0,r)}.\label{eq:Cp2(t)}
\end{align}
In our numerical evaluations, we do not find negative values of $m_n$. However, we do find that $b_n^2$ can be negative, which indicates that $C_{p2}(t,r)$ is also not a good function for the usual inner product. 

We believe that this issue comes from the fact that we stop the perturbative expansion at sub or sub-subleading order, given that the exact bulk propagator is a good function for the usual inner product, as we showed in Sec. \ref{ssec:Lanczos and complexity}, both by analytic expressions in particular regimes (i.e. when the radial location is the asymptotic boundary or the near horizon region), and through numerical evidence more generally.

\begin{figure}
         \centering
        \includegraphics[width=0.45\textwidth]{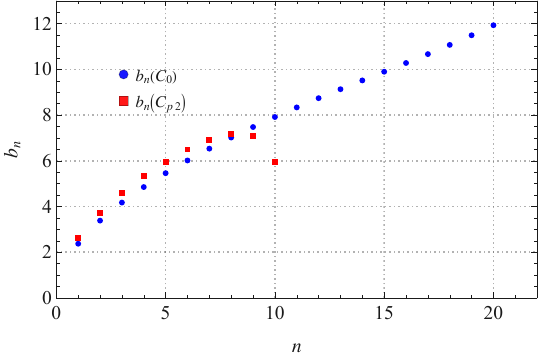}
      \includegraphics[width=0.45\textwidth]{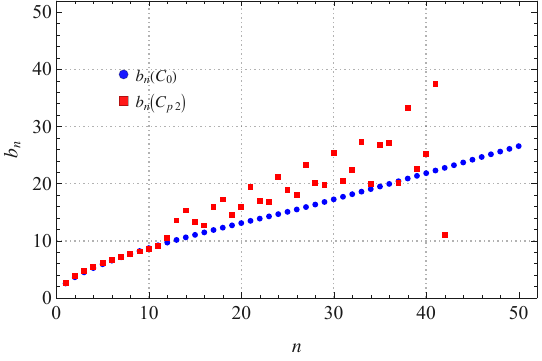}
         \caption{Comparison of the Lanczos coefficient $b_n$ for the exact bulk propagator (\ref{eq:autocorrelation function main}) (blue) with the perturbative expansion $C_{p_2}(t,r)$ (red) for $\Delta=15$, $d=4$, and: \emph{Left}: $r=1.5$, for which $b_n$ becomes imaginary at $n=11$. \emph{Right}: $r=1.9$, where $b_n$ becomes imaginary at $n=44$.}\label{fig_bnbulkpropagator15}
\end{figure}

It seems that the larger $r$ is, the larger $n$ for which $b_n^2$ is negative. And, the smaller $\Delta$ is, the larger $n$ for which $b_n^2$ is negative. This is because the subleading and sub-subleading terms become small. We refer the interested reader to Fig. \ref{fig_bnbulkpropagator15} (App. \ref{app:figs}) for a comprehensive collection of our numerical results for the Lanczos coefficients.

In Figures \ref{fig_bnbulkpropagator15}, we display the $b_n$ corresponding to the bulk-to-bulk propagator, where the blue dots represent the results given $C_{p2}(t,r)$ (\ref{eq:Cp2(t)}) and the red dots are $b_n$ for the exact autocorrelation function (\ref{eq:autocorrelation function main}).

\subsection{Lanczos coefficients of the perturbed CFT propagator}

In this subsection, we compare the previous results for the Lanczos coefficients computed from $O(1/r^{2})$ corrections of the bulk-to-bulk propagator~\eqref{eq:lanczpertbulkprop}, with the Lanczos coefficients computed from the T\texorpdfstring{$\overline{\text{T}}$}{}-deformed 2$d$ CFT autocorrelation function found in Eq. (3.13) of~\cite{Chattopadhyay:2024pdj}, where, crucially, the thermal correlation function is only known from a perturbative expansion in the deformation parameter of the holographic boundary theory (which formally corresponds to an $\mathcal{O}(1/r^{2})$ expansion in the bulk propagator at a finite cutoff~\cite{Kraus:2018xrn} for the relevant sign choice in the deformation parameter).

\begin{figure}
         \centering
        \includegraphics[width=0.45\textwidth]{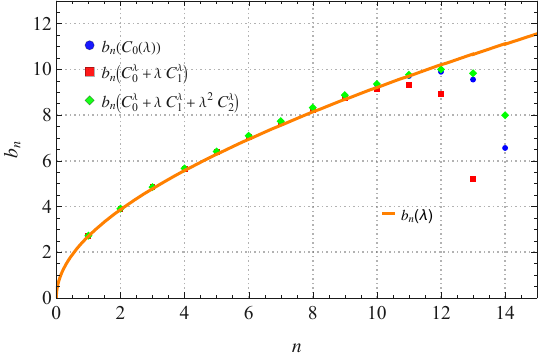}
       \includegraphics[width=0.45\textwidth]{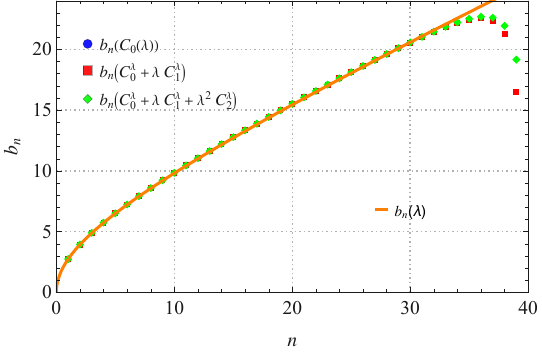}
       \caption{Comparison of the Lanczos coefficients from moments (computed using Eq.~\eqref{eq:autocserieslambda}) for $\beta=2\pi, \Delta=15, \epsilon=1, \mu=1/100$ for $C_{0}(t,\lambda)$ (blue), $C^{\lambda}_{0}(t)+\lambda C_{1}^{\lambda}(t)$ (red) and $C^{\lambda}_{0}(t)+\lambda C_{1}^{\lambda}(t)+\lambda^{2}\lambda C_{2}^{\lambda}(t)$ (green) for \emph{Left}: $\lambda=1/1000$ and \emph{Right}: $\lambda=1/10000$. In the first case, the Lanczos coefficients become imaginary for $n=15$, whereas in the second case they become imaginary for $n=40$. The orange curve represents the analytic Lanczos coefficients for Eq.~\eqref{eq:autocserieslambda}, found in~\cite{Chattopadhyay:2024pdj}.}
       \label{fig_bnttbarprop3}
\end{figure}

The relevant autocorrelation function in~\cite{Chattopadhyay:2024pdj} is given as
\begin{align}\label{eq:ttbarautoc}
    C^{\lambda}_{0}(t)=\frac{1}{\left(\cosh\left(\frac{\pi t}{\beta}\right)\right)^{2\Delta}}+\tilde{\lambda}\left\lbrace f_{1}(\nu_{1},\beta,t)-f_{2}(\nu_{2},\beta,t)\right\rbrace~.
\end{align}
It is important to properly normalize it,
\begin{align}\label{eq:ttbarautocnorm}
    C_{0}(t,\lambda)=\frac{ C^{\lambda}_{0}(t)}{ C^{\lambda}_{0}(0)}~,
\end{align}
where
\begin{align}
    f_{1}(\nu_{1},\beta,t)=\frac{\pi \Delta^{2}}{\left(\cosh\left(\frac{\pi t}{\beta}\right)\right)^{2\Delta+2}}\left(\nu_{1}+\frac{1}{2}\log\left(\cosh\left(\frac{\pi t}{\beta}\right)\right)^{2}\right)~,
\end{align}
\begin{align}
    f_{2}(\nu_{2},\beta,t)=\frac{\pi \Delta^{2}}{\left(\cosh\left(\frac{\pi t}~,{\beta}\right)\right)^{2\Delta}}\left(\nu_{2}+\frac{1}{2}\log\left(\cosh\left(\frac{\pi t}{\beta}\right)\right)^{2}\right)~,
\end{align}
\begin{align}
\tilde{\lambda}=\left(\frac{2\pi}\beta{}\right)^{2}\lambda~,\quad\quad h=\overline{h}=\frac{\Delta}{2}~,
\end{align}
\begin{align}
    \nu_{1}=\frac{1}{4}\left(\frac{4}{\epsilon}+2\gamma-5+2\log(\pi \mu)+3\log(4)\right)~,
\end{align}
\begin{align}
    \nu_{2}=\frac{1}{2}\left(\frac{2}{\epsilon}+\gamma-3+\log(\pi \mu)+2\log(4)\right)=\nu_{1}-\frac{1}{4}~.
\end{align}
Here $\mu$ is a mass scale appearing from dimensional regularization, $\gamma$ is The Euler--Mascheroni constant ($\gamma\approx 0.57721...$), $\epsilon$ is the dimensional regularization parameter $d\mapsto d+\epsilon$, and $\lambda$ is the coupling parameter in the T\texorpdfstring{$\overline{\text{T}}$}{} deformation of the CFT action
\begin{align}
    S(\lambda)=S_{0}+\lambda \int \sqrt{\vert g\vert}\,\textrm{d}^{2}\mathbf{x}\,(T\overline{T})_{0}+\ldots
\end{align}
The autocorrelation function~\eqref{eq:ttbarautoc} comes from the first order correction to the two-point function of primary operators $\mathcal{O}(w_{1},\overline{w}_{1})$, $\mathcal{O}(w_{2},\overline{w}_{2})$ of a $2d$ CFT on a cylinder at finite temperature, which can be expressed as
\begin{align}
\begin{split}\label{app:expvalTTbar}
    \langle \mathcal{O}(w_{1},\overline{w}_{1})\mathcal{O}(w_{2},\overline{w}_{2})\rangle_{\beta,\lambda}&=\langle \mathcal{O}(w_{1},\overline{w}_{1})\mathcal{O}(w_{2},\overline{w}_{2})\rangle_{\beta,0}+\\
    &+\lambda\int \textrm{d}^{2}\mathbf{w}\,\langle T(w)\overline{T}(\overline{w}) \mathcal{O}(w_{1},\overline{w}_{1})\mathcal{O}(w_{2},\overline{w}_{2})\rangle_{\beta,0}~.
    \end{split}
\end{align}
Then, the desired autocorrelation function~\eqref{eq:ttbarautoc} can be obtained as
\begin{align}
\begin{split}
   &C_{0}^{\lambda}(t)=\langle \mathcal{O}(t-i\beta /2)\mathcal{O}(0)\rangle_{\beta,\lambda}=\\
   &=\langle \mathcal{O}(t-i\beta /2)\mathcal{O}(0)\rangle_{\beta,0}+\lambda \langle \mathcal{O}(t-i\beta /2)\mathcal{O}(0)\rangle_{\beta}^{\lambda}
    \end{split}
\end{align}
where $\langle \mathcal{O}(t-i\beta /2)\mathcal{O}(0)\rangle_{\beta}^{\lambda}$ is the first order correction to the autocorrelation function. Then, the normalized autocorrelation function $C_{0}(t)~$\eqref{eq:ttbarautocnorm} has the power-series form in terms of $\lambda$
\begin{align}\label{eq:autocserieslambda}
\begin{split}
    &C_{0}(t,\lambda)=\sum_{i=0}^{\infty}\lambda^{i}C_{(i)}(t)\approx\frac{1}{\left(\cosh\left(\frac{\pi t}{\beta}\right)\right)^{2\Delta}}+\\
    &+\lambda\frac{\pi^{3}\Delta^{2}\sinh\left(\frac{\pi t}{\beta}\right)^{2}}{\beta^{2}\cosh\left(\frac{\pi t}{\beta}\right)^{2(1+\Delta)}}\left(-\frac{4}{\epsilon}-\left(-5+2\gamma + 2\left(\log(16\pi \mu)+\log\left(\left(\cosh\frac{\pi t}{\beta}\right)^{2}\right)\right)\right)\right) +O(\lambda^{2})=\\
    &=\frac{1}{\left(\cosh\left(\frac{\pi t}{\beta}\right)\right)^{2\Delta}}\\
    &+\sum_{k=1}^{\infty}P_{k}(\lambda)\ast\left(\frac{\pi^{3}\Delta^{2}\sinh\left(\frac{\pi t}{\beta}\right)^{2}}{\beta^{2}\cosh\left(\frac{\pi t}{\beta}\right)^{2(1+\Delta)}}\left(-\frac{4}{\epsilon}-\left(-5+2\gamma + 2\left(\log(16\pi \mu)+\log\left(\left(\cosh\frac{\pi t}{\beta}\right)^{2}\right)\right)\right)\right)\right)~,
    \end{split}
\end{align}
where $P_{k}(\lambda)$ is a polynomial of order $k$ in $\lambda$.
Note that the terms with higher order in $\lambda$ in~\eqref{eq:autocserieslambda} come from imposing the normalization at $t=0$~\eqref{eq:ttbarautocnorm}. Also, note that if one does not normalize ~\eqref{eq:ttbarautoc}, then
\begin{align}
    C^{\lambda}_{0}(0)=1+\frac{\pi^{3}\Delta^{2}\lambda}{\beta^{2}}~.
\end{align}
It should be noted that the divergences coming from $\epsilon$ and $\mu$ should be removed from~\eqref{eq:ttbarautoc}, as they are dependent on the regularization scheme of the integral involving the four-point function with insertions of the stress-energy tensor~\eqref{app:expvalTTbar}. As mentioned in Sec.~\ref{sec:disTTbar}, to the best of our knowledge, it is not clear how to appropriately regularize~\eqref{eq:ttbarautoc}, thus, we keep them in our expressions following~\cite{Chattopadhyay:2024pdj}.
We can compute the Lanczos coefficients numerically from the moments of the full normalized autocorrelation function, following Eq.~\eqref{eq:Alt Lanczos} in App.~\ref{app:app1}. Given this information, one can compare the Lanczos coefficients numerically with analytic formulas found in~\cite{Chattopadhyay:2024pdj}, which are shown in Fig.~\ref{fig_bnttbarprop3}. This Fig. serves as a comparison of our results in \eqref{eq:ttbarautocnorm} with analytic expression Eq. (3.17) in~\cite{Chattopadhyay:2024pdj}.

\bibliographystyle{JHEP}
\bibliography{references.bib}

\end{document}